%% file: main.tex
\documentclass[conference,10pt]{IEEEtran}   

\usepackage{amsthm}
\usepackage{algorithm}
\usepackage{algpseudocode}
\usepackage{textcomp}
\usepackage{amsmath}
\usepackage{amssymb}
\usepackage{listings}
\usepackage{stmaryrd}
\usepackage{graphicx}
\usepackage{syntax}
\usepackage{fancyhdr}
\usepackage{textcomp}
\usepackage{flushend}
\usepackage{semantic}
\usepackage{marvosym}
\usepackage{subfigure}
\usepackage[short,nodayofweek,level,12hr]{datetime}  

\usepackage{rotating}

\usepackage{xy}

\theoremstyle{plain} 
\newtheorem{thm}{Theorem}[section]

\newtheorem{definition}[thm]{Definition}

\reservestyle{\command}{\textbf}
\command{let,in,where,otherwise,if,then,else,template}

\newcommand{\var}[1]{\underline{\texttt{\textit{#1}}}}
\newcommand{\func}[2]{\overline{\textsf{\textit{#1}}}^{\textit{\textsf{\begin{tiny}#2\end{tiny}}}}}
\newcommand{\pplus}{+\hspace{-1.0ex}+}
\newcommand{\for}{\begin{large}for \end{large}}

\newcommand{\myrules}[3]{
 #1
 \begin{tabular}{c}{}
  #2\\
  \hline
  #3
 \end{tabular}
}

\newcommand{\myglsentry}[1]{\texttt{\textit{#1}}.\\}
\newcommand{\myglsdesc}[1]{\textit{#1}}

\begin{document}
\xyoption{all}

\title{Narrowing down XML Template Expansion\\ and Schema Validation\\
}

\author{
  \IEEEauthorblockN{Ren\'{e} Haberland}
  \IEEEauthorblockA{
  Technical University of Dresden, Free State of Saxony, Germany\\
  Saint Petersburg State University, Saint Petersburg, Russia\\
  (translation into English from 2007)
  }
}

\maketitle

\thispagestyle{plain}
\pagestyle{plain}

\newcommand{\switchXML}{\lstset{language=XML,tabsize=5}}
\newcommand{\switchXSD}{\lstset{language=XML,tabsize=3}}
\newcommand{\switchXSLT}{\lstset{language=XML,tabsize=3}}
\newcommand{\switchXTL}{\lstset{language=XML,tabsize=3}}
\newcommand{\switchPascal}{\lstset{language=Pascal,tabsize=2}}
\newcommand{\switchProlog}{\lstset{language=Prolog,tabsize=7}}
\newcommand{\switchJava}{\lstset{language=Java,tabsize=3}}
\newcommand{\switchScheme}{\lstset{language=Lisp,tabsize=1}}
\newcommand{\switchHaskell}{\lstset{language=Haskell,tabsize=4}}

\begin{flushleft}
\textbf{Keywords.}
\textit{\textbf{XML, XML-template, XML-schema, language unification, XSLT, DTD, XSD, RelaxNG, PHP, template language, template engine, tree automaton, document instantiation.}}
\end{flushleft}

\IEEEpeerreviewmaketitle

\section{Introduction}
\label{section:Introduction}

This section gives an overview of template expansion and schema validation.
In the beginning, a motivational example is given to introduce the topic of this thesis.
Existing contributions are presented.

\subsection{Motivation}

In XSLT, an application developer is often confronted with two fundamental problems related to XML documents: First, an XSLT-\textit{style sheet} needs to be instantiated.
Second, a given XML document, one part of a more complex XML processing pipeline, needs to be validated against a known XML schema.
The automatic generation of XML documents can be very diverse and may include numerous and even distributed applications, and so is the validation of such.
For example, the instantiation of the following XSLT-stylesheet fragment:

\switchXSLT
\lstset{tabsize=2}
\lstset{basicstyle=\ttfamily\small}
\lstset{numbers=none}
\lstset{numberstyle=\tiny}
\lstset{stepnumber=5}
\lstset{showstringspaces=false}
\lstset{captionpos=b}

\begin{lstlisting}
<xsl:template match="/">
  <foos>
    <xsl:value-of select="//page"/>
    <bar/>
  </foos>
</xsl:template>
\end{lstlisting}

may be transformed into the next document by corresponding queries to the source document:

\switchXML
\lstset{tabsize=2}
\lstset{basicstyle=\ttfamily\small}
\lstset{numbers=none}
\lstset{numberstyle=\tiny}
\lstset{stepnumber=5}
\lstset{showstringspaces=false}
\lstset{captionpos=b}
\begin{lstlisting}
<foos>
  1<bar/>
</foos>
\end{lstlisting}

In order to check validity, a RelaxNG schema may be required, for instance:

\switchXML
\lstset{tabsize=2}
\lstset{basicstyle=\ttfamily\small}
\lstset{numbers=none}
\lstset{numberstyle=\tiny}
\lstset{stepnumber=5}
\lstset{showstringspaces=false}
\lstset{captionpos=b}
\begin{lstlisting}
<element name="foos">
	<group>
		<data type="int"/>
		<element name="bar">
			<empty/>
		</element>
	</group>
</element>
\end{lstlisting}

Hence, apart from an XSLT-stylesheet an XML-schema is required a posteriori, which on the first view does not look similar to the stylesheet.
If only a schema could automatically be derived from a given stylesheet, both processes could be simplified. Even better, a unified view could serve this. Saying this, the total efforts saved is vital, significantly when dropping document meta-information of approximately half with quite a diverse document structure.\\

This approach is tracked by the minimalistic template-language XTL \cite{XTLSpec:2007} used for XML documents.
Within this work, an instantiator and validator for XTL are implemented.
Apart from that, unification, in general, is investigated.
Mainly, the following issues are discussed:

\begin{itemize}
	\item \textbf{Formalisation of instantiation and validation.}

	How can both semantics be formalised?\\[0.3cm]
	Which commonalities and differences do both semantics have?
	
	\item \textbf{Demonstration of examples.}
	
	How do prototypical implementations look?\\[0.3cm]
	Which other properties may be revealed?
	
	\item \textbf{Requirements for unification.}

	Does unification disproportionately restrict either template expansion or schema validation? -- How can such restrictions be overcome, if any.\\[0.3cm]
	Which restrictions are tolerable and which are not?
	
	\item \textbf{Comparison of instantiation and validation.}
	
	Why does a comparison always have to focus on schema validation first?\\[0.3cm]
	What are reasonable criteria for comparison? Which schema languages are suitable for comparison?\\[0.3cm]
	What are the limitations of XTL, and what would be reasonable extensions?\\[0.3cm]
	What about usability towards XTL?
\end{itemize}

\subsection{Preparations}

This section introduces basics and related work as well as bordering disciplines.

\subsubsection{Existing work}

\textbf{RelaxNG-validator.}

\cite{Cla:2007} proposes the algorithm which is used currently in validating RelaxNG-documents.
For instance, based on Clark’s approach, Torben Kuseler \cite{Kus:2005} developed the Haskell-XML-Toolbox \cite{HXT:04}.

\textbf{Transformation of regular expressions into an automaton.}
\cite{Tho:2000} provides practical instructions for the construction in Haskell for a string recognising determined automaton.

\cite{Brz:64} presents the construction of so-called \textit{Glushkov}-automata, which are stepwise built up by deriving transitions from previously set states.

The paper \cite{Ant:00} widens the definitions of a partial derivation according to mathematical analysis and proposes a transition calculation parametrised in comparison to \cite{Brz:64}.
	
\textbf{Tree Automata.}
In \cite{Mur:1999} and \cite{Mur:1995}, Makoto Murata introduces tree automata's syntax applied to XML documents.
Essential terms, particularly tree grammars and languages, as well as their properties, are provided.

\cite{Chi:2000} discusses the transformation of schema documents into regular tree expressions. The construction of tree automata is discussed in the context of database applications.

\cite{Com:2005} represents a compendium of techniques on tree automata and their applications.

\textbf{Comparisons and Field Studies.}
Murata, Lee and Mani \cite{Mur:2001} introduce a classification of XML-schema languages.
In \cite{Lee:2000}, Lee and Chu classify expressibility for selected schema languages.

Rahm and Bernstein \cite{Rahm:01} propose a classification of schema-matchers.

\textbf{Others.}
Both \cite{All:1988} and \cite{Ten:1976} provide a survey on denotational semantics.
\cite{Tho:1999} gives an introduction to Haskell. Additional material on Haskell and functional programming may be found in \cite{Pey:1987} and \cite{Pey:1992}.

\subsubsection{Related Work}
template engines are introduced in \cite{Par:2006} and \cite{Par:2004}, where moreover \cite{Par:2004} also provides a closer look from a Model-View-Controller perspective.
Evaluation improvements are discussed in  \cite{Ken:2005} on template expansion and in \cite{Bal:2004} on schema validation.
\cite{Ata:2004} deals with XML typing in general, and particularly with \textit{type isomorphism} in Haskell. The overall meaning of polymorphism in terms of XML is discussed in \cite{Hos:04}.\\

Lazy evaluation of XML-documents, refactorings of functional programs \cite{Coe:1992}, \cite{Lae:2000}, \cite{Tho:2001}, \cite{Cza:1999} and monads/arrows \cite{Lec:2005}, \cite{Hug:1991}, \cite{Erw:2004}, \cite{Hud:2002},
\cite{Hug:2000}, \cite{Wad:1995} are closely related to Haskell.
Particularly, \cite{Ken:2005}, \cite{Nog:2002} deals with lazy parsing of XML-documents, and \cite{Bol:2006} deals with XML-document updates.

\subsubsection{Cross-bordering disciplines}
Besides the just mentioned topics, this work's topic (cross-)borders with further disciplines:
\begin{itemize}
	\item document and schema transformation,
	\item parsing,
	\item functional programming and
	\item XML data binding.
\end{itemize}

\subsection{Structure of this work}
Section 2 introduces basics, e.g. instantiation and validation.
Section 3 investigates instantiation and validation as function.
New requirements for programs to be developed and to a unification process are formulated.

Semantics for both instantiation and validation of XTL is defined in section 4.
Properties are discussed.
The desired data model is introduced which is to simplify semantics.
Section 5 presents the software implementation of the previously defined semantics and data models in Haskell.
A design for an object-oriented implementation to be continued is discussed.
According to previously defined criteria, section 6 compares template expansion and schema validation in general and specifically for a selected set of XML schema languages.
The focus is the unification of both.
Special attention is paid to the semantics and syntax of schema languages.
Complexity is considered, primarily, w.r.t. practical needs.\\

"`XML-Schema"' denotes W3C's proposed XML-schema language, which is often better known as "`XSD"'.

All terms are described in the glossary attached and in the following text before its first occurrence.

If a term appears in italics for the second time, this indicates its glossary meaning may deviate depending on its context.
Used acronyms appear in italics on their first occurrence in the text.
References to the appendices are linked in the text.
Only stereotypes/patterns in sections sect.\ref{section:Analysis}, \ref{section:Implementation} may appear in italics and do not require further explanation, which is following roles as proposed by \cite{fowl:2000} and \cite{ker:2006}.

At the beginning of each section, a short overview and methodology are provided.
Questions raised are either answered immediately or within the following section(s) and are indented.
References, except mentioned differently in the text, are always related to prior statements.

\newpage
\section{Basics}
\label{section:Basics}
This section introduces instantiation and validation, defines basic terms with illustrative examples, and introduces the template-language XTL and theoretical foundations.
The modelling of XML documents is described based on a tree grammar as well as the translation.

\subsection{Instantiation}
\label{section:BasicsInstantiation}

\begin{figure}[h]
  \begin{center}
 \begin{minipage}{2cm}
  \xymatrixrowsep{15pt}
  \xymatrixcolsep{15pt}
  \xymatrix{
    *+[F]\txt{XML\\Template} \ar@<1ex>@2{=>}[rr] \ar@<1ex>@2{-->}[rr];[]  && *+[F]\txt{Template\\Engine} \ar@<1ex>@2{=>}[rr] \ar@<1ex>@2{-->}[rr];[] && *+[F]\txt{XML\\Instance} \\
    && *+[F]\txt{Instantiation Data} \ar@<1ex>@2{=>}[u] &&
  }
  \end{minipage}
  \end{center}
  \caption{Instantiation and Validation}
  \label{figure:InstantiationNValidation}
\end{figure}
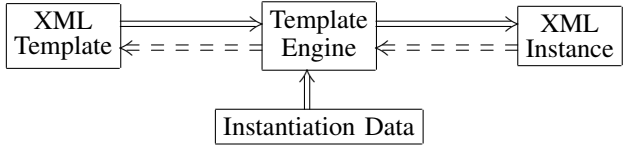

The \textit{instantiation} of XML documents is when a \textit{template engine} generates an \textit{instance} of a \textit{template} based on some \textit{source language} with some \textit{instantiation data}.
The result is called an \textit{instance} and follows a \textit{target language}'s constraints (see fig.\ref{figure:InstantiationNValidation}).
At this moment, it is agreed templates are present in XML.
It is further agreed on instances have to be in well-formed XML format.\\


A template consists of an arbitrary number of \textit{tags} and \textit{slots} \cite{Par:2006}, which also may be nested.
In general, \textit{slots} have to be unique among all other nodes.
Slots bind instantiation data, which may differ in general.
Instantiation data maybe XML or form a relation.
A slot may at most refer to one source.
Within an instantiation step, common tags are copied to the instance document, where the template engine evaluates a slot.
Due to its tree structure, XML documents are usually processed top-down.
So, the instantiation of templates is processed in pre-order (cf. sect.\ref{section:Trees}).
W.l.o.g. it is agreed that during an instantiation step, access to already instantiated nodes is prohibited due to the transparency of the chosen model (see sect.\ref{section:TreeAutomataMatchers}).

Slots are interfaces. Its purpose is to load data from external repositories into the instance.
The loading is controlled by queries that come from the instantiation data.
Queries are part of the slots. Neither its syntax nor its semantic have to follow XML or any a priori rules whatsoever. For example, a slot may contain queries in \textit{XPath} or \textit{SQL}.
Some source languages allow more precise queries than XPath does.
For instance, the returned value has to be formatted.
Settings, e.g. for XTL, will be interpreted by \textit{Placeholder-Plugins} (\textit{PHPs}; see sect.\ref{section:mainXTL}).
That is why every template (e.g. a PHP website) has to follow previously agreed interfaces that control instantiation data access.\\

For distinguishing a document instantiation from, for instance, object instantiation, the term \textit{template expansion} is sometimes used \cite{Par:2006},\cite{Par:2004}. The term "`expansion"' tries to illustrate something is replaced by something more extensive.
However, the replacement does not always have to be longer.
The opposite may be true.
A slot is replaced by an arbitrary number of nodes.
The replacement can be interpreted as reduction rather than expansion if the node(s) to be inserted are in total shorter than the slot.
Hence, it is agreed that both instantiation and template expansion may be used as synonyms.\\

For demonstration purpose, the instantiation of the following XSLT-stylesheet with XPath \cite{DeR:1999} as embedded query language is shown:

\switchXSLT
\lstset{tabsize=1}
\lstset{basicstyle=\ttfamily\small}
\lstset{numbers=right}
\lstset{numberstyle=\tiny}
\lstset{stepnumber=1}
\lstset{showstringspaces=false}
\lstset{captionpos=b}

\begin{lstlisting}
<?xml version="1.0" encoding="UTF-8"?>
<xsl:stylesheet xmlns:xsl="http://..." ..>
  <xsl:template match="/">
    <publications>
      <xsl:for-each select="//book">
        <title>
          <xsl:value-of select="@title"/>
        </title>
      </xsl:for-each>
    </publications>
  </xsl:template>
</xsl:stylesheet>
\end{lstlisting}

This template generates (once applied with appropriate instantiation data) for each \texttt{<book/>}-node \texttt{<title/>}-nodes with matching titles as text in its children nodes directly underneath \texttt{<publications/>}.
Line 4 matches against the top-level element node --- this means the whole document is considered.
In line 6, all \texttt{<book/>}-nodes starting from the top of the document are determined.
Now suppose the associated source document is:

\switchXML
\lstset{tabsize=2}
\lstset{basicstyle=\ttfamily\small}
\lstset{numbers=right}
\lstset{numberstyle=\tiny}
\lstset{stepnumber=1}
\lstset{showstringspaces=false}
\lstset{captionpos=b}

\begin{lstlisting}
<bibliography>
  <book author="Simon Thompson" 
    title="Haskell - The Craft ..."/>
  <magazin title="Informatik-Spektrum..."/>
  <book author="Joshua Kerievsky"
    title="Refactoring to Pa ..."/>
  <url title="XSD specification 1.0"/>
</bibliography>
\end{lstlisting}

In line 6 of the stylesheet, the nodes from lines 2 and 4 matches.
Each of these nodes is successively considered the source document itself (see lines 7-9).
While processing, first, the template element node \texttt{<publications/>} is copied, and second the instantiation of its child nodes is continued.

\switchXSLT
\lstset{tabsize=2}
\lstset{basicstyle=\ttfamily\small}
\lstset{numbers=right}
\lstset{numberstyle=\tiny}
\lstset{showstringspaces=false}
\lstset{captionpos=b}

\begin{lstlisting}
<xsl:for-each select="//book">
  <title>
    <xsl:value-of select="@title"/>
  </title>
</xsl:for-each>
\end{lstlisting}

instantiates for the first element of the set:

$C_{1}$ = 
\texttt{[\parbox[t]{17cm}{
 <book title="\/Haskell..." \ ... />,\\<book title="\/Refact..." \ ... />]}}

the node:

\switchXSLT
\begin{lstlisting}
<publications>
  <title>Haskell...</title>
  <xsl:for-each select="//book">
    <title>
      <xsl:value-of select="@title"/>
    </title>
  </xsl:for-each>
</publications>
\end{lstlisting}

In a second step the remaining set $C_{2} = $\texttt{[<book title="Refactoring..." \ ... />]}
is applied to the nested loop. The next node results:

\switchXSLT
\begin{lstlisting}
<publications>
	<title>Haskell...</title>
	<title>Refactoring...</title>
	
	<xsl:for-each select="//book">
		<title>
			<xsl:value-of select="@title"/>
		</title>
	</xsl:for-each>
</publications>
\end{lstlisting}

with $C_{3} = \emptyset$ as the remaining empty set.
The loop condition does not hold anymore for the instantiation of loops in XSLT (cf. \cite{Cla:1999},\cite{Wad:2000}). So, the nested loop is skipped. Since no more following nodes exist, the instantiation terminates and returns this node:

\switchXML
\lstset{tabsize=2}
\lstset{basicstyle=\ttfamily\small}
\lstset{numbers=none}
\lstset{numberstyle=\tiny}
\lstset{showstringspaces=false}
\lstset{captionpos=b}

\begin{lstlisting}
<publications>
  <title>Haskell - The Craft ...</title>
  <title>Refactoring to Patterns</title>
</publications>
\end{lstlisting}

\subsection{Validation}

\textit{Validation} checks if a given XML document is an instance of a template (see fig.\ref{figure:InstantiationNValidation}).
If the answer is "`yes"', then the given document is "`valid"', otherwise not.
Templates generate instances, which schemas may describe.
That means template and schema de facto describe the same instance document.
If we try to unify both processes, then both descriptions also need to be unified somehow.
Otherwise, not the same template or corresponding schema may be expressed.
One big problem in doing so is that both syntax and semantic differ in both cases.
The template engine would need to be reconfigured, so documents are not expanded but schema-validated.
In practice, this means validation is triggered rather than a template engine.\\

The \textit{schema} describes the set of all valid instances of a template.
Schemas often are tagged and, as such, are XML-dialects.
Represents of such are \textit{RelaxNG} \cite{Cla:2001} and \textit{W3C's XSD} \cite{Fal:2004}, \cite{Tho:2004}, \cite{Bir:2004}.
\textit{DTD} represents a schema language for XML documents, but it is not XML.
Schema languages in XML can be processed the same way templates are processed (cf. sect.\ref{section:Analysis}).
XML-schemas consist of \textit{literals}.

Literals are childless element and text nodes as well as composed element nodes. Composed element nodes have at least one child node and any number of commando tags in arbitrary order as children.
\textit{Commando tags} denote either a loop or a selection and may be decorated with \textit{constraints}.\\

Different to instantiation, validation does not consider instantiation data.
If instantiation data was complete, an instantiation could be performed.
In that case, validation would be equal to checking the given instance document's equality with the instantiated document.
If, however, instantiation was not complete, then conclusions would not always be correct.
Validation could then be reduced to the problem of \textit{document reconstruction} \cite{Hab:2006}.
In practice, however, document reconstruction would induce too many burdensome restrictions, s.t. instantiation of a template would have to be invertible, for instance.
That is why it is better not to make any assumption about the instantiation data instead.\\

%
The validation of instances may be considered as a \textit{word problem} (cf. \cite{Chi:2000}, \cite{Mur:1999}).
In that case, the instance documents act as word and the template document as grammar, which generates all valid instance documents.
In case the template contains cycles, the set of all valid instances becomes infinite, and instantiation becomes one possible derivation for grammar, which is the template document.
If the document is interpreted as a term, the instance document represents a standard form if all slots are reduced until only element and text nodes remain.

\subsection{XTL}
\label{section:mainXTL}

Until now, schemas could not (satisfactory) be represented in the language of templates.
A schema of some template language is too often too complicated to read or too complex.
That is why XSLT templates require many helper functions and rewritings.
Because of the properties an instance has to have, schemas can barely be automatically be inferred (cf. sect.\ref{section:ComparisonMain}).
Despite that disadvantage, schema languages are still being used that do not unify smoothly with templates.
As already mentioned, the differentiation between template and schema languages is why there is a considerable increase in maintenance and heterogeneous program systems.
Another disadvantage may be additional efforts in learning new schema languages.

That is why the following points are required to unify template and schema documents:

\begin{center}
\begin{itemize}
 \item keep schemas short and simple
 \item each element in the template shall correspond with a similar element in the schema, and vice versa.
\end{itemize}
\end{center}

The XML-template language XTL was defined as part of the \textit{SNOW}-project \cite{Snow:2005}. One of its goals was to investigate, if tractable at all, whether documents may be reused for template expansion and schema validation. XTL in version 1.0 currently counts seven intrinsic tags to be explained in more detail next. The comparison between template and schema nodes is taken out in sect.\ref{section:ComparisonMain}.\\

\textbf{xtl:attribute}\\[0.3cm]
The insertion of new attributes can be expressed like this:

\begin{center}
\texttt{<xtl:attribute name="}\textit{name}\texttt{" \/ select="}\textit{expression}\texttt{"/>}
\end{center}

The tag does not have children and is under an element node \cite{XTLSpec:2007}.
Both \lit{\texttt{name}} and \lit{\texttt{select}} have to be specified as attributes.
Attributes always have to be under an element node.
In case the same attribute name has already been defined, the new value \textit{expression} replaces the attribute.
The definition order of attributes is arbitrary and still defines the same element node.
W.l.o.g. it is agreed that attributes are in \textit{canonised} form.
It implies attribute assignment pairs are ordered ascending by the attribute name in lexicographic order --- attributes reference instantiation data by \lit{\texttt{select}}.
In JXPath as template engine and PHP as a template language, the \textit{expression} denotes a path expression.

\switchXTL
\lstset{tabsize=2}
\lstset{basicstyle=\ttfamily\small}
\lstset{numbers=none}
\lstset{numberstyle=\tiny}
\lstset{showstringspaces=false}
\lstset{captionpos=b}
\begin{lstlisting}
<book id="1">
 <xtl:attribute name="author"
  select="//book[position()==1]/@author"/>
 <xtl:attribute name="id" select="999"/>
</book>
\end{lstlisting}

For this example, sect.\ref{section:BasicsInstantiation} as instantiation data and JXPath as PHP, the node \texttt{<book author="\/Simon Thompson"\/ id="999"/>} is instantiated.
The first attribute definition inserts a new entry to the attributes list.
Since XPath maps integers on themselves \cite{DeR:1999}, the new attribute's value is \texttt{"999"}.\\

\textbf{xtl:text}\\[0.3cm]
The childless tag for text inclusion is

\begin{center}
\texttt{<xtl:text select="}\textit{expression}\texttt{"/>}
\end{center}

\textit{Expression} is passed to the placeholder plugin during instantiation, which will handle it further, such as XPath-expression.
The resulting nodes list is converted by implicit coercion in XPath \cite{DeR:1999} into a string, which finally is concatenated.
The concatenated text replaces the tag for XTL-text-inclusion.

The expansion of two neighbouring \lit{\texttt{xtl:text}} nodes as children is of interest, especially for validation.
This interest comes from asking how to split a common string best when there are no markers that indicate boundaries.
That is why the separation may become ambiguous.
Hence, strategies are wanted, which allow recognising \lit{\texttt{xtl:text}} nodes uniquely.
In contrast to this, \lit{\texttt{xtl:attribute}}-nodes do not have this problem.\\

\texttt{<xtl:text select="/"/>} instantiates the empty string for the example from sect.\ref{section:BasicsInstantiation}, because the source document is traversed in pre-order and occurrences of text nodes are accumulated and concatenated.\\

\textbf{xtl:include}\\[0.3cm]
The following childless tag can achieve an arbitrary element node enriched by instantiation data:

\begin{center}
\texttt{<xtl:include select="}\textit{expression}\texttt{"/>}
\end{center}

The PHP returns either one well-formed element node or none.
If multiple nodes match with \textit{expression}, PHP chooses only the first occurrence and drops all others \cite{XTLSpec:2007}.\\

The attribute \textit{expression} equals \texttt{"//url"} returns for the example from sect.\ref{section:BasicsInstantiation} the element node:

$$\texttt{<url title="XSD specification 1.0"/>}.$$

\textbf{xtl:if}\\[0.3cm]
\label{section:XTLIf}
Conditions in XTL have this form:

\begin{center}
\texttt{<xtl:if select="}\textit{expression}\texttt{"\/>...</xtl:if>}
\end{center}

If \textit{expression} evaluates to "`true"', then the evaluation continues with its children \lit{\texttt{...}}.
Otherwise, the children nodes of \lit{\texttt{xtl:if}} are dropped, and evaluation proceeds with the following siblings.\\

An example determining the second book of a bibliography, if any, looks like this:

\switchXTL
\lstset{tabsize=2}
\lstset{basicstyle=\ttfamily\small}
\lstset{numbers=none}
\lstset{numberstyle=\tiny}
\lstset{showstringspaces=false}
\lstset{captionpos=b}
\begin{lstlisting}
<xtl:if select="//book[position()==2]">
	<xtl:include select="//book"/>
</xtl:if>
\end{lstlisting}

For the element node \texttt{<bibliography/>} from sect.\ref{section:BasicsInstantiation} it retrieves the following node:

\begin{lstlisting}
<book author="Joshua Kerievsky"
  title="Refactoring to Patterns"/> .
\end{lstlisting}

\textbf{xtl:for-each}\\[0.3cm]
The tag for cycles is

\begin{center}
\texttt{<xtl:for-each select="}\textit{expression}\texttt{"\/>...</xtl:for-each>}
\end{center}

The evaluation of \textit{expression} by the PHP returns a nodes list. This list is iterated successively, and each node is propagated as context for the instantiation of children nodes (cf. sect.\ref{section:XTLIf}).
The instantiation of children nodes with the first element from the evaluation of \lit{\texttt{select}} returns an instantiated children lists.
The same goes for the ongoing instantiation.
Those are linked together until no more context exists.

The use of a context does not restrict the reachability of axes because every node remains reachable.
For example, nodes located in the upper section of a source document may by XPath \cite{DeR:1999} be addressed using \lit{\texttt{ancestor}}.
Contexts are using also used in XSTL \cite{Cla:1999} for the sake of usability.\\

\textbf{xtl:macro}\\[0.3cm]
Macros are defined in XTL as following:

\begin{center}
\texttt{<xtl:macro name="}\textit{ncname}\texttt{"\/>...</xtl:macro>}
\end{center}

Macros are \textit{symbols}, which bind arbitrary sequences of command, element and text nodes, except further macro definitions.
The macro is defined by a fully qualified name \textit{ncname} which must be unique among all macros within a template.
In a template, all macro definitions must be contiguous and before a sequence of non-macro definitions right underneath the top element node \cite{XTLSpec:2007}.\\

\textbf{xtl:call-macro}\\[0.3cm]

The macro call without children is defined as

\begin{center}
\texttt{<xtl:call-macro select="}\textit{ncname}\texttt{"\//>}
\end{center}

A macro call is similar to a function call without parameters. The macro call retrieves a list of element nodes.
During expansion and validation, the macro call is replaced by the right-hand side of element nodes from its definition. Recursive calls are permitted. Termination conditions need to be specified within \lit{\texttt{select}}-expressions in XTL command tags.\\

\textbf{Summary:}\\

The following fragment of a XTL-schema demonstrates validation with macros and cycles.

\switchXTL
\lstset{tabsize=2}
\lstset{basicstyle=\ttfamily\small}
\lstset{numbers=left}
\lstset{numberstyle=\tiny}
\lstset{stepnumber=1}
\lstset{showstringspaces=false}
\lstset{captionpos=b}

\begin{lstlisting}
<xtl:macro name="TDs">
 <td>
  <xtl:text select="@title"/>
 </td>
 <td>
  <xtl:text select="@author"/>
 </td>
</xtl:macro>

<table col="#FF0000">
 <th>
  <td>Title</td>
  <td>Author</td>
 </th>
	
 <xtl:for-each select="//book">
  <xtl:if select="position() mod 2=0">
   <tr col="#333300">
    <xtl:call-macro name="TDs"/>
   </tr>
  </xtl:if>
  <xtl:if select="position() mod 2=1">
   <tr>
    <xtl:call-macro name="TDs"/>
   </tr>
  </xtl:if>
 </xtl:for-each>
 <tr>
  <td>XSD specification 1.0</td>
  <td/>
 </tr>
</table>
\end{lstlisting}

A corresponding well-formed instance would be:

\switchXML
\lstset{tabsize=2}
\lstset{basicstyle=\ttfamily\small}
\lstset{numbers=left}
\lstset{numberstyle=\tiny}
\lstset{stepnumber=1}
\lstset{showstringspaces=false}
\lstset{captionpos=b}
\begin{lstlisting}
<table col="#FF0000">
 <th>
  <td>Title</td>
  <td>Author</td>
 </th>
 <tr>
  <td>Haskell - The Craft of Functio ...</td>
  <td>Simon Thompson</td>
 </tr>
 <tr col="#333300">
  <td>Refactoring to Patterns</td>
  <td>Joshua Kerievsky</td>
 </tr>
 <tr>
  <td>XSD specification 1.0</td>
  <td/>
 </tr>
</table>
\end{lstlisting}

\texttt{}\\

First, validation stores the macro \lit{\texttt{TDs}} defined on lines 2-7 and continues from line 10.
This line matches with line 1 of the instance. The child node at lines 11-14 entirely matches with the child node of lines 2-5 from the instance.
At lines 16-27, there is a non-deterministic decision to be made on whether child nodes match for a cycle.
It is impossible to determine how much further a cycle needs to be unrolled to match the schema in the instance document at line 6 without checking the following nodes.
If the cycle \lit{\texttt{xtl:for-each}} is left too early or too late, then the \texttt{<tr/>}-node from lines 28-30 may not exactly match with the expected number of nodes from the instance.
According to the instance document, \lit{\texttt{xtl:for-each}} may have no, one, two or three iterations.
It is necessary to continue on fails with alternatives, if there are any, to guarantee a correct validation
Only after all alternatives fail, validation fails.

In the previous example, the correct number of iterations, which is two, is guessed, s.t. both \texttt{<tr/>}-nodes from lines 6-9 and 10-13 from the instance match consecutively with lines 7-16 from the schema.
\lit{\texttt{select}}-expressions from the conditions are ignored here.
Consequently, shuffling \texttt{<tr/>}-nodes in the instance document, but also a sequence of colored \texttt{<tr/>}-nodes lead to a true validation.

The instantiation of the cycle is interpreted as unrolling all books from the instantiation data. Instantiation continues with the following tags. In analogy to that, validation tests if the last node of the instance from lines 14-17 matches \texttt{<tr/>}-nodes from the schema at lines 28-31. As this is the case and no other nodes follow, the validation quits successfully.

\subsection{Theoretic Foundations}
\label{section:TreeAutomataMatchers}

This section introduces theoretic foundations.
First, the tree-structured data model "`hedge"' is defined, then regular tree grammars and languages. Later a short overview is given on regular automata. Examples illustrate definitions.

\subsubsection{Trees}
\label{section:Trees}

The theories introduced later in this section may be applied to tree-structured objects, like XML schemas, XML instances and instantiation data.
XML documents can be represented as trees since nodes of an XML document are in a hierarchy, and there are no cycles included all through ascending edges leading from leaves to the root element node.
Keys as used to describe relations of a schema are not of interest regarding a schema's syntax.

XML documents are multi-way trees with child-rich elements as nodes and childless elements, and text nodes as leaves.
Attributes of element nodes may be transformed into element nodes with new child nodes that represent such attributes.
For example, the node \texttt{<a id="1"/>} can be transformed into \texttt{<a><id>1</id><sep/></a>}, where \texttt{<sep/>} separates attribute nodes from original child nodes not representing former attributes.\\

Before trees and their properties are introduced, their practical meaning is recapitulated.

XML documents can either be interpreted as unstructured, namely as text, or structured.
By interpreting XML as unstructured text, important information vanishes, for instance, newlines or ordering.
A replacement of element nodes in trees by symbols returns trees.
Replacements may cause shorter nodes sequences.
That is the reason why existing string-grammars are going to be extended and reused (see sect.\ref{section:TreeGrammars}, cf. \cite{Mur:1999}, \cite{Com:2005}).

In structured XML-interpretations tags emphasise text regions.
Tags denote meta-information and are not part of the document text.
Hence, XML documents are predestined for structural interpretation, for both instantiation with command tags, and validation with a common element and text nodes.
For example, the unstructured interpretation of:

 \texttt{<a>hello<b>world<c/></b></a>}
 
does not allow a simple processing neither by a user nor by a program. The interpretation of its structure makes access to the documents' content easy.\\[-0.3cm]

\definition{A \textit{hedge} (after Murata \cite{Mur:1999}) is defined over a finite symbol set $\Sigma$ and a finite variable set $X$ as following:}

\begin{center}
\begin{tabular}{c l}
	$\varepsilon$	 & .. the empty hedge\\
	\texttt{x}	 & .. variable with \lit{\texttt{x}}$\in X$\\
	\texttt{a<u>}	 & .. element node \lit{\texttt{a}}$\in \Sigma$ with hedge \lit{u}\\
	\texttt{u v}	 & .. concatenation of hedges \lit{u} and \lit{v} \qquad $._\square$
\end{tabular}
\end{center}

The two XML-element nodes \texttt{<a/>} and \texttt{<b><b/>x</b>} are representable as hedge
\texttt{a<$\varepsilon$>b<b<$\varepsilon$>x>} with the symbol set
$\Sigma$=$\{a,b\}$ and variable set $X$=$\{x\}$.
$\Sigma$ does not oblige any restriction.
Element names may have any prefix and suffix.
Hence, every XML document is representable as a hedge, even XTL-templates and XSLT stylesheets.

Based on this model, each navigation operator over trees can be defined (see \cite{Mur:1995}, \cite{Hab:2006}, \cite{DeR:1999}).
For instance, the function \texttt{subtree} \cite{Mur:1995} distinctively determines a predecessor node for a given number-encoded path.

\subsubsection{Regular Tree Grammars}
\label{section:TreeGrammars}

As previously mentioned, \textit{string grammars} are not sufficient to describe trees.
Hedges do not only grow in width by concatenation, but they also grow into depth by insertion of child nodes.
\textit{Regular} tree grammars are suggested as one way to resolve this issue.

\definition{A regular hedge-grammar (\textit{RHG}, after Murata \cite{Mur:1999}) is a grammar $G=(\Sigma,X,N,P,n_{f})$ with}

\begin{center}
\begin{tabular}{r l}
	$\Sigma$ & .. finite symbol set\\
	$X$		 & .. finite variable set\\
	$N$		 & .. finite non-terminal set\\
	$P$		 & .. production rules\\
	$n_{f}$  & .. final state set.
\end{tabular}
\end{center}

A production rule from $P$ has either the form $n \rightarrow x$, where $n \in N$, $x \in X$, or $n \rightarrow$\texttt{ a<}$r$\texttt{>}, where \texttt{a} $\in \Sigma$ and $r$ is a regular expression over $N \cup X$ herewith.
$n_{f}$ denotes a regular expression over $N$, which is accepted by the grammar.
\qquad ${}_\square$

\definition{Regular expression over hedges.}

Let $r, r_{1}, r_{2}$ be regular expressions over a finite set of non-terminals.
Then the following expressions are also regular:

\begin{center}
\begin{tabular}{r l}
 $r_{1} \cdot r_{2}$	& .. concatenation\\
 $r_{1} \mid r_{2}$	& .. alternative\\
 $(r)$			& .. parantheses\\
 $r^{*}$		& .. repetition \qquad ${}_\square$
\end{tabular}
\end{center}

Regular expressions are better for XML-schemas than relations or rigid associations, as demonstrated by \cite{Chi:2000}.
Relations do have a fixed amount and order of arguments. Contrary to this, regular expressions allow short but flexible expressions.

The form of its productions also demonstrates the regularity of RHS.
The set of valid instances for a XTL-schema:

\switchXTL
\lstset{tabsize=2}
\lstset{basicstyle=\ttfamily\small}
\lstset{numbers=none}
\lstset{numberstyle=\tiny}
\lstset{showstringspaces=false}
\lstset{captionpos=b}
\begin{lstlisting}
<xtl:if select="//checked">
 <a>
  <xtl:for-each select="//person">
   <x/>
  </xtl:for-each>
 </a>
</xtl:if>
\end{lstlisting}

recognises the regular tree language 

$$L(G)=\{\varepsilon, \texttt{a}<\varepsilon >,\texttt{a}<x >, \texttt{a}<xx>, \texttt{a}<xxx>, ... \}$$

(cf. sect.\ref{section:RegularTreeLanguages}).
A corresponding grammar $G$ is $G=\{\Sigma, X, N, P, n_{1}?\}$ with $\Sigma$=$\{a\}$, $X$=$\{x\}$, $N$=$\{n_{1},n_{2}\}$ and 
%
\begin{tabular}[t]{l l}
 P: & $n_{1} \rightarrow a<n_{2}^{*}>$\\
    & $n_{2} \rightarrow x$ .
\end{tabular}

RHG differs from string-grammars in variables, which can be considered terminals and a final state set, describing the accepted language.
Productions are similar to regular productions, whereas the right-hand side of each rule may contain further non-terminals of the form $e_{0} \cdot e_{1} \cdot ... \cdot e_{n}$, where $e_{n}$ is a non-terminal and all other $e_{j}$ for $\in [0 .. (n-1)]$ denote terminals.
Terminals stand solely or to the left of a non-terminal.

Derivations for regular tree grammars work similarly to string-grammars.
Non-terminal symbols are derived left-to-right until the derived expression does no more contain non-terminals.

The derivation of a regular tree grammar corresponds to the instantiation of an XTL-template.

The derivation of regular grammars may be non-deterministic due to multiple rules for selection.
If there exist at least two derivations of a non-terminal, then an implemented automaton may not decide in general without a stack (cf. \cite{Chi:2000}, \cite{Tho:2000}, \cite{Brz:64}).\\

Regarding grammars, the question concerning expressibility emerges, for instance, if a tree grammar is context-free or context-sensitive.
Context-free grammar does not leave open questions from a practical standpoint (see sect.\ref{section:Analysis}).
However, context-sensitive grammars are not that easy.
That is why, often, contextual information is transmitted by a different mechanism.
Turing-mighty tree grammars are not further considered here.\\

Murata, Lee, and Mani \cite{Mur:2001} categorise regular tree grammar's expressibility as following:

$$local \subseteq single-type \subseteq ranked-competing \subseteq regular$$

The ordering is due to the level of non-determinism of the production rules (\textit{ambiguity}).
Local tree grammars are the weakest. Regular tree grammars are the most powerful.
More powerful grammars entirely contain weaker grammars. Moreover, more powerful grammars always contain non-empty cases which are not covered by the weaker grammars \cite{Mur:2001}.
Four tree grammars may be categorised as follows:

\begin{description}
	\item[Local:] A terminal may not occur in more than one rule.
	
	\item[S.-t.:] Non-terminals of children nodes do not compete with each other, which means $\pi(e_{i}) \cap \pi(e_{j})$ is empty for each two distinct nodes $e_{i}$ and $e_{j}$.
        $\pi(e_{j})$ determines the set of possible beginnings for some node $e_{j}$.
\end{description}
\begin{description}
	\item[R.-c.:] A hedge $r$ is uniquely decomposable. 
	It means $\forall U,V,W \in N:$  
$r \nvdash_{*}  UAV$ and $r \nvdash_{*}  UBW$ for competing non-terminals $A$, $B$ in $r$.
	
	\item[Reg.:] All regular grammars which are not ranked-competing.
\end{description}

\texttt{}\\[-0.3cm]
The following two examples explain the membership of a specific tree grammar:

\begin{description}
		\item[Ex 1] Let the grammar $G_{1}$ have the following productions:\\
		
		$Doc \rightarrow doc(Para1, Para2^{*})$\\
		$Para1 \rightarrow para(Pcdata)$\\
		$Para2 \rightarrow para(Pcdata)$\\
		$Pcdata \rightarrow pcdata \ \varepsilon$\\
		
		$Para1$ and $Para2$ compete with each other in the first production. Hence, for a ranked-competing tree grammar no $U$, $V$, $W$ may exist, s.t. $r \vdash_{*}  U\ Para1 \ V$ and $r \vdash_{*}  U\ Para2 \ W$, where $r = Para1 \ Para2^{*}$. Since $U$ must be different in the first derivation from the second, we just found a contradiction. Because no other decompositions exist, $G_{1}$ is ranked-competing and therefore regular too.\\

		\item[Ex 2] Let the grammar $G_{2}$ have this productions:
		
		$Doc \rightarrow doc(Para1^{*},Para2^{*},Pcdata)$\\
		$Para1 \rightarrow para(Pcdata)$\\
		$Para2 \rightarrow para(Pcdata)$\\
		$Pcdata \rightarrow pcdata \ \varepsilon$\\
		
		$Para1$ competes with $Para2$ in $$r = Para1^{*} \ Para2^{*} \ Pcdata$$.
		Hence, no $U$, $V$, $W$ exist, s.t.
		$r \vdash_{*}  U\ Para1 \ V \quad$ and $\quad r \vdash_{*}  U\ Para2 \ W$.
                But, there exists the valid decomposition $U=\varepsilon$, $V=V' \ Pcdata$, $W=Pcdata$.
		That is why this grammar is not ranked-competing.
\end{description}

\subsubsection{Regular Tree Languages}
\label{section:RegularTreeLanguages}

\textit{Regular Tree Languages} are formal languages generated by RHG, hedge-regular expressions and deterministic and non-deterministic hedge-automata (see \cite{Mur:1995}).

The transformation between the models mentioned above is very similar to those in string-based formal languages.
Hedge and tree models and grammar and expressions are from a computability perspective equivalent --- this is shown in \cite{Mur:1995}.
Trees are specialised hedges, and a hedge is a tree with an empty root node.

It is worth mentioning that element and text nodes can be treated nearly the same (cf. sect.\ref{section:Trees}).
As shown earlier, attributes may be simulated by element nodes.
Examples to each of the mentioned models in this section may also be found in \cite{Mur:1995}, \cite{Mur:1999}, \cite{Chi:2000}, \cite{Ant:00}.

\subsubsection{Finite Tree Automata}
\label{section:TreeAutomata}

Regular automata are being addressed in \cite{Mur:1999}, \cite{Com:2005}.
Next, only such automata are characterised, which correspond to the expressibility of regular tree grammars.
The taxonomy proposed in \cite{Mur:2001} is used to achieve this goal, particularly the determinism and evaluation order is of utmost interest and is summarised in tab.\ref{table:TaxonomyTreeAutomata}.
Bottom-up automata can recognise ranked-competing grammars.

\begin{table}[h]
\begin{center}
\mbox{
  \begin{tabular}{|c||c|c|}
   \hline
   & deterministic	& non-deterministic \\
   \hline
   \hline
   top-down & local, single-type &	regular\\
   \hline
   bottom-up	&	ranked-competing	&	regular\\
   \hline
  \end{tabular}}
\end{center}
 \caption{Taxonomy of Tree Automata}
 \label{table:TaxonomyTreeAutomata}
\end{table}

According to \cite{Mur:2001}, grammars whose languages are recognised by non-deterministic top-down and bottom-up automata are reducible to the same.
Non-deterministic grammars and recognition algorithms are often significantly shorter in their description than equivalent deterministic grammars.
However, those algorithms may be more complex and involve extensive backtracking.
In contrast to top-down automata, Bottom-up automata are considerably more complex and, therefore, more challenging to maintain.
Especially, error messaging may become more complicated by far since, in general, all alternatives need to be checked before deciding if validation fails.

Moreover, the "`real"' reason would have to be tracked somehow among alternating backtraces.

\textit{Derivatives} \cite{Brz:64} is one approach, which maps regular expressions over hedges onto finite \textit{regular tree automata}.
During a document validation, a given schema and regular expression are reduced towards an instance until both sides cannot be reduced any further.
If both expressions are empty, then validation would succeed.
Otherwise, validation fails.
After each derivation step, a new state is introduced.
Since the incoming schema is finite and symbols are not allowed (see sect.\ref{section:mainXTL}), the algorithm terminates.

\textit{Partial-Derivatives} \cite{Ant:00} is an improved approach that calculates derivatives only when needed (see app.\ref{appendix:PartialDerivativesAlgorithm}).
Once calculated, solutions are not determined a second time.
Similar parsing approaches increase performance by 70\% in XML-documents \cite{Ken:2005} and 80\% in instantiating those \cite{Nog:2002}.

The additional cost to be paid on XML-parsers is only causing approximately 10\% of overhead.
This algorithm has a best-case complexity of $\theta(n)=n$ and $O(n)=n^{2}$ for the worst case, where $n$ is the given regular expression length.
For the reasons mentioned, this approach is of interest for practical implementations.
It can be stated that tree automata describe functions over XML documents, particularly the template expansion and schema validation.

\section{Analysis}
\label{section:Analysis}

This section investigates instantiation and validation.
Both functions are considered and requirements formulated for an unification on document level.

\subsection{Current Situation}
\label{section:AnalysisStateOfArt}

This paragraph XTL is analysed w.r.t. language features. Then instantiation and validation are investigated closer, e.g. properties of XML-template and schema languages.

\subsubsection{XTL}
\label{section:StatOfArtAnalysisXTL}

\textbf{First Considerations}

Represents of template languages are \textit{JSP}, \textit{ASP}, XSLT and XTL.
Prolog may also be considered for instantiation of XML-document \cite{Hab:2006}.

The generated target language may distinguish template languages.
If both template and target language unite, as may be the case with XML, it is obvious that unification may be easier.

Its intention may also distinguish template languages.
For instance, template languages with variables and functions are more appropriate for programming than for document processing.
In any case, it is worth, to separate the program from the document, especially when it comes to validation (cf. sect.\ref{section:StateOfTheArtInstantiation}).\\

Parr \cite{Par:2004} proposes template languages to have the following minimal asset of template commands, which should also count for document processing:

\begin{center}
\begin{tabular}{l}
 1. Attribute references,  \\
 2. Conditions,  \\
 3. Recursive Template Calls, \\
 4. Conditional Template Inclusion. \\
\end{tabular}
\end{center}

XTL, which is also a template language, already has these features (cf. sect.\ref{section:mainXTL}).
\lit{\texttt{xtl:attribute}} express attributes, conditions by \lit{\texttt{xtl:if}}, cycles by \lit{\texttt{xtl:for-each}}, text inclusions  by \lit{\texttt{xtl:text}} and element inclusions by \lit{\texttt{xtl:include}}.
\textit{Applications} can be expressed, but only without parameters (done by \lit{\texttt{xtl:call-macro}}).

XTL instantiates free of side effects.
So, queries to instantiation data do not alter the source nor the template.
This way, \textit{referential transparency} is guaranteed.

If PHPs allow access to documents, then XTL indeed is also favourable for documents.
Helper functions should be banned from XTL in general and moved to external sources, which may be referenced by \lit{\texttt{select}}.

\textbf{Separation of Concerns}

\begin{figure}[h]
 \begin{minipage}{2cm}
  \xymatrixrowsep{50pt}
  \xymatrixcolsep{50pt}
  \xymatrix{
\fbox{Model} \ar[r]^{instantiation}_{data} & \fbox{Renderer} \ar[r] & \fbox{View}\\
& \parbox[r]{2.3cm}{\fbox{Controller}\\\textit{\small{template engine}}} \ar[ul]^{template} \ar@{.>}[u] \ar[ur]_{instance} &
  }
  \end{minipage}
  \caption{Model-View-Controller for Instantiation}
  \label{figure:MVCInstantiation}
\end{figure}
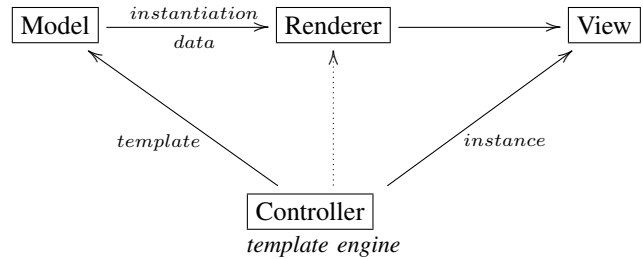

Czarnecki and Eisenecker \cite{Cza:2000} insist a transformation is taken out by referencing only a minimal set of operators.
That also goes for instantiations because these are transformations too.
Both suggest restricting ourselves to loops and selections instead.
In XTL, this is done by the command tags \lit{\texttt{xtl:for-each}}, \lit{\texttt{xtl:text}}, \lit{\texttt{xtl:include}}, \lit{\texttt{xtl:attribute}}, and the attribute \lit{\texttt{select}}.
However, functions shall better be removed from instantiation and be passed to a module responsible for that particular task.
For the same reason, calculations shall also be removed from templates.
It also affects arithmetic expressions, e.g. in XPath, which are not needed for navigating instantiation data.

Furthermore, Parr \cite{Par:2004} insists template languages separate between concerns of a given instantiation problem.
Fig.\ref{figure:MVCInstantiation} shows a concern separation for XTL using the \textit{Model-View-Controller} meta-pattern after Reenskaug.
The resulting instance document shall be independent of the instantiation, s.t. all calculations and existing constraints fit within the chosen model.
Output in View often requires formatting. Therefore, the Controller triggers the Renderer, who passes information further from Model to View. By doing so, the strict separation remains.\\

The separation of \textit{command language} from the template language by \lit{\texttt{select}} and common tags lead to an increasing similarity between instance and template.
It is the foundation for the unification of instantiation and validation (see sect.\ref{section:ComparisonCriteria}).

The separation between template and command languages causes the command language has to provide conventional interfaces for text and attribute access, inclusions, conditions and loops.
Since XTL does not allow direct access to instantiation data initially, it is not Turing-mighty.
The optional attribute \lit{\texttt{realm}} was introduced. In XTL command tags, it may be inserted by  a \lit{\texttt{select}}-attribute in order to allow XTL to access different instantiation data.

\textbf{Type Safety}

XTL guarantees a well-formed document during instantiation of a template if it is sound w.r.t. its specification \cite{XTLSpec:2007}.
Each XTL-template is well-formed XML.
The only restriction is the top root node may not be a command-tags according to its specification. It must be an element node.
All nodes located underneath are well-formed because those nodes are composed of text and element nodes only.
In some cases \lit{\texttt{select}}-queries are not sound.
The affected XTL-command are left empty by the XTL template engine.

PHP functions assert type safety.
Hence, instances are always type-safe here.
Static types of PHP functions make instantiation more predictable before running.
It means command languages become exchangeable, and so the source language becomes more flexible.

In XTL, \textit{term-evaluator}  \cite{Hart:2006} and \textit{instantiation data evaluator} have been introduced to evaluate \lit{\texttt{select}}-expressions.
The instantiation data evaluator checks the types of a solution with the inferred type or an optional type.
Instantiation data has a polymorph type and so have query results.
In order to process arbitrary instantiation data, polymorphic data needs to be transformed into non-interpretable text.
That is done by the Renderer, which knows about desired output formatting.

The following table shows the types of XTL-commands, which need to be known to the Renderer before building the instance document:

\begin{center}
 \begin{tabular}{r c l}
  \texttt{xtl:attribute} & ::	&	\texttt{String $\rightarrow$ [a] $\rightarrow$ String} \\
  \texttt{xtl:include}   & ::	&	\texttt{String $\rightarrow$ [a] $\rightarrow$ XML} \\
  \texttt{xtl:text} 	   & ::	&	\texttt{String $\rightarrow$ [a] $\rightarrow$ String} \\
  \texttt{xtl:if}		   & ::	&	\texttt{String $\rightarrow$ [a] $\rightarrow$ Bool} \\
  \texttt{xtl:for-each}  & :: &   \texttt{String $\rightarrow$ [a] $\rightarrow$ [a]}
 \end{tabular}
\end{center}

The first type denotes an \texttt{select}-expression. The second type denotes the instantiation type, which is polymorph. The typing of PHP-functions is explained in more detail:

\switchXTL
\lstset{tabsize=2}
\lstset{basicstyle=\ttfamily\small}
\lstset{numbers=none}
\lstset{numberstyle=\tiny}
\lstset{showstringspaces=false}
\lstset{captionpos=b}

\begin{center}
\begin{tabular}{ll}
\begin{lstlisting}
<book>
  <xtl:attribute  
    name="title"
    select="..."/>
</book>
\end{lstlisting} \qquad & $\longleftarrow$ \parbox{3cm}{\texttt{SELECT title \\	FROM books \\	WHERE id='1'}}
\end{tabular}
\end{center}

It is assumed, a PHP function determines for a given SQL-query a relation with exactly one result.
This result is converted by the Renderer in a representable string and placed into an instance document.
The example generates a \texttt{<book/>}-node with attribute \texttt{title="Haskell..."} related to the example from sect.\ref{section:Introduction}.
Though, the resulting nodes are always well-typed and well-formed for any instantiation data and PHP functions.
So, every XML document created is type-safe.

\textbf{Variability}

It means the exchange of the command language.
PHP has standard interfaces and concrete implementations.
By doing so, the internal organisation can be hidden from the user \cite{Bol:2006}.
The processing of heterogeneous data is managed by previously agreed interfaces only.

The template engine restricts access to instantiation data.
The template may not grant access.
Although this rigid restriction may not always be desired, as demonstrated by the following fragment:

\switchXTL
\lstset{tabsize=2}
\lstset{basicstyle=\ttfamily\small}
\lstset{numbers=none}
\lstset{numberstyle=\tiny}
\lstset{showstringspaces=false}
\lstset{captionpos=b}

\begin{lstlisting}
<book>
  <xtl:include select="document('a.xml')
   //title"/>
</book>
\end{lstlisting}

The difficulty is the template addresses external documents during instantiation.
The function \lit{\texttt{document}} is evaluated with the path expression following, even if the function is not XPath \cite{DeR:1999}.
The source is passed immediately to the instantiator.
Alternatively, all referenced spots are moved to a separate document to resolve this problem --- this is tractable only if there are multiple sources.
In the template, those entries are referenced.
Moreover, the extraction of all needed sources by additional templates is in preparation.

\textbf{Style}

Besides the formal grammar type, schema languages can also be characterised \cite{Lee:2000} by the corresponding grammar style.

A language has a \textit{grammar style} if the associated schema language is described shorter by a grammar than by a corresponding regular expression \cite{CLA:06} (see following sections). Otherwise, the language is in a \textit{pattern-based style}.

A schema language allows many different tags at different locations.
Schemas that only have a few restricting symbols only are more often in pattern-based style than in grammar-style.

Macro definitions can easily be described in XTL in grammar style.
Right-hand sides are equal to a not necessarily right-ranked hedge (cf. sect.\ref{section:DesignMain}).
All remaining XTL-tags may be described in pattern-based style as well as in grammar style.

In contrast, we have schemas from rigid associations/relations whose elements need to be placed separately.
Relations do not insist on strict ordering.
However, related entities must obey certain conventions. For example, relations need to be defined a priori, which determine uni-directional element and bi-directional attribute relations.
It also causes a whole graph is represented.

\textbf{Regularity}

Since RHG may be described by XML-schema languages \cite{Chi:2000}, \cite{Mur:1999}, a corresponding representation exists consisting of regular hedge-expressions.
If \lit{\texttt{select}}-expressions, which denote the number of repetitions, in \lit{\texttt{xtl:for-each}}-loops are kept arbitrary, then context-sensitive and context-regular expressions become regular expressions for the price of further abstraction. Arbitrary repetitions become Kleene's star operators (cf. sect.\ref{section:AnalysisValidation}).
Conditions in XTL are represented either by $\varepsilon$, hedges, or text and element nodes as literals.

Macro definitions and macro calls require special treatment because right-hand sides of macro definitions may contain an arbitrary number of further macro calls anywhere within the hedge.
Both extend the expressibility of regular tree grammars.
Therefore, macros need to be investigated when it comes to judgements about expressibility.
After all, and for the sake of simplicity, it still makes sense to regard XTL as a regular schema language.
A more detailed investigation of the grammar class follows.\\

\switchXTL
\lstset{tabsize=2}
\lstset{basicstyle=\ttfamily\small}
\lstset{numbers=none}
\lstset{numberstyle=\tiny}
\lstset{showstringspaces=false}
\lstset{captionpos=b}

\begin{table}
\begin{center}
\begin{tabular}{l}
\begin{lstlisting}
<xtl:macro name="M">
 <a/>
</xtl:macro>
<a/>
\end{lstlisting} \\
\end{tabular}\\
\small{(a)}
\end{center}
\begin{center}
\begin{tabular}{l}
\begin{lstlisting}
<xtl:macro name="A"/>
	<a/>
	<xtl:call-macro name="C"/>
</xtl:macro>

<xtl:macro name="B"/>
	<a/>
	<xtl:call-macro name="D"/>
</xtl:macro>

hello
<xtl:call-macro name="A"/>
world
<xtl:call-macro name="B"/>
!
\end{lstlisting}
\end{tabular}\\
\small{(b)}
\end{center}
\begin{center}
\begin{tabular}{l}
\begin{lstlisting}
 ...
 
 <xtl:macro name="A">
	<xtl:for-each select="//A/book"> 
		<a/>
	</xtl:for-each>
 	<xtl:call-macro name="C"/>
 </xtl:macro>

 <xtl:macro name="B">
	<xtl:for-each select="//B/book">
		<a/>
	</xtl:for-each>
	<xtl:call-macro name="D"/>
 </xtl:macro>

 hello
 <xtl:call-macro name="A"/>
 <xtl:call-macro name="B"/>
 world!
\end{lstlisting}
\end{tabular}\\
\small{(c)}
\end{center}
	\caption{Expressibility of regular schemas in XTL}
	\label{table:AnalysisRegularSchemas}
\end{table}

Example (a) from tab.\ref{table:AnalysisRegularSchemas} shows that XTL is not only local because \texttt{<a/>} occurs in the body of the macro and in the hedge.
It means the corresponding grammar has two productions with the exact right sides.
Example (b) shows XTL is not only of single-type. Macros \lit{\texttt{A}} and \lit{\texttt{B}} are competing -- both contain \lit{\texttt{<a/>}} as a starting symbol.
Hence, $\pi(\texttt{A}) \cap \pi(\texttt{B})$ is not empty.
Example (c) shows XTL is not only ranked-competing because decompositions cannot always be found due to uneven macros \lit{\texttt{C}} and \lit{\texttt{D}}.
Hence, XTL is as expressible as the class of languages generated by regular tree grammars.

Only regular tree languages are enclosed under union, intersection and complement (cf. \cite{Chi:2000}) as \textit{string}-grammars are.
That is particular of interest when extending XTL by command-tags composed of existing tags.\\

Those tags oppose \textit{non-monotone} operators \cite{Hab:2006} because those may change instantiation data fragments and those do not necessarily merge existing instantiation data.
Those operators are advantageous and compact when an instance shall have many details and when the difference between instance and source documents is relatively small.
Rather than requesting much information to build up the document from scratch by filling numerous slots, it may be more efficient to copy the document instead nearly.
Although non-monotone operators can reduce a template significantly, violated closure properties may cause disturbance in the separation of concerns between instance and instantiation data because validation can not make assumptions about the instantiation data.
That is why non-monotone operators must be restricted in regular languages a priori not to hinder the unification of instantiation and validation.

\textbf{Context-free Tree Languages}

Schemas can be defined in two ways (see tab.\ref{table:AnalysisCFLanguages}) to recognise the context-free tree language $L(a^{n} b^{n})$.

\begin{table}[h]

\begin{center}
\begin{tabular}{l}
\textbf{Variant 1:}\\
\begin{lstlisting}
<xtl:macro name="S">
 <xtl:if select="...">
  <a/>
   <xtl:call-macro name="S"/>
  <b/>
 </xtl:if>
</xtl:macro>

<xtl:call-macro name="S"/>
\end{lstlisting}
\end{tabular}
\end{center}
\begin{center}
\begin{tabular}{l}
\textbf{Variant 2:}\\
\begin{lstlisting}
<xtl:for-each select="//book">
  <a/>
</xtl:for-each>
...
<xtl:for-each select="//book">
  <b/>
</xtl:for-each>
\end{lstlisting}
\end{tabular}
\end{center}
\caption{Context-free schemas}
\label{table:AnalysisCFLanguages}
\end{table}

Variant 2 is universal since $L(a^{n} c b^{n})$ could be generated.
That cannot be done with variant 1.
Furthermore, $L(a^{n} u b^{n} v c^{n})$ and $L(x a^{n} v b^{n} v c^{n} wd^{n}y)$ can only be recognised by variant 2.
So, the difficulty is to express exactly instances that shall be recognised during validation.
For instance, if \lit{\texttt{select}}-expressions describe $L(a^{n} c b^{n})$, then this language would be context-free.
In contrast to regular language recognition, the recognition of programming languages, especially those defined by a LL(k) or LR(k)-grammar \cite{grun:91}, is much more complex.
Therefore, the recognition of context-free and context-sensitive tree languages, in general, might be much more complex and is doable but only with much more efforts to be spent.
XTL is regular if only expressions of the command languages are not considered exactly during validation.
Variant 1 describes a part of context-free schemas. However, if $L(a^{n}cb^{n})$ is to be recognised, then variant 2 would need to be chosen.
Nevertheless, context-free languages may not be validated exactly.
This means \texttt{<a/><a/><a/>x<b/><b/>}  is enclosed by $L(a^{n}xb^{n})$ and is validated by variant 2, although the input is not context-free.
Hence, context-free schemas are not going to be considered further.

\textbf{Termination}

If in the body of macro definitions appear macro calls, recursion may appear.
\textit{Fixpoints} may appear in the template and schema. Fixpoints may be formulated with the command language and remain unreachable due to \textit{left-recursions} -- so instantiation and validation do not terminate.
However, the recognition of left-recursion is in general not decidable for XTL-documents due to the Halting problem.

In contrast, it may effectively be decided whether, for instance:

\switchXTL
\lstset{tabsize=2}
\lstset{basicstyle=\ttfamily\small}
\lstset{numbers=none}
\lstset{numberstyle=\tiny}
\lstset{showstringspaces=false}
\lstset{captionpos=b}

\begin{center}
\begin{tabular}{c}
\begin{lstlisting}
<xtl:macro name="M">
  <xtl:call-macro name="M"/>
</xtl:macro>
\end{lstlisting}
\end{tabular}
\end{center}

contains a non-terminating cycle.
However, this does not work for arbitrary XTL-document before executing the program.

\textbf{Functions}

XTL has a small vocabulary.
Because XTL does not have parameters, formal functions cannot be defined.
In XPath, it is not possible to define functions with an arbitrary arity. It is also not possible to call such a function from other tags.
A locally defined function may not violate the template engine's referential transparency because of the strict separation between template and instantiation data.
So, many functions, mainly $\mu$-recursive and tail-recursive functions, cannot be expressed within XTL.

Tail-recursive functions can be simulated in XTL but only under additional restrictions.
For example, counters may be expressed by special \lit{\texttt{select}}-expressions.
The function \lit{\texttt{position()}} allows accessing the actual counter in XPath.
The amount of loop iterations is limited in \lit{\texttt{xtl:for-each}} by a constant value evaluated by \lit{\texttt{select}}.
Cycles simulate tail-recursive functions in XTL but without an argument list.

Due to the lack of defining and composing functions, XTL is not primitive-recursive --- even when while-loops are replaced by recursion with a preceding \lit{\texttt{xtl:if}} or when tail-recursive functions without arguments are mimicked.
Due to the separation of concerns, document processing does not require sophisticated arithmetic nor logical functions.

\subsubsection{Instantiation}
\label{section:StateOfTheArtInstantiation}

Instantiators can be interpreted as term rewriting systems.
The $\lambda$-calculus provides a mechanism for doing so (see \cite{Bar:81}).
Slots can be represented as variables and nodes as terms.
Then instantiation equals a derivation.

However, terms need to be adequately modelled.
It may be needed to assign each element node with a certain arity its semantics.
An evaluated node may denote a node with a qualifying functor, which encodes the number of children.
Otherwise, a concatenation of nodes may not be injective.

Furthermore, empty nodes, attributes and hedges need to be mapped, which may require additional handling.

Values of \lit{\texttt{select}}-expressions are bound to variables. Within terms of $\lambda$-abstractions, this is done by applying expressions to slot variables. The internal evaluation by PHP functions remains invisible.

Termination is equal to reaching a normal-form.
Reduction is strictly monotone.
Once evaluated, nodes are not reverted.
The number of evaluation steps is polynomially bound, except macros.
Macros may lead to self-application because macro bodies may reify variables.
In that case, no normal-form could be found, so the template engine would not terminate.
The evaluation sequence of a hedge does not matter since all hedge nodes have to be evaluated and independent.
XTL conditions terminate when evaluated outside-in.
However, they do not terminate in general in the opposite direction.

As already mentioned, slots return strings and nodes.
That is why types over nodes and slots make sense.

Atomic element nodes represent their type themselves and do not require additional conventions.
Atomic nodes may directly be passed to instance document and do not cause any side effects.
Hence, a formal description of the instantiation benefits from \textit{denotational semantics}.

The simple untyped $\lambda$-calculus is not sufficient here.
The typed $\lambda$-calculus is needed for a formal description of instantiation.
It is because \textit{types} and constructors are helpful to the description of element and text nodes.
Constructors denote parametrised types whose type variable is typed again.
Types are composed of other types.
On the other side, the simple $\lambda$-calculus does not provide a compact notation for function calls.
Fixpoint \textit{combinators} are the only possibility to mimic recursion.

In contrast to that, the denotational semantics allows us to express constructors, types and formal function in a meta-language.
So, it becomes possible to express pattern-matching compactly and to use combinators too.
The construction of nodes should be done according to minimisation criteria \cite{Hab:2006}.

\subsubsection{Validation}
\label{section:AnalysisValidation}

In analogy to parsers, validators check for a given programming language if an incoming XML document (program) is a valid instance of a given schema (grammar).
It is more appropriate to refer to \textit{matchers} \cite{Rahm:01} when validation is meant.
Compilers have an invariant set of rules but are applied to ever-changing incoming programs.
When considering validators not only do the input data (instance) change and the set of rules (schema).
Moreover, there is no translation going on, but a boolean value is calculated.

An XTL-validator checks an instance document node-wise against a regular schema.
In \cite{Rahm:01}, the classification of schema-matchings is proposed.
The XTL-validator is hence a class on its own and is schema-centric.
The validation operates on hierarchic XML-nodes and therefore is structure-centric.
Both \lit{\texttt{xtl:for-each}} and \lit{\texttt{xtl:if}} are meta-operators that influence the processing of an instance document.
On a lingual level, they are constraints, which have nothing really in common with \lit{\texttt{select}}-expressions.
That is why the validator belongs to the class of \textit{graph-matchers}.

The description of validation can either be described as textual or graphical.
The relation between matching documents can be expressed by $\cong (s_{1},s_{2})$, where $s_{1}$ denotes a schema node and $s_{2}$ an instance node.
Schema nodes matching with some instance node $s_{2}$ generate a set $S_{1}$, which in general is not singular.
If $S_{1}$ is indeed not singular, then validation becomes valid and non-deterministic.
$S_{1}$ cannot be empty since it at least contains $s_{2}$.\\

The validation problem can be interpreted as \textit{typing problem} \cite{Bar:81}.
In \cite{Wall:99}, typing is proposed as a static validation approach using Haskell’s built-in type system.
If a document validates, then a type can be inferred.
Otherwise, Haskell shows up a typing mismatch.
The validation takes place without an actual validation algorithm by doing so.
This approach only works with dedicated constructors used for constructing in Haskell a whole XML document.

The formal notation can be based on the $\lambda$-calculus, but the formalisation for validation is problematic since there seems to be no good representation adequacy.\\

Context-free languages are not recognised, except the feature discussed in sect.\ref{section:StatOfArtAnalysisXTL}.
It means hedges are invalid as soon as they appear twice.
The number of opening and closing brackets is only of minor interest -- this is different from programming languages.
A context-free schema may still be recognised by moving a sequence $c$ to a prefix from the language $L(a^{n}cb^{n})$ or moving $c$ to a suffix of $a^{n}b^{n}$.
In node $c$ $n$ may not occur.
Otherwise, the given schema is not context-free.

To improve the runtime behaviour towards non-deterministic decisions determining a tree automaton requires a regular schema whose recognition has a complexity of $O(2^{n})$. Herewith, $n$ is the cardinality of the set of states (see \cite{Brz:64},\cite{Ant:00}, cf. sect.\ref{section:TreeAutomata}).
For validation to be used only once, these may mean too high costs.
However, if many documents are going to be validated against the same schema, then the situation becomes different, as it may be the case with database triggers.\\

As seen in sect.\ref{section:Basics}, tree automata fit only for schemas that do not change over time.

Clark \cite{Cla:2007} proposes a top-down non-deterministic algorithm for RelaxNG-schemas.
Non-deterministic matchings are resolved by so-called \textit{interleavings} \cite{Cla:2007}.
He proposes rules for element nodes for all possible occurring cases.
Inclusions denote variable symbols which arbitrary nodes may replace.
It causes nodes to appear valid, although they do not occur at corresponding positions in the instance document.
So, instead of an $\varepsilon$-node, its successor may be taken for validation, and if validation fails, then everything from there backwards needs to be analysed manually, which is quite laborious.
It may be more efficient to track all possible nodes during validation and decide when to include the next time.
Unfortunately, even small documents generate such a vast search space, so it becomes not doable due to an exponential rise in complexity.
The \textit{extensive search} \cite{Cla:2007} should if used at all, massively reduce invalid nodes.
There is no optimal solution for this problem.
However, there exist a few heuristics which may overcome the practical problem for previous domain ranges, such as the strategy "`try all valid states until a contradiction occurs"'.


Antimirov \cite{Ant:00} proposes an algorithm turning a regular expression into a \textit{non-deterministic finite automaton} (NFA).
Hence, XML schemas can be considered as regular expressions.
Antimirov's approach matches regular tree expressions also.
Non-deterministic finite automata are dual to deterministic finite tree automata (cf. \cite{Mur:1995},\cite{Mur:2001}).
In contrast to \cite{Brz:64}, needed derivatives only are calculated, and those are calculated only once.
In procedural and object-oriented programming languages, the determination can be achieved by merging non-determined states or balancing non-determinism, e.g. by backtracking.\\


An alternative to validation (with XTL) is the \textit{transformation} (of XTL) into another already existing schema language, like RelaxNG.
Problems that need to be addressed could, but do not necessarily need to be:  restriction of expressibility (of XTL) or coverage of the schema-language to be replaced.
A schema transformation would also require additional well-defined schema languages, which is not part of this work.

\subsubsection{Properties}
\label{section:AnalysisProperties}

Instantiation, as introduced in sect.\ref{section:Basics}, is a mapping whose co-domain XML is entirely covered.
Instantiation can be interpreted as \textit{endomorphism} because both domain and co-domain denote the same set, namely XML.
The well-formedness of XTL itself guarantees this.
Therefore, instantiation is an enclosed operation (cf. \cite{Hart:2005}).\\

If instantiation is indeed considered an operation, then associativity does not hold.
Commutativity also does not hold because instantiation of a slot-containing template document has XML as a result. The result syntactically does not match in general with the origin template.
Instantiation of an XML document without slots is idempotent for any instantiation data.\\

For validation, particularly a derivation of a regular expression, \textit{homomorphism} holds.
It follows from this equation from \cite{Ant:00}, which also holds for schemas:
$$val(x \cdot y) = val(x) \odot val(y)$$

The operator \lit{$\cdot$} concatenates two regular expressions and \lit{$\odot$} logically ANDs two interleavings.
So, some regular expression $val(x \cdot y)$ is congruent to $val(y)$ modulo $val(x)$ (see sect.\ref{section:DesignNFA}).
Therefore, the order does not matter, whether first regular subexpressions $x$ and $y$ are evaluated and concatenated second, or whether first those expressions are concatenated and second $x$ and $y$ are evaluated.
It might be helpful when at least one subexpression may be dropped, for instance, for seeking optimal solutions.

\subsubsection{Arrows and Filters}
\label{section:AnalysisArrowsFilters}

An \textit{arrow} is a generalised monad (see \cite{Hug:2000}, \cite{Lec:2005}, \cite{HXT:04}).
It encapsulates functions as parameter.
An arrow in Haskell is an instance of the class \texttt{arr} with two functions:\\

\begin{tabular}{l}
\texttt{arr::(a$\rightarrow$b)$\rightarrow$arr a b} \\
\texttt{>>>::arr a b$\rightarrow$arr b c$\rightarrow$arr a c}
\end{tabular}\\

Like variables in programming languages, functions may also be made available in dedicated environment scopes and namespaces.
Particularly for lazy parsing and serialisation, certain sets may be evaluated partially, allowing higher usability.

Instantiation may use macro definitions instead of arrows.
However, during validation, macro definitions shall be avoided because apart from a macro environment, further semantic fields may be required, e.g. a list of all valid element nodes --- which was decided not to research further for the sake of previous outcomes in this work.\\

\textit{Filters} denote functions having a polymorph type \texttt{a$\rightarrow$[a]}, where $a$ denotes a type variable.
Filters are functions with an input vector and an arbitrary output vector.
They can be classified according to their behaviour, for instance, by common combinators (cf. \cite{Wall:99}) and functions not in typical combinator representation.
The class mentioned second are functions whose head is specified using pattern matching.
If possible, implementations should make use of pattern matching --- the same as semantics make.
By doing so, redundant iterations of trees are avoided, and structural definitions can be reused. Both effects increase readability.
The elimination of multiple iterations cost high efforts (cf. \cite{Tho:2001}), causing an increase in complexity.
The gap between denotational semantics on the one side and implementations on the other side diminishes by specifying pattern matching.\\

Unfortunately, both XML-parsing and serialisation, violate referential transparency, but this must be since files can neither be read nor written to without side effects.
Luckily, these are the only places where this is required, and there is no other place having this effect.
As an alternative to arrows, multi-paradigmal programming may be an option (cf. \cite{Ricci:2001},\cite{Hab:2006}).
By violating the absence of side-effects, read and write operations increase the flexibility of a function in general.
Input and output operations are no longer restricted to a certain location in a function but now can be located and used anywhere.
Multi-paradigmal programming shall be used, s.t. input and output operations are implemented by machine-dependent instructions, and where an abstract programming language implements instantiation and validation.
Here, a violation of encapsulation would increase usability.
Named functions shall be strictly typed and be implemented as \textit{super-combinators}.

\subsection{Requirements}
This work aims to attempt to unify both views, instantiation, and validation (see sect.\ref{section:Basics}).
In consequence, functionality increases.

For demonstration purposes, denotational semantics would be required for both instantiation and validation.
Algorithms based on it, implementation and an object-oriented design with test cases are prepared.

\subsubsection{Limitations}

No assumptions on a concrete command language are agreed upon that.
So, no information on the syntax nor internal states, nor properties are known to the instantiator. No assumptions on the structure of instantiation data are made.
Only PHP functions with previously agreed interfaces are to be considered (cf. sect.\ref{section:DesignInstantiationSemantic}).
Communication is established exclusively by these interfaces.\\

Validation does not interpret \lit{\texttt{select}}-expressions.
\textit{Bypass}- and \lit{\texttt{realm}}-attributes are not treated on instantiation.
The support of \lit{\texttt{bypass}}-attributes is optional during instantiation because this requires an the interference of several expansions within one template.
The effect of \lit{\texttt{bypass}} can be simulated by running several templates sequentially.
\lit{\texttt{bypass}}-attributes are then redirected as command tags into the instance document or as another \lit{\texttt{bypass}}-attribute with a smaller value on the total amount of phases to be run \cite{XTLSpec:2007}.

No precautions are made w.r.t. detect of non-terminating loops since this problem is in general undecidable (cf. sect.\ref{section:mainXTL}).

Encountered problems of XTL in comparison to other schema languages are to be examined, and improvements shall be shown.
The goal herewith is a compatible syntax extension (cf. sect.\ref{section:RegularTreeLanguages}).\\

Another helpful tool for XTL is a semi-automated schema-generator, which generates a schema from an instance document (so-called "'validation by instance"'-approach) \cite{FITZ:07}.
The implementation of a schema-generator for practical use would go far beyond the goal of this work.
Therefore, it is not considered here further (cf.  \cite{FITZ:07}, \cite{DTDGen:07}). Here are some reasons why:

\begin{itemize}
 \item \textbf{Parameter:}
A schema appears useful to the user whenever it is fine-grained and recognises many regular substitutions in the instance.

However, since it is not obvious if a hedge may be replaced by a sequence of nodes or by a cycle with conditions, some criterion must be defined regarding granularity.
So, complex substitutions could reduce an instance by a line, for instance, but the obtained instance would be tough to check by the user.

 \item \textbf{Ordering:}
Is the ordering within a children list fixed, or may it permute?

The ordering of a hedge is either explicit or selective.
The ordering of nodes in a hedge, also referring to the following hedges, can be assigned by any attribute.
Childless nodes do not necessarily need to be defined in a schema to be childless.
Child-rich nodes may confirm in the following nodes the exception.
Such differences rest exclusively on the user and may not be considered automatically.

 \item \textbf{Quantity} 
Multipliers for specifying elements that occur several times are hardly available.
Neither makes it sense to specify multiplicity on each occurring hedge.
Instead, only those hedges should be quantified, which, for instance, occur two or more times.

 \item \textbf{Configuration:}
All presented constraints must be configurable on the needs of a user.
Rules should be assigned to \textit{member functions}.
Based on those rules, for instance, an expert system may derive optimal decisions.
\end{itemize}

\subsubsection{Instantiation}

Inputs are a well-formed XTL template, as well as one or more instantiation sources.
Further formal and non-formal requirements not mentioned in sect.\ref{section:StatOfArtAnalysisXTL} are:

\begin{enumerate}
 \item The implementation is to be done in Haskell.
The Haskell-Toolbox for XML-processing \cite{HXT:04} shall be used (see sect.\ref{section:Implementation}).
		
 \item The implementation should essentially not deviate from the denotational semantics.

The data model and rules should be simple.
Invalid XTL-tags should be treated as atomic element nodes.
\end{enumerate}

\subsubsection{Validation}

XTL-conform schemas and XML instance document count as validation input.
The result of validation is a "'yes"'/"'no"' answer.
The requirements to a validator are as following:

\begin{enumerate}
 \item Definition of an appropriate data model.

 \item Rules have to be minimal w.r.t. amount and length.	

The premise of each matching case contains one node for the schema and one node, for instance.
The validation of any node from the schema may only refer to the actual corresponding instance node, references to the following nodes are not allowed.

The next rule to be applied shall be non-deterministic. No additional assertions on the selection are allowed.
This approach is similar to instantiation.

 \item Before a validation fails, all other alternatives need to be checked first.\\

The error message should \textit{trace} the actual error location and reason.
If validation succeeds, a console notification should be emitted.
\end{enumerate}

\subsubsection{Unification}
One main goal of unification of XTL-instantiation and validation is the lingual unification of both processes (cf. sect.\ref{section:Introduction}).
Apart from that, the reuse of templates as schema shall be examined.

A rule-based approach would be desired for a better understanding and a qualitative investigation (see sect.\ref{section:ComparisonMain}).
Here agreed data models should be used for both processes.
Helper functions should be reused as much as possible.
It requires those functions to be as generic as possible.
Pre-defined functions shall be reused (cf. \cite{Coe:1992}).

\subsubsection{Implementation in Java}

The programs to be written in Haskell shall later be implemented in Java.
Therefore, at least an object-oriented design and a translation of the denotational semantics into Java are needed (see sect.\ref{section:Basics}).
Here several questions emerge:

\begin{enumerate}
 \item How is polymorphism \cite{Card:1999} and functionals implemented accordingly in Java?
 \item Can the non-deterministic top-down automata remain as is?
 \item What are appropriate class candidates? What do associations look like between them?
 \item Which roles can be abstracted, and how do these roles interact?
 \item How will most generic implementations look?
 \item How do the architectural design patterns look?
\end{enumerate}

In this work, two programs are implemented, one for instantiation and one for validation.
A test suite is introduced.
Existing XSD schemas guarantee the well-formedness and validity of XTL-templates.
Existing frameworks for XML processing are being used where appropriate.

\section{Design}
\label{section:DesignMain}

In this section, introduced data models and semantics for instantiation and validation are presented.

Haskell is used as a programming language.
Haskell's functional character allows a straight transformation from denotational semantics (cf. sect.\ref{section:Introduction}).

\subsection{Data models}
\label{section:DesignDataModels}

The goal of data models introduced in this section is easy denotational semantics.\\

The features set of HXT is relatively tiny.
Usability is medium -- so compromises must be made.
So, the simpler the description gets, the more valuable the simplification becomes.
In conclusion, the need arises to transform models.
It takes small efforts instead when it comes to validation since transformation back again is not needed.
In contrast to this, instantiation requires both transformation directions.

Furthermore, the data model transformation completeness and correctness need to be assured by covering both domains and co-domains.
Those coverings may be used as a test suite for the implementations.

\begin{figure}[h]
\begin{center}
 \begin{minipage}{2cm}
  \xymatrixrowsep{30pt}
  \xymatrixcolsep{30pt}
  \xymatrix{
*+[F]\txt{HXT} \ar[r]_{f_1} & *+[F]\txt{XTL} \ar[l]_{f^{-1}_1} \ar[r]^{f_2} & *+[F]\txt{Reg} \ar@{.>}[r]^{f_3} & *+[F]\txt{NFA}
  }
  \end{minipage}
\end{center}
  \caption{Data models for Instantiation and Validation}
  \label{figure:DataModels}
\end{figure}
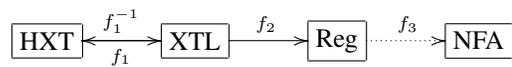

Fig.\ref{figure:DataModels} shows the transformation for all the data models considered in this section.    
The data models \texttt{HXT} and \texttt{XTL} represent document nodes.
These nodes can be transformed by the functions $f_1$ and its inverse $f_{1}^{-1}$ into each other.
Both found the formal semantics for instantiation and validation.
The composed mapping \texttt{HXT $\mapsto$ XTL $\mapsto$ HXT} describes instantiation.
The first transformation transfers a parsed \texttt{XmlTree} into a simple fine-grained XTL-representation.

In conclusion, the instantiated XTL document is transformed back to \texttt{HXT} (see sect.\ref{section:DesignInstantiation}).
In addition to that, validation turns both XTL representations, for schema and instance, into the \texttt{Reg}-model representing regular expressions (see sect.\ref{section:DesignValidationSemantic}).
A transformation back to \texttt{HXT} is not needed.\\

The models \texttt{HXT, XTL} and \texttt{Reg} are concretisations. All these models have equivalent expressibility, but each model has unique elements characteristic only for those.
The \texttt{HXT}-model is only partially considered.
All concretisations obey the following strict ordering:

$$\texttt{HXT} \ll \texttt{XTL} \ll \texttt{Reg}$$

The model \texttt{NFA} is only sketched.
It is only to be an alternative proposition.
Besides the disadvantages of sect.\ref{section:DesignNFA} NFAs are not compatible by default with hedge-based data models like \texttt{HXT} and \texttt{XTL}.
That is because NFAs are hard to represent as a tree.
After all they are graphs in general.

The coverage of the domain determines the completeness of a model transformation.
Coverage is considered separately in each section.

\subsubsection{HXT}

The denotational semantic uses functions from the Haskell XML Toolkit \cite{HXT:04}, solely for input and output processing of XML documents (see sect.\ref{section:Implementation}).
The contained data structure \lit{\texttt{XmlTree}} is the foundation for further processing using the toolkit.
An XML-node \lit{\texttt{XmlTree}} and a hedge \lit{\texttt{XmlTrees}} are defined in Haskell as follows:

\switchHaskell
\begin{center}
\begin{tabular}{l}
\begin{lstlisting}
type XmlTree  = NTree XNode
	 XmlTrees = [XmlTree] .
\end{lstlisting}
\end{tabular}
\end{center}

The \textit{type constructor} \lit{\texttt{NTree}} represents multi-way trees in general and is defined in \textit{GHC} as

\switchHaskell
\begin{center}
\begin{tabular}{l}
\begin{lstlisting}
type NTree a = NTree a [NTree a] .
\end{lstlisting}
\end{tabular}
\end{center}

Here, the node type \lit{\texttt{XmlTree}} explicitly denotes \lit{\texttt{XNode}}, so child nodes may only be one of these:

\switchHaskell
\begin{center}
\begin{tabular}{l}
\begin{lstlisting}
XText String
| XAttr QName
| XTag QName XmlTrees
\end{lstlisting}
\end{tabular}
\end{center}

\lit{\texttt{QName}} denotes a qualified name. It may occur in element nodes.
In \textit{HXT}, these are composed of the type constructor \lit{\texttt{QN}}, a namespace prefix, a local identifier and a URI.
The ordering is defined as:

\switchHaskell
\begin{center}
\begin{tabular}{l}
\begin{lstlisting}
type QName = QN ns local uri
\end{lstlisting}
\end{tabular}
\end{center}

So, the XML-node \texttt{<a id="1"/>} is represented as \texttt{XmlTree} as following:

\switchHaskell
\begin{center}
\begin{tabular}{l}
\begin{lstlisting}
NTree (XTag (QN "" "a" "") 
  [NTree (XAttr (QN "" "id" "") 
    [NTree (XText "1") []])])
\end{lstlisting}
\end{tabular}
\end{center}

The disadvantages of HXT are obvious.
Even simple \lit{\texttt{XmlTrees}} are very long and heavily loaded with brackets.
So, it would be quite hard to experience the benefits of pattern-matching here.
Since a clear and simple denotational semantic of a node is a precondition for simple processing semantic in general, composed combinators (see sect.\ref{section:Analysis}) are not equivalent.

Although the usability of the \lit{\texttt{XmlTree}}-model in HXT is difficult at least, the amount of features in HXT is quite big (see sect.\ref{section:Implementation}).

Multiple functions and constructs overlay and are usable in some special cases only.

Implicit assumptions often cannot be seen by a function's name.
For example, the type constructor \lit{\texttt{XAttr}} and the constructor function  \lit{\texttt{xattr}} can build up attribute nodes.
\lit{\texttt{xattr}} implicitly insists attributes are specified first for children node constructions.
This circumstance and unintentional constructor errors cause tree constructions to become hard to read and bloated quickly.
Another issue is that all data types, type definitions and functions are loaded into their environment.
The loading takes place as soon as an HXT-module is imported.
A restricted module import command can resolve this problem.
However, it requires a high level of awareness and can become very easy uncontrollable even with few imports.

Another severe problem is the too lax syntax of an XML node.
So, many syntactic correct nodes may be generated which, however, are semantically not sound.
Semantic mistakes may only be detected while serialising a document, only by throwing an exception.
Otherwise, they will remain unnoticed.
For example, attributes could accidentally be mistaken for element nodes because they have the \lit{\texttt{XNode}} too.
Also, the definition of a node in HXT does not prohibit an element node as an attribute node. 
The localisation of non-matching functions for that reason is a significant flaw.
Often errors may only be localised manually by analysing the call stack. However, this is not sufficient.
Mainly, due to a lack of good tool support for Haskell despite current attempts (cf. for instance with \cite{GHC}, \cite{VHask}), the motivation rises even more to make data models and function as simple as possible.

\subsubsection{XTL}
\label{section:DesignXTL}

The algebraic data type \texttt{XTL} is defined as

\switchHaskell
\begin{center}
\begin{tabular}{l}
\begin{lstlisting}
data XTL = XAtt String String
 | XTxt String
 | XInclude String
 | XMacro String [XTL]
 | XCallMacro String
 | XIf String [XTL]
 | XForEach String [XTL]
 | ElX String [(String,String)] [XTL]
 | TxtX String
\end{lstlisting}
\end{tabular}
\end{center}

Here, \lit{\texttt{XAtt}} defines an attribute entry consisting of name and value.
\lit{\texttt{XTxt}} denotes an XTL text node (see sect.\ref{section:mainXTL}).
The type constructors \lit{\texttt{XTxt}}, \lit{\texttt{XInclude}}, \lit{\texttt{XIf}},
\lit{\texttt{XForEach}} have \lit{\texttt{select}}-expression as its first \texttt{String}.
\lit{\texttt{XMacro}}, \lit{\texttt{XIf}}, \lit{\texttt{XForEach}} denote a macro definition,
condition and cycle. All of those have an \lit{\texttt{ElX}} with \texttt{[XTL]} as a hedge.
\lit{\texttt{TxtX}} denotes an arbitrary XML text node.
\lit{\texttt{ElX}} is an XML node composed of a name, a list of attribute entries, and a child node hedge.\\

The mapping of HXT-nodes also affects XTL-tags (see sect.\ref{section:Basics}), element and text nodes (see tab.\ref{table:DesignHXN2XTL}).
Element nodes and XTL-tags differ, for instance, when control should (not) depend on data.
Comments and  \textit{Processing-Instruction}-nodes are not considered.
XTL-tags not matching with command-tags should match with rule (El) and therefore should be transformed into usual element nodes.
\lit{\texttt{children2}} on the right-hand side of the rules (M), (If), (FE) and (El) denotes recursive continuation of the mapping onto the hedge \lit{\texttt{children}}.
\var{qn2} denotes a qualified name as a string, which \var{qn} generates.
Because of the restriction of the HXT-model, the mappings are not injective, but they are surjective because all elements of \lit{\texttt{XTL}} are covered.
Therefore an inverse mapping exists, which recursively applied to \lit{\texttt{children2}} results in \lit{\texttt{children}}.

The node \texttt{<a id="1"\/><b/></a>} is represented in XTL as:

$$\texttt{ElX "\/a\/" \/ [("\/id\/","\/1\/")] [ElX "\/b\/" \/ [] []]}$$

\begin{table} \quad
\parbox{9cm}{
\begin{tabular}{l}
\/ \\

\myrules{$(@)$}{\parbox{6.7cm}{\var{a} = \texttt{QN "xtl" \/ "\/attribute" \/ _}\\
\var{n} = \texttt{NTree (XAttr (QN "\/" \/ "name" \/ "\/"))}\\
          \qquad \texttt{\qquad \qquad  [NTree (XText \var{name}) []]}\\
\var{s} = \texttt{NTree (XAttr (QN "\/" \/ "\/select" \/ "\/"))}\\
          \qquad \texttt{\qquad \qquad [NTree (XText \var{select}) []]}\\
\texttt{NTree (XTag \var{a} [\var{n}, \var{s}]) [] }}}
{\texttt{XAtt \var{name} \var{select}}}\\\\

\myrules{$(\#)$}{\parbox{6.3cm}{\var{a} = \texttt{QN "xtl\/" \/ "text" \/ _}\\
\var{s} = \texttt{NTree (XText \var{select}) []}\\
\texttt{NTree (XTag \var{a} [NTree}\\
  \qquad \texttt{\qquad \qquad (XAttr (QN "\/"  \/ "\/select" \/ "\/")) [\var{s}]]) []}}}
{\texttt{XTxt \var{select}}}\\\\

\myrules{$(I)$}{\parbox{7.5cm}{\var{a} = \texttt{QN "xtl" \/ "\/include" \/ _}\\
\var{s} = \texttt{NTree (XText \var{select}) []}\\
\texttt{NTree (XTag \var{a} [}\\
  \qquad \texttt{\qquad \qquad NTree (XAttr (QN "\/" \/ "\/select" \/ "\/")) [\var{s}]]) []}}}
{\texttt{XInclude \var{select}}}\\\\

\myrules{$(M)$}{\parbox{6cm}{\var{a} = \texttt{QN "xtl" \/ "macro" \/ \_}\\
\var{t} = \texttt{NTree (XText \var{mname}) []}\\
\texttt{NTree (XTag \var{a} [NTree (XAttr (QN "\/" \/ "name" \/ "\/")) [\var{t}]]) \var{children}}}}
{\texttt{XMacro \var{mname} \var{children2}}}
\end{tabular}
}\\
\parbox{9cm}{
\begin{tabular}{l}
\/ \\
\myrules{$(C)$}{\parbox{6.3cm}{\var{a} = \texttt{QN "xtl" \/ "\/call\-macro" \/ \_}\\
\var{t} = \texttt{NTree (XText \var{mname}) []}\\
\texttt{NTree (XTag \var{a} [NTree}\\
  \qquad \texttt{\qquad \qquad (XAttr (QN "\/" \/ "name" \/ "\/")) [\var{t}]]) []}}}
{\texttt{XCallMacro \var{mname}}}\\\\

\myrules{$(If)$}{\parbox{7.5cm}{\var{a} = \texttt{QN "xtl" \/ "\/if" \/ _}\\
\var{t} = \texttt{NTree (XText \var{select}) []}\\
\texttt{NTree (XTag \var{a} [NTree}\\
  \qquad \texttt{\qquad \qquad (XAttr (QN "\/" \/ "\/select" \/ "\/")) [\var{t}]]) \var{children}}}}
{\texttt{XIf \var{select} \var{children2}}}\\\\

\myrules{$(FE)$}{\parbox{7.5cm}{\var{a} = \texttt{QN "xtl" \/ "for-each" \/ _}\\
\var{t} = \texttt{NTree (XText \var{select}) []}\\
\texttt{NTree (XTag \var{a} [NTree}\\
 \qquad \texttt{\qquad \qquad (XAttr (QN "\/" \/ "\/select" \/ "\/")) [\var{t}]]) \var{children}}}}
{\texttt{XForEach \var{select} \var{children2}}}\\\\

\myrules{$(El)$}{\parbox{5cm}{\texttt{NTree (XTag \var{qn} \var{atts}) \var{children}}}}
{\texttt{ElX \var{qn2} \var{atts2} \var{children2}}}\\\\

\myrules{$(Txt)$}{\parbox{3.7cm}{\texttt{NTree (XText \var{text}) []}}}
{\texttt{TxtX \var{text}}}
\end{tabular}
}\\
\caption{Mapping HXT $\mapsto$ XTL}
\label{table:DesignHXN2XTL}
\end{table}

The XTL data model is used for instantiation and validation.
Although instances do not have command tags, it is beneficial for unified semantics to express instances by XTL.
It may also be used as an automated schema generator (see sect.\ref{section:DesignReg}) or as a schema parser (see sect.\ref{section:ComparisonMain}).

\subsubsection{Reg}
\label{section:DesignReg}

The regular data model \lit{\texttt{Reg}} is defined in Haskell as shown in fig.\ref{figure:DataReg}:

\begin{figure}[h]
\begin{center}
\begin{tabular}{l}
\begin{lstlisting}
data Reg = MacroR String
 | AttrR String String
 | TextR String
 | IncludeR String
 | ElR String [(String,String)] Reg
 | TxtR String
 | Epsilon
 | Or Reg Reg
 | Then Reg Reg
 | Star Reg
\end{lstlisting}
\end{tabular}
\end{center}
 \caption{Data model \texttt{Reg}}
 \label{figure:DataReg}
\end{figure}

\lit{\texttt{Reg}} follows the model presented in sect.\ref{section:DesignXTL}.
\lit{\texttt{AttrR}}, \lit{\texttt{TextR}}, \lit{\texttt{IncludeR}}, \lit{\texttt{TxtR}} and
\lit{\texttt{ElR}} represent literals.
Although element nodes are recursive, each can still be considered literal, especially when they are empty hedges or processed.
The type constructor \lit{\texttt{AttrR}} has the same structure as \lit{\texttt{XAtt}}, 
\lit{\texttt{TextR}} the same as \lit{\texttt{XTxt}} and \lit{\texttt{IncludeR}}
as \lit{\texttt{XInclude}}.
\lit{\texttt{Epsilon}} denotes the empty word, \lit{\texttt{Or}} denotes selection, \lit{\texttt{Then}} denotes concatenation and \lit{\texttt{Star}} denotes arbitrary repetition (cf. \cite{Tho:2000}).

In contrast to sect.\ref{section:DesignXTL}, regular expressions may not contain arbitrary macro calls.
Otherwise, this could be considered as a non-right congruent derivation --- this would be a context-free derivation.
Regular expressions still can contain macro calls and can be unrolled initiated by a caller.
In the following only those macros shall be considered whose derivation terminates.
This statement means whose expression is regular.
Replacement of regular expressions by other regular expressions preserves regularity according to the definition of tree grammars (see sect.\ref{section:TreeGrammars}).

The node

\switchXTL
\lstset{tabsize=2}
\lstset{basicstyle=\ttfamily\small}
\lstset{numbers=none}
\lstset{numberstyle=\tiny}
\lstset{showstringspaces=false}
\lstset{captionpos=b}

\begin{center}
\begin{tabular}{c}
\begin{lstlisting}
<a id="1">
 <xtl:attribute name="title" 
   select="//AAA" />
 <b/>
</a>
\end{lstlisting}
\end{tabular}
\end{center}

matches \lit{\texttt{Reg}}:

\switchHaskell
\begin{center}
\begin{tabular}{c}
\begin{lstlisting}
ElR "a" [("id","1")]
	Then (AttrR "title" "//AAA")
		 (Then (ElR "b" [] Epsilon) Epsilon)
\end{lstlisting}
\end{tabular}
\end{center}

\begin{figure}[h]
\begin{flushleft}
\begin{minipage}{9cm}
\parbox{8.7cm}{
	\mbox{
	 \subfigure[]{$Macro_{mname}$} \qquad
	 \subfigure[]{$Text_{select}$} \qquad
	 \subfigure[]{$\#_{text}$} \qquad
	 \subfigure[]{$Include_{select}$}
        }\\

        \mbox{
         \subfigure[]{\begin{minipage}{2cm}\xymatrixrowsep{20pt}\xymatrixcolsep{30pt}\xymatrix{e_{name,[(.,.)]} \ar[d]\\Reg}\end{minipage}} \qquad
         \subfigure[]{\begin{minipage}{2cm}\xymatrixrowsep{20pt}\xymatrixcolsep{5pt}\xymatrix{&Then \ar[ld]\ar[rd] &\\Reg && Reg}\end{minipage}} \qquad
         \subfigure[]{\begin{minipage}{2cm}\xymatrix{Star \ar[d]\\Reg}\end{minipage}} \qquad
        }\\
	
	\mbox{
        \subfigure[]{\begin{minipage}{2cm}\xymatrixrowsep{20pt}\xymatrixcolsep{5pt}\xymatrix{&Or \ar[ld]\ar[rd] &\\Reg && Reg}\end{minipage}} \qquad
        \subfigure[]{\begin{minipage}{2cm}\xymatrixrowsep{20pt}\xymatrixcolsep{5pt}\xymatrix{&@ \ar[ld]\ar[rd] &\\name && value}\end{minipage}} \qquad
        \subfigure[]{$\varepsilon$}
        }
}
\end{minipage}
\end{flushleft}
  \caption{Graphical Representation of \texttt{Reg}-nodes}
  \label{figure:DesignRegTrees}
\end{figure}
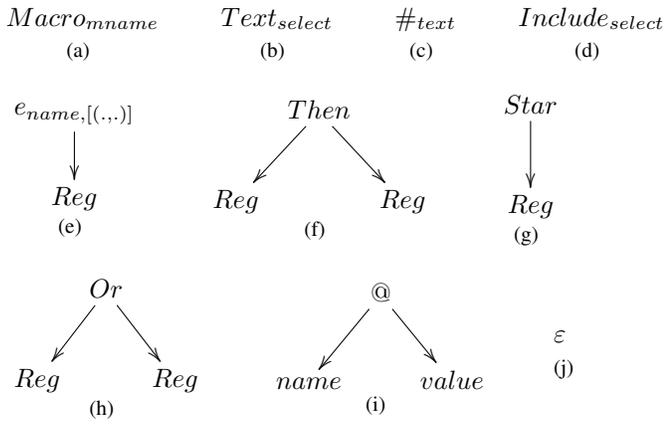


The introduced regular expressions can be interpreted as \textit{OBDD}.
OBDDs are graphical representations of terms consisting of variables and terms again, constants and binary descriptors.
When extending the graphical notation proposed in fig.\ref{figure:DesignRegTrees}  by the unary functors \lit{\texttt{Star}} and \lit{\texttt{e}}, an OBDD's representation is fully covered together with \lit{\texttt{Then}}, \lit{\texttt{Or}} and \lit{\texttt{@}}.
Element nodes have a non-empty empty.
Element nodes are already tree-structured and therefore can also be interpreted as nodes in OBDDs.
Childless element nodes are leaves in a tree.
This interpretation is isomorphic to boolean terms.
\lit{$\land$} and \lit{$\lor$} represent the binary functors, and boolean variables represent leaves.

The consideration of OBDDs has mainly two advantages here.
First, the difference between nodes and hedge vanishes.
A node representing a hedge with precisely one child is represented the same as a hedge with multiple children by \texttt{Then}.
Second, the set of alternatives is represented the same way (cf. \cite{Tho:2000}, \cite{Cla:2007}).
One advantage of a binary tree over a multi-way tree is a simpler specification of nodes.

Furthermore, the graphical notation presented makes functions safer because fewer exceptions and cases need to be distinguished, and in conclusion there are fewer places to commit an error by the developer.
There is either an \lit{\texttt{Epsilon}} or a \lit{\texttt{Then}}, where it is agreed \lit{\texttt{Then}} may not have an \lit{\texttt{Epsilon}} as its left child.
Same as lists, the OBDD-notation allows \textit{lazy evaluation}, so infinite OBDDs are a meaningful completion to the test cases from sect.\ref{section:Implementation}.

\begin{figure}[h]
\begin{center}
	\subfigure[]{
 \begin{minipage}{2cm}
  \xymatrixrowsep{10pt}
  \xymatrixcolsep{10pt}
  \xymatrix{
     &  & Then \ar[dl] \ar[dr] & \\
     & Then \ar[dl] \ar[d] &  & Then \ar[dl] \ar[d]\\
    a_0 & a_1 & ... & Then \ar[dl] \ar[d] \\
     &  & a_{n-1} & Then \ar[dl] \ar[d] \\
     &  & a_n & \varepsilon
  }
  \end{minipage}
        }\\
	\subfigure[]{
 \begin{minipage}{2cm}
  \xymatrixrowsep{10pt}
  \xymatrixcolsep{10pt}
  \xymatrix{
     &  & Then \ar[dl] \ar[dr] & \\
     & Then \ar[dl] \ar[d] &  & Then \ar[dl] \ar[d]\\
    a_0 & Then \ar[dl] \ar[d] & a_{n-1} & Then \ar[dl] \ar[d] \\
    a_1 & ... & a_n & \varepsilon
  }
  \end{minipage}
        }
\end{center}
  \caption{Two different orderings of \texttt{Then}}
  \label{figure:DesignReg3Thens}
\end{figure}
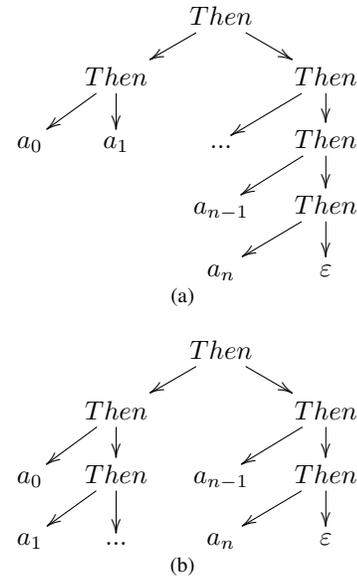

A set of alternatives $\{a_{0},...,a_{n}\}$ can within a \lit{\texttt{Then}} be arranged in different ways (see fig.\ref{figure:DesignReg3Thens}).
For instance, the \lit{\texttt{XTL}}-node:

$$\texttt{ElX "\/a\/" \/ [("\/id\/","\/1\/")] [ElX "\/b\/" \/ [] []]}$$

is transferred into the \lit{\texttt{Reg}}-node:

$$\texttt{ElR "\/a\/" \/ [("\/id\/","\/1\/")] Then (ElR "\/b\/" \/ []}$$
$$\texttt{ Epsilon) Epsilon}$$

The composed node of XTL-command tags:

\switchXTL
\lstset{tabsize=2}
\lstset{basicstyle=\ttfamily\small}
\lstset{numbers=none}
\lstset{numberstyle=\tiny}
\lstset{showstringspaces=false}
\lstset{captionpos=b}

\begin{center}
\begin{tabular}{c}
\begin{lstlisting}
<book>
 <title>Haskell</title>
  <xtl:for-each select="//authors">
   <author>
     <xtl:text select="."/>
   </author>
 </xtl:for-each>
</book>
\end{lstlisting}
\end{tabular}
\end{center}

is transformed into a \lit{\texttt{Reg}}:

\switchHaskell
\begin{center}
\begin{tabular}{c}
\begin{lstlisting}
ElR "book" []
  Then
    (ElR "title" []
      Then (TxtR "Haskell") Epsilon)
    Then
      (Star 
        (ElR "author"
         Then (TextR ".") Epsilon))
      Epsilon
\end{lstlisting}
\end{tabular}
\end{center}

The normalisation of regular expressions can simplify the validation.
Arbitrary orderings of \lit{\texttt{Then}} and \lit{\texttt{Or}} are disallowed, significantly reducing the number of rules to be considered.
It is agreed that a regular expression is in normal form, if:

\begin{itemize}
 \item No two text nodes are neighbours. Text inclusions are exempted.

Strings can be separated only with difficulties.
It is because it is hard to decide with the truncation of a string is correct.
The intention of two \lit{\texttt{xtl:text}} could be the merge into one given string. That is why a truncation may not make sense.

 \item \lit{\texttt{Then}} and \lit{\texttt{Or}} are right-associative.

It implies that there is a regular expression on both the left and the right, and the left expression does not match with the parent node

 \item \lit{\texttt{XAtt}} directly underneath a \lit{\texttt{Then}} as the first contiguous sequence in the corresponding hedge \lit{\texttt{ElR}} (cf. \cite{XTLSpec:2007}).
\end{itemize}

Using OBDDs qualitative properties may be probed (see sect.\ref{section:Implementation}, cf. \cite{Cla:2000}) because fewer base cases require consideration.\\

The mapping \texttt{XTL} $\mapsto$ \texttt{Reg} is total.
Each \lit{\texttt{XTL}}-element is injectively assigned to an element from \lit{\texttt{Reg}}.
If \lit{\texttt{select}}-expressions are ignored (see sect.\ref{section:Analysis}) and arbitrary non-fixed repetitions can be introduced using \textit{Kleene's star operator}.
The mapping is from now on, then no more invertible, s.t. the origin is reproducible exactly.
\lit{\texttt{XMacro}} can be mapped onto empty since it is just added to a macro environment, and no real regular expression is associated with it.

\begin{table}[h]
\begin{center}
\begin{tabular}{c}
\parbox{8cm}{
\begin{tabular}{r c l}
\texttt{XIf _ l} & $\mapsto$ & \texttt{Or Epsilon l2}\\
\texttt{XForEach _ l} & $\mapsto$ & \texttt{Star l2}\\
\texttt{XAtt name value} & $\mapsto$ & \texttt{AttrR name value}\\
\texttt{XTxt select} & $\mapsto$ & \texttt{TextR select}\\
\texttt{XInclude select} & $\mapsto$ & \texttt{IncludeR select}\\
\texttt{TxtX text} & $\mapsto$ & \texttt{TxtR text}\\
\texttt{ElX name atts l} & $\mapsto$ & \texttt{ElR name atts l2}\\
\texttt{XCallMacro mname} & $\mapsto$ & \texttt{MacroR mname}
\end{tabular}}
\end{tabular}
\end{center}
 \caption{Mapping XTL $\mapsto$ Reg}
 \label{table:DesignXTL2Reg}
\end{table}

The exact mapping is in tab.\ref{table:DesignXTL2Reg}.

\lit{\texttt{l2}} denotes hedges recursively generated by the hedge \lit{\texttt{l}}.
The inverse mapping from \lit{\texttt{Reg}} onto \lit{\texttt{XTL}} is not possible, not even by introducing "'\textit{don't-care}"' variables 
\lit{_} or referring to implicit macro environments.
However, validation does not insist on it.

\subsubsection{Non-deterministic Finite Automaton}
\label{section:DesignNFA}

As already mentioned in sect.\ref{section:TreeAutomataMatchers}, regular tree automata just perfectly fit when it comes to recognising tree-structured regular input data, for instance, as XML documents are (cf. \cite{Mur:1999}).
The construction of a Non-deterministic Finite Automaton (NFA) over hedges is similar to that over strings.
It follows the so-called "`toolbox"'-principle.
This principle states all nodes are applied nodes successively, for instance, during concatenation.
After application, the end-points of one component are connected, so the resulting automaton grows (cf. \cite{Tho:2000}).\\

The partial-derivatives algorithm \cite{Brz:64} derives only symbols lying in $\pi(t)$, where $\pi$ denotes the set of all good beginnings, and $t$ denotes an arbitrary yet to be determined regular expression.
Due to the homomorphism for regular expressions (see sect.\ref{section:AnalysisProperties}), the calculations can be placed within the remainder field modulo a literal.
In the congruency $a \equiv b$ \texttt{mod ($c$)}, $a$ stands for an initial regular expression, $b$ stands for the abstracted congruency reduced by $c$, and $c$ is the remainder partition or, with other words, another possible beginning of $a$, so $c \in \pi(a)$.

The construction of the corresponding NFA always depends on the current derivation.
All calculated derivations are put into a hashing table.
Every time an expression is derived, it is first checked whether this derivation is already in the hashing table.
For the original schema, $x^{*} \cdot (xx+y)^{*}$, either $x$ (4.1) or $y$ (4.3) can be derived.
Due to the star-operator (4.1) can either have the same expression again or an $x$, because $x^{*}$ would have been removed from $x^{*} \cdot (xx+y)^{*}$ and the second subexpression $(xx+y)^{*}$ $x$ would follow after $x$, which, however, would follow another $(xx+y)^{*}$ (4.2).
The congruencies (4.4)-(4.6) can be obtained after further derivations.
Until (4.6) all right sides are determined. So, the corresponding NFA has no new transitions to be added (cf. app.\ref{appendix:PartialDerivativesAlgorithm}).

\begin{center}
\begin{tabular}{l}
\parbox{8.3cm}{
\begin{tabular}{rlcll}
$x^{*} \cdot (xx+y)^{*}$ & $\equiv$ & $x^{*} \cdot (xx+y)^{*}$ & $\texttt{mod(}x\texttt{)}$ & (4.1) \\
$x^{*} \cdot (xx+y)^{*}$ & $\equiv$ & $x \cdot (xx+y)^{*}$ & $\texttt{mod(}x\texttt{)}$ &  (4.2) \\
$x^{*} \cdot (xx+y)^{*}$ & $\equiv$ & $(xx+y)^{*}$ & $\texttt{mod(}y\texttt{)}$ & (4.3) \\
$x \cdot (xx+y)^{*}$ & $\equiv$ & $(xx+y)^{*}$ & $\texttt{mod(}x\texttt{)}$ & (4.4) \\
$(xx+y)^{*}$ & $\equiv$ & $x \cdot (xx+y)^{*}$ & $\texttt{mod(}x\texttt{)}$ & (4.5) \\
$(xx+y)^{*}$ & $\equiv$ & $(xx+y)^{*}$ & $\texttt{mod(}y\texttt{)}$ & (4.6)
\end{tabular}}
\end{tabular}
\end{center}

Such an approach is appropriate for complex schemas, which may be reused (cf. sect.\ref{section:Analysis}).
The state-based approach is also appropriate for error location.
However, multiple transitions leaving a terminal and $\varepsilon$-transitions may make recognition not determined.
Determination of the NFA (see \cite{Du:2001}) comes for Rabin-Scott's powerset construction cost.
Validation would be beneficial only if the automaton were determined. The additional cost pays off if multiple instances are validated with the same DFA.

On the one side, there is the direct approach as described in sect.\ref{section:DesignInstantiationSemantic}, \ref{section:DesignValidationSemantic}.
On the other side, there is the graph-based approach.
The validation problem can be interpreted as a \textit{path problem} in a graph.

\subsection{Instantiation}
\label{section:DesignInstantiation}

Before presenting the denotational semantics for instantiator and validator, a brief discussion on semantics should sum up the pros and cons.

\subsubsection{Semantics Form}
As already described in sect.\ref{section:Analysis}, the untyped $\lambda$-calculus and attribute grammar are not appropriate for describing the instantiation semantics.

A logical model seems reasonable at first glance since a matcher algorithm would be needed to be designed (cf. sect.\ref{section:Analysis}).
A first implementation may even allow backtracking until an optimisation could be found.

The structure to be matched against can undoubtedly be represented as term expression (cf. \cite{Hab:2006}).
Schema, instance, and helper functions can be transformed into Horn-clauses.
Relations can only poorly express functions since logical programming mainly interprets terms and relations.
Functions, however, have "`only"' static mappings.
Apart from that, future implementations in an object-oriented programming language should guarantee referential transparency, and the evaluation ought to proceed forward.
Concepts like backtracking and cuts in a logical programming language, however, disallow this --- to mention some on Prolog, for instance.\\

For these reasons, the semantics of instantiation and validation should be applied to the functional paradigm (cf. \cite{All:1988}, \cite{Ten:1976}).

The semantics shall be as easy as possible using type constructors, so a comprehending implementation in Haskell matches the denotational semantics.
Node specifications using type constructors describe interfaces and attributes of classes (cf. sect.\ref{section:ImplementationHaskell2Java}).

\subsubsection{Semantic}
\label{section:DesignInstantiationSemantic}

The complete semantics for instantiation and validation are enclosed in app.\ref{appendix:DenotationalSemantics}
Instantiation turns a template into an instance. The instance document can be represented as XTL-term (cf. sect.\ref{section:DesignXTL}).
The denotational semantics is described in Haskell.
Next, some helper functions will be described in the denotational semantic:

The functions:

\begin{center}
\begin{tabular}[t]{l}
\texttt{filter::(a$\rightarrow$Bool)$\rightarrow$[a]$\rightarrow$[a]} and \\
\texttt{concatMap::(a$\rightarrow$[b])$\rightarrow$[a]$\rightarrow$[b]}
\end{tabular}
\end{center}

are defined in the GHC-package \texttt{GHC.List}.
The function \lit{\texttt{filter}} filters any given list using a predicate (means a function returning a boolean value here).
So, for instance, \texttt{filter (odd) [1..10]} returns as result the list \texttt{[1,3,5,7,9]}.
The function \lit{\texttt{concatMap}}
applies for any given list each element in some given mapping, where for each list element, a list type is returned as an element of the co-domain.
Later all created sub-lists are concatenated with each other.
\texttt{concatMap (\textbackslash x->['a']) [1..10]} returns \texttt{"\/aaaaaaaaaa"}.

The function \lit{\texttt{qSort}} of type \lit{\texttt{Ord a$\Rightarrow$[a]$\rightarrow$[a]}} is a generalised (so-called lifted) function sorting a list of any comparable type.
\lit{\texttt{qSort}} is used for canonisation.

The functions $f_0^1, f_0^2, f_0^3, f_0^4$ denote the agreed PHP functions (cf. sect.\ref{section:Analysis}).
These functions describe in the given order access to external texts, the filtering mode if there are multiple solutions, the checking on satisfiability and the return of an element node.
The corresponding types can be found in tab.\ref{table:DesignPHPFunctionsTyping}.
The type \lit{\texttt{a}} denotes arbitrary instantiation elements, for instance, element nodes for XML documents.
Be aware of \lit{\texttt{XInclude}}. It always produces an \lit{\texttt{XmlTree}} to the instance document.
The result of function $f_0^2$ is \lit{\texttt{[a]}}.
That means the polymorphic results are being passed through as context to the loop body.

\begin{table}
\begin{center}
\begin{tabular}{|ccl|}
	\hline
	Funktion & XTL-Konstruktor & Typung\\
	\hline
	$f_0^1$ & \texttt{XTxt, XAtt} & \texttt{String$\rightarrow$a$\rightarrow$String}\\
	$f_0^2$ & \texttt{XForEach} & \texttt{String$\rightarrow$a$\rightarrow$[a]}\\
	$f_0^3$ & \texttt{XIf} & \texttt{String$\rightarrow$a$\rightarrow$Bool}\\
	$f_0^4$ & \texttt{XInclude} & \texttt{String$\rightarrow$a$\rightarrow$XmlTree}\\
	\hline
\end{tabular}
\end{center}
  \caption{Placeholder-Plugin functions $f_0^1,f_0^2,f_0^3,f_0^4$}
  \label{table:DesignPHPFunctionsTyping}
\end{table}

The helper function \texttt{getMacro} of type \lit{\texttt{[XTL]$\rightarrow$[XTL]}} from (A3) returns for a macro environment $\mu$ the corresponding macro body.
It is further assumed, a matching macro is defined before calling it.
Otherwise, an according to error message shall be dumped.
The surrounding rule by name $\var{m1}$ transfers the name of the wanted macro definition.

The abbreviation (S) stands for the starting rule of instantiation, (E) stands for eliminating XTL-attributes, which are located directly under the top-level element node.
Rules (I1)-(I3) initiate instantiation by filtering all macro-definitions first and passing those to the instantiation second.
Rules (A1)-(A7) are the core instantiation rules for XTL-tags and non-XTL-tags.

The helper function $\varepsilon^{MA}$ of type \lit{\texttt{XTL}$\rightarrow$\texttt{Bool}} checks for a given \lit{\texttt{XTL}}-node if it is, in fact, an XTL-attribute or not.
In analogy, $\varepsilon^{MM}$ checks for macro-definitions.
The semantic fields include macro-environment and context (see app.\ref{appendix:DenotationalSemantics}).
Here, a macro environment $\mu$ is built up in (I3).
$\mu$ consists of the mapping "`Macro-Name$\times$XTL-Hedge"'.
The hedge represents a list.
The macro-environment $\mu$ is an invariant part in $\varepsilon^{\alpha}$ (see app.\ref{appendix:DenotationalSemantics}).
Tab.\ref{table:DesignMacroEnvironmentExample} demonstrates the macro environment's evaluation (cf. sect.\ref{section:mainXTL}) until the selected row.

\begin{table}
\begin{center}
\parbox{8.3cm}{
\begin{tabular}{|c|l|l|}
	\hline
	\textbf{Line}	&	\textbf{Document} & $\mu$ \\
	\hline
	\hline
	0 & ...										&	$\emptyset$ \\
	1 & \texttt{<xtl:macro name="x"\/>} 		& 	$\{x/\texttt{<a/>}\}$ \\
	2 & \texttt{\quad <a/>}						& 	\\
	3 & \texttt{</xtl:macro>}					& 	\\
	4 & \texttt{<xtl:macro name="y"\/>}			& 	$\{x/\texttt{<a/>}\}$ \\
	5 & \texttt{\quad <b/><b/>}					& 	\\
	6 & \texttt{</xtl:macro>}					& 	$\{x/\texttt{<a/>}, y/\texttt{<b/><b/>}\}$ \\
	\hline
\end{tabular}}
\end{center}
  \caption{Macro Environment $\mu$ for an example}
  \label{table:DesignMacroEnvironmentExample}
\end{table}

Context has multiple purposes, one of which is a simplification.
For instance, the simplification of XPath-expressions (see sect.\ref{section:Basics}, \ref{section:Analysis}).
It returns a node or a value of arbitrary type and is used while calling PHP functions as the instantiation data source.

The mappings use the domains from tab.\ref{table:DesignDomainInstantiation}.
\lit{\texttt{XTL}} describes the domains in more detail in sect.\ref{section:mainXTL}.
\lit{\texttt{[XTL]}} denotes a hedge of type \lit{\texttt{XTL}}.
The hedge type is a list in terms of denotational semantic.
It is worth noting the type is \textit{monadic}.

\begin{table}
  \begin{center}
    \begin{tabular}{|cl|}
	\hline
	Domain & Meaning \\
	\hline
	\texttt{Bool} & \{\texttt{True, False}\} \\
	\texttt{XTL} &  XTL data type \\
	\texttt{[XTL]} & hedge of \texttt{XTL} \\
	\hline
    \end{tabular}
  \end{center}
  \caption{Domains of the Instantiation}
  \label{table:DesignDomainInstantiation}
\end{table}

The considered denotational semantic variables are used quite often.
Anonymous and "`don't-car"' variables are marked as \lit{\texttt{_}}.
These variables often stand in constructor specification and cannot be addressed in a rule's body.
There are underlined variables written in italics -- these specify XTL-nodes and fragments of it.
These can also be memoised.
For example, XTL-attributes in (E) can be interrupted and escaped before reaching (E)'s end.
The same counts for non-XTL-attributes in \var{nodes} too.
Non-underlined variables with a Greek letter $\pi$ and $\mu$ denote each PHP-tuple ($f_0^1, f_0^2, f_0^3, f_0^4$) and also the macro environment.
Functions are written in italics, and each has an upper index, which denotes the arity of that function.
The only exceptions are doubly-indexed PHP-function from $f_0^1$ to $f_0^4$.
The arities of those functions are slightly different and are fully listed in tab.\ref{table:DesignPHPFunctionsTyping}.

\textbf{Rules of the denotational semantic}

The rule

(S) \quad $\eval[Start]{\var{x}}{\var{s} \pi}$ :=
 $\eval{\var{x2}}{\var{s} \pi}$ \for \var{x2} = $\eval[r]{\var{x}}{}$ 
 
initiates instantiation.

\var{x} denotes the template, \var{s} denotes the context, and $\pi$ denotes the placeholder-function 4-tuple.
The given template is a non-empty \texttt{XTL}-node.
The context is the source document at the beginning.
Because \var{x} may also have XTL-attribute in the hedge, \var{X} is reduced first of all with $\eval[r]{}{}$.

(E) \/ \parbox[t]{14cm}{$\eval[r]{ElX \var{n} \var{a} \var{c}}{}$ := \\ 
\mbox{\quad \parbox{10cm}{
 \<let> \parbox[t]{14cm}{
		\var{attDefs} = $\func{filter}{2} ( \lambda \var{child}. \eval[MA]{\var{child}}{}) \var{c}$, \\
        \var{nodes} = $\func{filter}{2} ( \lambda \var{child}. \func{not}{1} \; \eval[MA]{\var{child}}{}) \var{c}$} \\
 \<in> \texttt{ElX \var{n} $(\func{qSort}{1} (\var{a} \pplus \var{attDefs})) \; \var{nodes}$}}
}}\\[0.3cm]
It is implicitly assumed that the top-level root node is an element node that may not be an XTL-tag.
The hedge is split into one hedge containing \lit{\texttt{xtl:attribute}} and another hedge containing all others.
\var{attDefs} binds \lit{\texttt{xtl:attribute}}-nodes.
\var{nodes} bind all others.
Then the original attributes \var{a} are united with \var{attDefs} and canonised last.
This list and the remaining nodes \var{nodes} and the unmodified element name \var{n} make up the reduced element node.
Afterwards, instantiation is triggered.\\

The pure instantiation shall be implemented as filter (see sect.\ref{section:AnalysisArrowsFilters}). It returns a list type.
The triggering $\eval[]{}{}$ requires one \lit{\texttt{XTL}} only:

(I1) \/ $\eval{XTxt \var{t}}{\_ \; \_}$ :=
  \texttt{XTxt \var{t}}
  
(I2) \/ $\eval{XAtt \var{n} \var{v}}{\_ \; \_}$ :=
  \texttt{XAtt \var{n} \var{v}}
  
(I3) \/ $\eval{ElX \var{n} \var{a} \var{c}}{\var{s} \pi}$ := \\[-0.5cm]
\begin{center}  \parbox{7.3cm}{\<let> \var{mdefs} = $\func{filter}{2} (\lambda \var{child}.\eval[MM]{\var{child}}{}) \var{c}$,\\
  \var{nodes} = $\func{filter}{2} (\lambda \var{child}.\func{not}{1} \; \eval[MM]{\var{child}}{}) \var{c}$ \\
  \<in> \texttt{ElX \var{n} \var{a} $(\func{concatMap}{2}\\
   \qquad (\lambda \var{node}.\eval[\alpha]{\var{node}}({\var{s}, \var{mdefs}, \pi})) \; \var{nodes})$}}
\end{center}

In both cases, (I1) and (I2) attributes and text nodes are considered.
Case (I3) evaluates one element node, which may have occurrences of macro calls in the case of a hedge.
These are filtered, similar to (E), and filtered afterwards to \var{mdefs}.
Nodes that are no macros are bound to \var{nodes}.
Later the instantiation $\eval[\alpha]{}{}$ proceeds for each non-macro node \var{node}.
$\func{concatMap}{2}$ concatenates obtained hedges.
The instantiation of each \var{node} is passed through referring to source \var{s} and PHP-tuple $\pi$.
The list of all macros for this rule serves in $\eval[\alpha]{}{}$ as macro environment $\mu$ and remains untouched during the remaining instantiation.

The instantiation of \lit{\texttt{XTL}}-nodes has the typing:

$$\eval[\alpha]{}{}\texttt{::XTL}\rightarrow\texttt{a}\rightarrow\texttt{[(String,[XTL])]}$$
$$\rightarrow\texttt{PHP a}\rightarrow\texttt{[XTL]}.$$

This means the hedge of type \texttt{[XTL]} establishes, after an XTL-node, instantiation data of type \lit{\texttt{a}} and a macro environment are passed.
The macro-environment consists of a list of tuples herewith, where each tuple has the mapping "`Macro-Name$\mapsto$Macro-Body"'.
The macro body is a hedge of type \lit{\texttt{[XTL]}} and may contain all XTL-tags except \lit{\texttt{XMacro}}.

Instantiation is described by rules (A1) until (A7).
Notably, the tag \lit{\texttt{XMacro}} is missing.
That is because the extraction of $\mu$ happens before.
Furthermore, all seven rules are complete.
W.r.t. XTL-tags have not imposed any further restrictions.
The rules are not prioritised.
Element nodes that are non-XTL-tags are handled as common element nodes in the semantic (cf. sect.\ref{section:DesignXTL}).

The handling of conditions is in (A1).\\[-0.3cm]

(A1) \/ $\eval[\alpha]{XIf \var{x} \var{gc}}{(\var{s},\mu,(\func{f}{1}_{0},\func{f}{2}_{0},\func{f}{3}_{0},\func{f}{4}_{0}))}$ := \\
\mbox{\qquad \qquad \qquad \parbox{5cm}{\<if> $(\func{f}{3}_{0} \var{x} \var{s})$ \<then> $\func{concatMap}{2}\\
   (\lambda \var{c}.\eval[\alpha]{\var{c}}{}(\var{s},\mu,(\func{f}{1}_{0},\func{f}{2}_{0},\func{f}{3}_{0},\func{f}{4}_{0}))) \var{gc}$\\
  \<else> \texttt{[]}}
}\\

So, the string \var{x} is passed to the PHP-function $f_0^3$ together with \var{s}, the instantiation data source.
If $f_0^3$ succeeds (cf. sect.\ref{section:Analysis}), it returns \lit{\texttt{True}}, \lit{\texttt{False}} otherwise.
The Haskell syntax implies that on success, instantiation continues with child node \var{c}.
In case of error, it returns an empty list.
However, an empty list is neutral w.r.t. list concatenation, and this is why no special handling is required on the caller's side.

The handling of loops is done in (A2).\\[-0.3cm]

(A2) \/ $\eval[\alpha]{XForEach \var{x} \var{gc}}{(\var{s},\mu,(\func{f}{1}_{0},\func{f}{2}_{0},\func{f}{3}_{0},\func{f}{4}_{0}))}$ := \\
\mbox{ \qquad \qquad \qquad  \parbox{5.3cm}{\<let> \var{sels} = $\func{f}{2}_{0} \var{x} \; \var{s}$ \\
  \<in> $\func{concatMap}{2} \; (\func{f3}{1}) \; \var{sels}$ \\
  \for $\func{f3}{1} = \lambda \var{c}. \func{concatMap}{2}\\
 (\lambda \var{c2}. \eval[\alpha]{\var{c2}}{(\var{c}, \mu, (\func{f}{1}_{0},\func{f}{2}_{0},\func{f}{3}_{0},\func{f}{4}_{0}))}) \var{gc}$}
}\\

The variable \var{sels} binds all nodes of type \lit{\texttt{a}}, which establish during the evaluation of the \lit{\texttt{select}}-expression \var{x}, the source of instantiation data \var{s}, and the PHP-function $f_0^2$.
In case \var{sels} equals an empty list, then function $\func{concatMap}{2}$ results in the empty set.
That is because Haskell has a \textit{non-strict} evaluation order.
Otherwise, first, instantiation continues for each child node \var{c2} of hedge \var{gc}, and second, all determined hedges are then concatenated together into eventually one resulting hedge.
The list \var{sels}\texttt{::[a]} is successively propagated through to $\func{concatMap}{2}$ using the bound variable \var{c}.
Every element in \var{sels} has type \lit{\texttt{a}} and is passed over to each child node \var{c2} for hedge \var{gc} as new instantiation data \var{c}.

The handling of macro calls is done in (A3).\\[-0.3cm]

(A3) \/ $\eval[\alpha]{XCallMacro \var{m1}}{(\var{s},\mu,\pi)}$ := \\
\mbox{\qquad \qquad \qquad  \parbox{6cm}{\<let> $\mu_{2}$ = $\func{getMacro}{1} \mu$ \\
  \<in> $\func{concatMap}{2} (\lambda \var{m}. \eval[\alpha]{\var{m}}{(\var{s},\mu,\pi)}) \; \mu_{2}$
}
}\\

The function \lit{\texttt{getMacro}} was mentioned initially.
It determines for a given macro name \var{m1} and a macro-environment $\mu$ the corresponding macro body $\mu_2$, a hedge.
The instantiation proceeds for each node sequentially from the hedge with the same initial context \var{s}, PHP-tuple $\pi$ and the same macro-environment $\mu$.
The resulting hedge as the list is concatenated, so this rule's returned value has the type \lit{\texttt{[XTL]}}.

The inclusion of text follows rule (A4).\\[-0.3cm]

(A4) \/ $\eval[\alpha]{XTxt \var{t}}{(\var{s},\mu,(\func{f}{1}_{0},\_,\_,\_))}$ :=\\
  \mbox{\qquad \qquad \qquad \texttt{[ XTxt $\func{f}{1}_{0} \; \var{t} \; \var{s}$ ]}}\\[-0.3cm]

Access is granted by the function $\func{f}{1}_{0}$ for each \lit{\texttt{select}}-expressions \var{t} and instantiation data \var{s}.
The instantiation result is a singular list of text nodes, which is determined by the evaluation.
Access to XML-nodes of instantiation data is granted by rule (A5), which is analogous to (A4) except that $f_0^4$ returns an XML-node.\\[-0.3cm]

(A5) \/ $\eval[\alpha]{XInclude \var{x}}{(\var{s},\mu,(\_,\_,\_,\func{f}{4}_{0}))}$ := \\
  \mbox{\qquad \qquad \qquad \texttt{[ $\func{f}{4}_{0} \; \var{x} \; \var{s}$ ]}}\\[-0.3cm]
  
Instantiation of common nodes is done by (A6)-(A7).\\[-0.3cm]

(A6) \/ $\eval[\alpha]{ElX \var{n} \var{a} \var{c}}({\var{s},\mu,\pi)}$ := \\
\mbox{\qquad \qquad
  \parbox{7.7cm}{\texttt{[ ElX \var{n} \var{a}\\ $(\func{concatMap}{2} (\lambda \var{child}. \eval[\alpha]{\var{child}}{(\var{s},\mu,\pi)}) \; \var{c})$ ]}}\\[-0.3cm]
}\\

(A7) \/ $\eval[\alpha]{TxtX \var{t}}{(\_,\_,\_)}$ := \texttt{[ TxtX \var{t} ]} \\[-0.3cm]

Element nodes are taken as-is.
Instantiation proceeds with children nodes with the same instantiation data, macro environment and PHP-tuple.
Text nodes are just copied unconditionally.

\textbf{Example}\\
Let the following document be given

\begin{verbatim}
<books>
  <xtl:for-each select="//book">
    <title>
     <xtl:text select="@title"/>
    </title>
  </xtl:for-each>
</books>
\end{verbatim}

where $\var{x}$ is the bibliography document wanted from sect.\ref{section:BasicsInstantiation}.

So, the following then holds

\begin{tabular}{rcl}
	$\pi$	   & = & $(\func{f}{1}_{0},\func{f}{2}_{0},\func{f}{3}_{0},\func{f}{4}_{0})$\\
	\var{s}	   & = & \parbox[t]{7cm}{\texttt{ElX "books" []}\\ \texttt{ [XForEach "//book" [ElX "title"}\\ \texttt{ [] [XTxt "@title"]]]}}\\
\end{tabular}\\

where\\

\begin{tabular}{rcl}
  \var{sel1} & = & \parbox[t]{6.3cm}{\texttt{ElX "book" \ [("\/author","\/Simon\\.."),("title","Has...")] []}}\\
  \var{sel2} & = & \parbox[t]{6.5cm}{\texttt{ElX "book" \ [("\/author","\/Joshua\\.."),("title","Re...")] []}}
\end{tabular}\\

denote element nodes from $\var{s}$.
These are reused as following:
The derivation during instantiation starts with $\eval[Start]{}{}$ and is interrupted on the first occurrence of \lit{...} by $\eval[r]{}{}$.
The remaining segments are composed in analogy to that and are also disrupted by minor calculations.

\begin{center}
\begin{tabular}{l}
\parbox[t]{8.7cm}{
$\eval[Start]{ElX "books" \ [] [...]}{\var{s} \pi}$

= $\eval[]{\var{x2}}{\var{s} \pi}$ \for \var{x2}=$\eval[r]{ElX "books" \ [] [...]}{}$

= ...

= $\eval[]{\var{x2}}{\var{s} \pi}$ \for 
	\var{x2}= \texttt{ElX "books" \ [] [XForEach "//book" \ [...]]}

= $\eval[]{ElX "books" \ [] [...]}{\var{s} \pi}$

= \parbox[t]{15cm}{\<let> \parbox[t]{13cm}{\var{mdefs} = $\func{filter}{2} (\lambda \var{child}.\\
 \eval[MM]{\var{child}}{}$)\texttt{[XForEach "//book" \ [...]]}, \\
								 \var{nodes} = $\func{filter}{2} (\lambda \var{child}. \func{not}{1} \ \eval[MM]{\var{child}}{})\\
$\texttt{[XForEach "//book" \ [...]]}}\\
  \<in> \texttt{ElX "books" \/ [] ($\func{concatMap}{2}$ $(\lambda \var{node}$.\\
\mbox{\qquad $\eval[\alpha]{\var{node}}{(\var{s},\var{mdefs},\pi)}) \ \var{nodes})$}}}

= \parbox[t]{15cm}{\<let> \parbox[t]{13cm}{\var{mdefs} = \texttt{[]}, \\
								 \var{nodes} = \texttt{[XForEach "//book" \/ [...]]}} \\
  \<in> \texttt{ElX "books" \/ [] ($\func{concatMap}{2} (\lambda \var{node}$.\\
\mbox{\qquad $\eval[\alpha]{\var{node}}{(\var{s},\var{mdefs},\pi)}) \ \var{nodes})$}}}

= \texttt{\parbox[t]{13cm}{ElX "books" \ [] \\ \quad $(\func{concatMap}{2} (\lambda \var{node}.\eval[\alpha]{\var{node}}{(\var{s},\texttt{[]},\pi)})$\\
$\texttt{[XForEach "//book" \ [...]]}  )$}}

= \texttt{ElX "books" \ [] $(\func{concat}{1}$\texttt{[}\\
\mbox{\qquad $\eval[\alpha]{XForEach "//book" \ [...]}{(\var{s},\texttt{[]},\pi)}$\texttt{]}$)$}}

= ...

= \parbox[t]{8cm}{\texttt{ElX "books" \ [] $(\func{concat}{1}$\texttt{ [[ \parbox[t]{8cm}{
	ElX "title" \ [] [TxtX "Has..."], \\
	ElX "title" \ [] [TxtX "Re..."]]]$)$}}}}

= \parbox[t]{8cm}{\texttt{ElX "books" \ [ElX "title" \ [] [TxtX "Has..."], ElX "title" \ [] [TxtX "Re..."]]  $\/_{\blacksquare}$}}}
\end{tabular}
\end{center}


\begin{center}
\begin{tabular}{l}
\parbox[t]{8.7cm}{
$\eval[r]{ElX "books" \ [] [...]}{}$

=\\[-0.4cm]
\mbox{\quad \parbox{15cm}{\<let> \parbox[t]{13cm}{\var{attDefs} = $\func{filter}{2} (\lambda \var{child}. \eval[MA]{\var{child}}{}$)$\\
$\texttt{[XForEach "//book" \ [...]]}, \\
		 \var{nodes} = $\func{filter}{2} (\lambda \var{child}. \func{not}{1} \eval[MA]{\var{child}}{})$\\
\texttt{[XForEach "//book" \ [...]]}}\\
  \<in> \texttt{ElX "books" \ $(\func{qSort}{1} (\texttt{[]} \pplus \var{attDefs})) \ \var{nodes}$}}}\\\\
= \parbox[t]{8cm}{\<let> \parbox[t]{13cm}{\var{attDefs} = \texttt{[]}, \\
		 \var{nodes} = \texttt{[XForEach "//book" \ [...]]}}\\
  \<in> \texttt{ElX "books" \ $(\func{qSort}{1} (\texttt{[]} \pplus \var{attDefs})) $\\
\mbox{\quad $\var{nodes}$}}}\\\\
= \parbox[t]{8cm}{\texttt{ElX "books" \ [] [XForEach "//book" \ [...]]}  $\/_{\blacksquare}$}}
\end{tabular}
\end{center}


\begin{center}
\begin{tabular}{l}
\parbox[t]{8cm}{
$\eval[\alpha]{ElX "title" \ [] [...]}{(\var{sel1}, [], \pi)}$

= \parbox[t]{7.5cm}{\texttt{[ElX "title" \ []} $(\func{concatMap}{2} (\lambda \var{child}.$\\
$ \eval[\alpha]{\var{child}}{(\var{sel1},\texttt{[]},\pi)})$ \texttt{[XTxt "@title"]]}}

= \parbox[t]{7.5cm}{\texttt{[ElX "title" \ []} $(\func{concat}{1}$\\
\texttt{[}$\eval[\alpha]{XTxt "@title"}{(\var{sel1},\texttt{[]},\pi)}$\texttt{]}$)$\texttt{]}}

= \parbox[t]{7.5cm}{\texttt{[ElX "title" \ []} $(\func{concat}{1}$ \texttt{[[TxtX $\func{f}{1}_{0}$\\
 "@title" \ \var{sel1}]]}$)$\texttt{]}}

= \parbox[t]{7.5cm}{\texttt{[ElX "title" \ []} $(\func{concat}{1}$ \texttt{[[TxtX}\\
\texttt{"Has..."]]}$)$\texttt{]}}

= \texttt{[ElX "title" \ [] [TxtX "Has..."]]}  $\/_{\blacksquare}$}
\end{tabular}
\end{center}


\begin{center}
\begin{tabular}{l}
\parbox[t]{8cm}{
$\eval[\alpha]{XForEach "//book" \ [...]}{(\var{s},\texttt{[]},\pi)}$

=\\[-0.4cm]
\mbox{\quad \parbox{15cm}{\<let> \var{sels} = $\func{f}{2}_{0}$ \texttt{"//book"} \var{s} \\
  \<in> $\func{concatMap}{2} (\func{f3}{1}) \ \var{sels}$ \\
  \for $\func{f3}{1}$ = $\lambda \var{c}. \func{concatMap}{2} (\lambda \var{c2}. \eval[\alpha]{\var{c2}}{(\var{c},\texttt{[]},\pi)})$\\
 \texttt{[ElX "title" \/ [] [...]]}}}\\
  
\mbox{= \parbox[t]{15cm}{\<let> \var{sels} = \texttt{[\var{sel1},\var{sel2}]} \\
  \<in> $\func{concatMap}{2} (\func{f3}{1}) \ \var{sels}$ \\
  \for $\func{f3}{1}$ = $\lambda \var{c}. \func{concatMap}{2} (\lambda \var{c2}. \eval[\alpha]{\var{c2}}{(\var{c},\texttt{[]},\pi)})$\\
 \texttt{[ElX "title" \/ [] [...]]}}}\\[-0.1cm]

= $\func{concat}{1}$ \texttt{[}
\parbox[t]{6cm}{
	$\func{concatMap}{2} (\lambda \var{c2}. \eval[\alpha]{\var{c2}}{(\var{sel1},\texttt{[]},\pi)})$\\
 \texttt{[ElX "title" \/ [] [...]]},\\
	$\func{concatMap}{2} (\lambda \var{c2}. \eval[\alpha]{\var{c2}}{(\var{sel2},\texttt{[]},\pi)})$ \texttt{[ElX "title" \/ [] [...]]}
	\texttt{]}}\\
	
= $\func{concat}{1}$ \texttt{[}
\parbox[t]{6cm}{
	$\func{concat}{1}\texttt{[}\eval[\alpha]{\texttt{ElX "title" [] [..]}}\\\\
{(\var{sel1},\texttt{[]},\pi)}\texttt{]}$,\\
	$\func{concat}{1} \texttt{[}\eval[\alpha]{\texttt{ElX "title" [] [..]}}\\\\
{(\var{sel2},\texttt{[]},\pi)}\texttt{]}$
	\texttt{]}}\\
	
= $\func{concat}{1}$ \texttt{[}
\parbox[t]{6cm}{
	$\eval[\alpha]{\texttt{ElX "title" [] [...]}}\\\\
{(\var{sel1},\texttt{[]},\pi)}$,\\
	$\eval[\alpha]{\texttt{ElX "title" [] [...]}}\\\\
{(\var{sel2},\texttt{[]},\pi)}$
	\texttt{]}}\\
	
= ...

= $\func{concat}{1}$ \texttt{[}
\parbox[t]{6cm}{ \texttt{[ElX "title" [] [TxtX "Has..."]]},\\
  \texttt{[ElX "title" [] [TxtX "Re..."]]}
  \texttt{]}}\\
	
= \parbox[t]{7.5cm}{\texttt{[ElX "title" [] [TxtX "Has..."]},\\
\texttt{ElX "title" [] [TxtX "Re..."] ]}  $\/_{\blacksquare}$}}\\\\
\end{tabular}
\end{center}


\begin{center}
\begin{tabular}{l}
\parbox[t]{8cm}{
$\eval[\alpha]{ElX "title" \ [] [...]}{(\var{sel1}, [], \pi)}$

= \parbox[t]{7.7cm}{\texttt{[ElX "title" []} $(\func{concatMap}{2} (\lambda \var{child}.\\
 \eval[\alpha]{\var{child}}{(\var{sel1},\texttt{[]},\pi)})$ \texttt{[XTxt "@title"]$)$]}}

= \parbox[t]{7.7cm}{\texttt{[ElX "title" \ []} $(\func{concat}{1}$\\
 \texttt{[}$\eval[\alpha]{XTxt "@title"}{(\var{sel1},\texttt{[]},\pi)}$\texttt{]}$)$\texttt{]}}

= \parbox[t]{7.7cm}{\texttt{[ElX "title" \ []} $(\func{concat}{1}$\\
 \texttt{[[TxtX $\func{f}{1}_{0}$ "@title" \ \var{sel1}]]}$)$\texttt{]}}

= \parbox[t]{7.7cm}{\texttt{[ElX "title" \ []} $(\func{concat}{1}$ \texttt{[[TxtX}\\
\texttt{ "Has..."]]}$)$\texttt{]}}

= \texttt{[ElX "title" \ [] [TxtX "Has..."]]}  $\/_{\blacksquare}$}
\end{tabular}
\end{center}


\begin{center}
\begin{tabular}{l}
\parbox{8cm}{
$\eval[\alpha]{ElX "title" [] [...]}{(\var{sel2}, [], \pi)}$

= \parbox[t]{7.7cm}{\texttt{[ElX "title" []} $(\func{concatMap}{2} (\lambda \var{child}. \eval[\alpha]{\var{child}}{(\var{sel2},\texttt{[]},\pi)})$ \\ \texttt{[XTxt "@title"]$)$]}}

= \parbox[t]{7.7cm}{\texttt{[ElX "title" []} $(\func{concat}{1}$\\
 \texttt{[}$\eval[\alpha]{XTxt "@title"}{(\var{sel2},\texttt{[]},\pi)}$\texttt{]}$)$\texttt{]}}

= \parbox[t]{7.7cm}{\texttt{[ElX "title" []} $(\func{concat}{1}$\\
 \texttt{[[TxtX $\func{f}{1}_{0}$ "@title" \ \var{sel2}]]}$)$\texttt{]}}

= \parbox[t]{7.7cm}{\texttt{[ElX "title" []} $(\func{concat}{1}$\\ \texttt{[[TxtX "Re..."]]}$)$\texttt{]}}

= \parbox[t]{7.7cm}{\texttt{[ElX "title" [] [TxtX "Re..."]]}  $\/_{\blacksquare}$}}
\end{tabular}
\end{center}


\newpage
\subsection{Validation}

Schema and instance are represented as regular expressions on validation (see sect.\ref{section:DesignReg}).
Instances, however, do not have alternatives, no star-operator, no XTL attributes and no text inclusions -- in contrast to schemas.
Instances can be described neatly by $\varepsilon$ and \lit{\texttt{Then}}.
In the normal form, the last node of a recursive node in an OBDD $\varepsilon$ is used.
\lit{\texttt{Then}} is recommended for node concatenation to a hedge.

It means a matcher (see sect.\ref{section:Analysis}) must recognise the cases from fig.\ref{figure:DesignInterleaving}.
The left side denotes the set of instance nodes, and the right side denotes the set of schema nodes.
All 32 cases of the bipartite graph need to be handled because XML instances a priori do not contain XTL-tags (cf. \cite{XTLSpec:2007})

XTL is well-formed and safe according to sect.\ref{section:Analysis}.
Except for bypasses and node inclusion, there is no other way to generate element nodes.
A matching of hedges with node inclusion is no longer considered here because the essential question researched here is w.l.o.g. if two generated languages are the same or not -- and for that, node inclusion does not matter in practice.
An equality solver is needed, which would determine solutions just from the relations to address this issue.
Here, a solution does not necessarily have to exist (see \cite{Du:2001}).
So, it is guaranteed that element nodes of the instance match only with tags that are not XTL-commands -- meaning only element nodes.

\subsubsection{Semantic}
\label{section:DesignValidationSemantic}

The validation bases on regular expressions \lit{\texttt{Reg}}, which are mapped onto boolean values $\mathbb{B}$.
Interpretations of $\mathbb{B}$ map onto \{\texttt{True, False}\}.
All XML instances and XTL-schemas are definable over \lit{\texttt{Reg}} (see sect.\ref{section:DesignReg}).
Macros contain a head determined by a macro name and a macro body, a hedge of element nodes.
The hedge is represented by a top-level \lit{\texttt{Then}} and which is an OBDD.
The information about which macro name is assigned which macro body must be available on validation as context information $\mu$.
The macro-environment $\mu$ has the typing \texttt{String$\rightarrow$Reg}.

The following semantic-rules variables will be used differently, namely for the specification of instance and schema nodes and or strings.
That allows a shorter rule representation than without, and by memoisation, it avoids redundant calculations.\\

The validation is done by

$$\eval[]{$\mathbb{I,S}$}{}\texttt{::Reg$\rightarrow$Reg$\rightarrow$Bool}$$

, where $\mathbb{I}$ denotes the instance and $\mathbb{S}$ the schema document.

Used helper functions are listed in tab.\ref{table:designValidationHelpers}.
The function \lit{\texttt{qSort}} is a lifted function of type \texttt{[a]$\rightarrow$[a]} and is used, particularly, for canonisation.
The functions $\func{frontSplits}{1}$ and $\func{splits}{1}$ divide non-deterministically a regular expression into two disjoint parts.
All parts of a \lit{\texttt{Reg}} are calculated lazily.
In contrast to $\func{splits}{1}$, $\func{frontSplits}{1}$ calculates one partition less --- it skips the trivial partition \texttt{(Epsilon,Reg)}.
The partition is considered here as an inversion of the \lit{\texttt{Then}}-concatenation.
It works in analogy to the non-deterministic partition of strings in $\func{frontSplitText}{1}$ and $\func{splitText}{1}$.

For instance, 
$\func{splitText}{1}$ \texttt{"\/ab"}
returns the list
\texttt{[("\/","\/ab"),("\/a","\/b"),("\/ab","\/")]}, but
$\func{frontSplitText}{1}$ \texttt{"\/ab"} only returns the last two tuples.

$\func{extractAttributes}{1}$ extracts from an element node given by a \lit{\texttt{Reg}} all attributes \lit{\texttt{AttrR}} and adds to preexisting ones into the element node.
The function $\func{getMacro}{2}$ scans $\mu$ and find for a given macro name the matching body, a \lit{\texttt{Reg}}.
It is assumed during validation that all macros are defined before running it.

The validation of a hedge is successful if all children of the schema successively match -- not necessarily with the same position index.
It is implied by the logical operator \lit{$\land$}.
Alternatives are evaluated by \lit{$\lor$}.

\begin{table}
\begin{center}
\begin{tabular}{|rcl|}
	\hline
	Function &  & Type\\
	\hline 
	\texttt{frontSplits}		& :: & \texttt{Reg$\rightarrow$[(Reg,Reg)]}\\
	\texttt{splits} 			& :: & \texttt{Reg$\rightarrow$[(Reg,Reg)]}\\
	\texttt{qSort}				& :: & \parbox[t]{3cm}{\texttt{[(String,String)]$\\\rightarrow$[(String,String)]}}\\
	\texttt{frontSplitText}	    & :: & \texttt{String$\rightarrow$[(String,String)]}\\
	\texttt{splitText}	    	& :: & \texttt{String$\rightarrow$[(String,String)]}\\
	\texttt{extractAttributes}	& :: & \texttt{Reg$\rightarrow$Reg}\\
	\texttt{getMacro}			& :: & \texttt{String$\rightarrow$[(String,Reg)]$\rightarrow$Reg}\\
	\hline
\end{tabular}
\end{center}
  \caption{Validation -- Helper Functions}
  \label{table:designValidationHelpers}
\end{table}

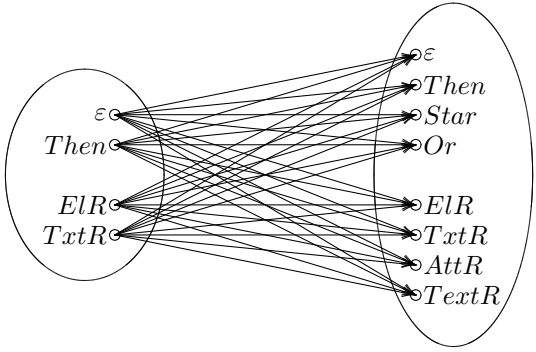
\begin{figure}
\begin{center}
\begin{tabular}{l}
\begin{minipage}{7cm}
\begin{xy}
 (0,0)   = "x1" *\cir<2pt>{} *+!R{\varepsilon},
 (0,-4)  = "x2" *\cir<2pt>{} *+!R{Then},
 (0,-12) = "x3" *\cir<2pt>{} *+!R{ElR},
 (0,-16) = "x4" *\cir<2pt>{} *+!R{TxtR},
 (-4,-8)  = "ellipse1" *\xycircle<30pt,40pt>{},
 (40,8)  = "y1" *\cir<2pt>{} *+!L{\varepsilon},
 (40,4)  = "y2" *\cir<2pt>{} *+!L{Then},
 (40,0)  = "y3" *\cir<2pt>{} *+!L{Star},
 (40,-4) = "y4" *\cir<2pt>{} *+!L{Or},
 (40,-12) = "y5" *\cir<2pt>{} *+!L{ElR},
 (40,-16) = "y6" *\cir<2pt>{} *+!L{TxtR},
 (40,-20) = "y7" *\cir<2pt>{} *+!L{AttR},
 (40,-24) = "y8" *\cir<2pt>{} *+!L{TextR},
 (45,-8)  = "ellipse2" *\xycircle<30pt,65pt>{},
 {\ar "x1";"y1"},
 {\ar "x1";"y2"},
 {\ar "x1";"y3"},
 {\ar "x1";"y4"},
 {\ar "x1";"y5"},
 {\ar "x1";"y6"},
 {\ar "x1";"y7"},
 {\ar "x1";"y8"},
 {\ar "x2";"y1"},
 {\ar "x2";"y2"},
 {\ar "x2";"y3"},
 {\ar "x2";"y4"},
 {\ar "x2";"y5"},
 {\ar "x2";"y6"},
 {\ar "x2";"y7"},
 {\ar "x2";"y8"},
 {\ar "x3";"y1"},
 {\ar "x3";"y2"},
 {\ar "x3";"y3"},
 {\ar "x3";"y4"},
 {\ar "x3";"y5"},
 {\ar "x3";"y6"},
 {\ar "x3";"y7"},
 {\ar "x3";"y8"},
 {\ar "x4";"y1"},
 {\ar "x4";"y2"},
 {\ar "x4";"y3"},
 {\ar "x4";"y4"},
 {\ar "x4";"y5"},
 {\ar "x4";"y6"},
 {\ar "x4";"y7"},
 {\ar "x4";"y8"}
\end{xy}
  \caption{Interleaving of cases}
  \label{figure:DesignInterleaving}
\end{minipage}
\end{tabular}
\end{center}
\end{figure}

\textbf{The rules of the denotational semantic}

The rules of the denotational semantic obey fig.\ref{figure:DesignInterleaving}.
Validation rules are not going to be explained just sequentially to illustrate the topic.

First of all, rules do not follow any order a priori here.
However, it is still agreed upon the prioritisation of listed rules.
The \textit{precedence} of interleaving rules raises with the rule number.
The following rules can so be described shorter.

Moreover, it is agreed upon in the matching relation $\cong$ (see sect.\ref{section:AnalysisValidation}) first argument in is the instance $\mathbb{I}$ to the left, and second argument the schema $\mathbb{S}$ to the right.
Hence, $\cong$ as denotational semantic can be divided into four partitions according to the \lit{\texttt{Reg}}-structure of the instance.\\[-0.3cm]

(E1) \/ $\eval{Epsilon, TxtR "\/"}{\mu}$ :=
 \texttt{True}
 
(E2) \/ $\eval{Epsilon, TxtR \_}{\mu}$ :=
 \texttt{False}
 
(E3) \/ $\eval{Epsilon, Epsilon}{\mu}$ :=
 \texttt{True}

(E4) \/ $\eval{Epsilon, ElR \_ \_ \_}{\mu}$ :=
 \texttt{False}

(E5) \/ $\eval{Epsilon, Star \_}{\mu}$ :=
 \texttt{True}

(E6) \/ $\eval{Epsilon, TextR \_}{\mu}$ :=
 \texttt{True}

(E7) \/ $\eval{Epsilon, Then \var{r1} \var{r2}}{\mu}$ :=\\
\mbox{\qquad \qquad \texttt{$\eval{Epsilon, \var{r1}}{\mu} \wedge \eval{Epsilon, \var{r2}}{\mu}$}}\\

In the first paragraph, empty instance nodes match with empty text nodes (E1) and (E6), and empty schema nodes (E3) matches with star-operator (E5).
Exemptions apply to element nodes (E4) and non-empty strings (E2).
The rules (E2) and (E1) overlap because (E1) is a particular case of (E2).
However, the relatively simple rule (E2) is preferred over an explicit representation.
That is why empty text nodes shall first match with (E1).
(E7) considers the case a schema has a concatenation, whose left \lit{\texttt{Then}}-branch is not $\varepsilon$ -- but this node could still be derived to $\varepsilon$.
That is why both branches of the schema-node must be derivable to $\varepsilon$.\\

The second passage treats concatenations of instance nodes --- this means hedges.
Since the normal form of OBDDs in instances excludes two consecutive $\varepsilon$ and text nodes (see sect.\ref{section:DesignReg}), and since the left branch of a \lit{\texttt{Then}} may not be empty, it can be inferred an instance-hedge is not derivable to $\varepsilon$ (Then1).

(Then 1) \quad $\eval{Then \_ \_, Epsilon}{\mu}$ :=
 \texttt{False}

Moreover, the normal form implies that a hedge either entirely contains a string in the left branch and the right branch validates against $\varepsilon$, or no validation is valid here (Then2).
The derivation of the right branch is needed because the left branch does not have to be syntactically identical -- it could also result from a hedge evaluation that requires attention.

\mbox{\parbox[t]{12cm}{
(Then 2) \/ $\eval{Then \var{r1} \var{r2}, TxtR \var{text}}{\mu}$ := \\
\mbox{\qquad \parbox{7cm}{
	\texttt{$\eval{\var{r1}, TxtR \var{text}}{\mu} \wedge \eval{\var{r2}, Epsilon}{\mu}$}}
}}}\\

This rule differs from\\[-0.3cm]

(Then 8) \/ $\eval{Then (TxtR \_) Epsilon, TextR \_}{\mu}$\\
\mbox{\qquad \qquad \qquad  := \texttt{True}}
 
(Then 9) \/ $\eval{Then \_ \_, TextR \_}{\mu}$ :=
 \texttt{False} \\[-0.3cm]

An \lit{\texttt{xtl:text}}'s instantiaion insists a text node in the instance document follows.
There is no other validation (Then9).

The validation of a hedge with the element node at the beginning (Then3) can only be successful against an element node in the schema if both element nodes are homomorphic w.r.t. validation and all remaining nodes of the hedge validate against $\varepsilon$.
All other cases lead to an unsound instance (Then4).\\[-0.3cm]

(Then 3) \/ $\eval{X_L, ElR \var{name2} \var{atts2} \var{r2}}{\mu}$ := \\
\mbox{\qquad \qquad $X_L=$ Then (ElR \var{name1} \var{atts1} \var{r1}) \var{r}}\\[0.1cm]
\mbox{\qquad \qquad \parbox{10cm}{
 \texttt{$\eval{ElR \var{name1} \var{atts1} \var{r1}, $X_R$}{\mu} \wedge \\ \eval{\var{r}, Epsilon}{\mu}\\ X_R=\texttt{ElR \var{name2} \var{atts2} \var{r2}} $}}
}\\
  
(Then 4) \quad $\eval{Then \_ \_, ElR \_ \_ \_}{\mu}$ :=
 \texttt{False}\\[-0.3cm]

Rule (Then5) attempts to match a repeating sequence from a hedge non-deterministically.
Here the given sequence is divided, s.t. the first part with hedge from the cycle, and the right part matches with the whole cycle, which may also be empty.
If there is a repetition in the given hedge, then the right \lit{\texttt{Then}}-branch may match the remaining hedge.
One corner case occurs when the left \lit{\texttt{Then}}-branch contains the whole cycle, so the remaining hedge $\varepsilon$ successfully matches with \texttt{Star \var{s}}.\\[-0.3cm]

(Then 5) \quad $\eval{Then \var{r1} \var{r2}, Star \var{s}}{\mu}$ := \\
\mbox{\qquad \qquad \qquad \qquad \parbox{10cm}{
 \texttt{\parbox{6cm}{$\vee$[True|\\
\parbox[t]{10cm}{(\var{s1},\var{s2})$\leftarrow \func{frontSplits}{1}$(Then \var{r1} \var{r2})\\ $\wedge \ \eval{\var{s1}, \var{s}}{\mu} \wedge \eval{\var{s2}, Star \var{s}}{\mu}$]}}}}
}\\

(Then 6) represents the case with a single child node, which according to the normalform occurs instead of a literal.
In this case, for example, there cannot exist an element node in \var{s2}.
Generally considered, the left branch \var{r1} must match with the whole hedge from \var{s} and \var{s2}.
This case is the base case for (Then 7), then (Then 7) initiates validation of both branches --- which without (Then 6) would not necessarily terminate if the right branch derives to $\varepsilon$.\\[-0.3cm]

(Then 6) \/ \parbox[t]{7cm}{$\eval{Then \var{r1} Epsilon, Then \var{s1} \var{s2}}{\mu}$:=\\
$\eval{\var{r1}, Then \var{s1} \var{s2}}{\mu}$}

(Then 7) \/ $\eval{Then \var{r1} \var{r2}, Then \var{s1} \var{s2}}{\mu}$ := \\
\mbox{\qquad \qquad \qquad \qquad\parbox{10cm}{
 \texttt{$\vee$[True|\\
\parbox[t]{6cm}{(\var{t1},\var{t2})$\leftarrow \func{splits}{1}$(Then \var{r1} \var{r2}) \\ $\wedge \ \eval{\var{t1}, \var{s1}}{\mu} \wedge \eval{\var{t2}, \var{s2}}{\mu}$]}}}}\\

The third passage considers text nodes.
Empty text nodes are derivable to $\varepsilon$ (\#2).
Non-empty texts, however, cannot be derived to $\varepsilon$ (\#3).
Star-operators are also derivable to $\varepsilon$, so empty text nodes are derivable to arbitrary star-operators (\#4).\\[-0.3cm]

(\#2) \/ $\eval{TxtR "\/", Epsilon}{\mu}$ :=
 \texttt{True}
 
(\#3) \/ $\eval{TxtR \_, Epsilon}{\mu}$ :=
 \texttt{False}
 
(\#4) \/ $\eval{TxtR "\/", Star \_}{\mu}$ :=
 \texttt{True}\\[-0.3cm]

It still is possible (\#6), that \lit{\texttt{xtl:text}} in the instance can generate arbitrary text as output.\\[-0.3cm]

(\#6) \/ $\eval{TxtR \var{text}, TextR \_}{\mu}$ :=
 \texttt{True}\\[-0.3cm]

Comparing two text nodes (\#7) is trivial, the same as validation of a text node against an arbitrary element node (\#8).\\[-0.3cm]

(\#7) \/ \parbox[t]{6cm}{$\eval{TxtR \var{text1}, TxtR \var{text2}}{\mu}$ :=\\
 \texttt{text1 == text2}}\\
 
(\#8) \quad $\eval{TxtR \var{text}, ElR \_ \_ \_}{\mu}$ :=
 \texttt{False} \\[-0.3cm]

The validation of a string against a hedge is similar to the validation of element nodes against a hedge (\#5).
A repeating pattern is searched.
If no non-deterministically obtained partitions matches, so the considered string may only derive to $\varepsilon$.\\[-0.3cm]

(\#5) \/ $\eval{TxtR \var{text}, Star \var{r}}{\mu}$ := \\
\mbox{\qquad \qquad \parbox[t]{7cm}{
\texttt{if (\var{h}==True) then True else $\eval{TxtR \var{text}, Epsilon}{\mu}$}\\[-0.5cm]

\parbox[t]{11cm}{
\for \var{h} == \texttt{$\vee$[True|\\
\qquad \qquad \parbox[t]{7cm}{(\var{s1},\var{s2})$\leftarrow \func{frontSplitText}{1} \var{text}\\ \wedge \ \eval{TxtR \var{s1}, \var{r}}{\mu}\\ \wedge \ \eval{TxtR \var{s2}, Star \var{r}}{} $]}}
}
}
}\\

The validation against a hedge is only valid if a sound, possibly empty partition of a hedge exists, so both texts concatenated equals the wanted text.\\[-0.3cm]

(\#1) \quad $\eval{TxtR \var{text}, Then \var{r1} \var{r2}}{\mu}$ := \\
\mbox{\qquad \qquad \qquad \parbox{6.5cm}{
\texttt{$\vee$[True|\\
(\parbox[t]{6cm}{\var{s1},\var{s2})$\leftarrow \func{splitText}{1} \var{text} \\
\wedge \ \eval{TxtR \var{s1}, \var{r1}}{\mu}\\
\wedge \ \eval{TxtR \var{s2}, \var{r2}}{\mu} \ ]$}}}
}\\\\

The fourth passage considers element nodes as an instance.
Element nodes are in contrast to text nodes atomic.
It means an element cannot be part of another element node at the same level.
This atomicity leads in (ElR8), (ElR9) and (ElR10) to schema-tags are excluded from the very beginning.\\[-0.3cm]

(ElR8) \/ \quad $\eval{ElR \_ \_ \_, TextR \_}{\mu}$ :=
 \texttt{False}

(ElR9) \/ \quad $\eval{ElR \_ \_ \_, Epsilon}{\mu}$ :=
 \texttt{False}

(ElR10) \quad $\eval{ElR \_ \_ \_, TxtR \_}{\mu}$ :=
 \texttt{False}\\[-0.3cm]

Case (ElR7) states an element node validates against a star-operator only if the subexpression underneath validates against the instance node.
So, the loop body's beginning and end nodes are derivable to $\varepsilon$, and there is only the element node between them.\\[-0.3cm]

(ElR7) \/ \parbox[t]{7cm}{$\eval{ElR \var{name} \var{atts} \var{r}, Star \var{s}}{\mu}$ :=\\
 \mbox{\qquad \texttt{$\eval{ElR \var{name} \var{atts} \var{r}, \var{s}}{}$}}}\\

Rule (ElR1) may only look trivial at first glance.
It validates an element node against another.
However, it is crucial to notice that besides equality of the element name, there is also the set of existing schema nodes that need to match after the canonisation.
In the case of schema nodes, \lit{\texttt{AttrR}} also need to be considered.
In any case, validation has to continue with the reduced schema hedge \var{r3}.\\[-0.3cm]

(ElR1) \/ $\eval{L, R}{\mu}$ := \\
\mbox{\qquad \qquad \parbox[t]{7cm}{
  \texttt{L=ElR \var{name1} \var{atts1} \var{r1}},\\
  \texttt{R=ElR \var{name2} \var{atts2} \var{r2}},\\
	\texttt{$(\var{name1} == \var{name2})\\
\wedge (\func{qSort}{1} \var{atts1} == \var{atts3})\\
\wedge \eval{\var{r1}, \var{r3}}{\mu}$} \\
          \for \texttt{(ElR \_ \var{atts3} \var{r3}) =}\\
               \texttt{$\func{extractAttributes}{1}$(ElR \var{name2} \var{atts2} \var{r2})}}
}\\

Validation against \texttt{Then} cannot just proceed.
It needs to answer first the question of which element ought to be validated first.
So, \mbox{$\pi(\mathbb{S})$} would need to be determined, which is not, at least not initially (see sect.\ref{section:Analysis}).
So the left branch of \lit{\texttt{Then}} needs to be checked.
Depending on that, the cases (ElR2), (ElR3), (ElR4) and (ElR6) result for all remaining schema nodes.\\[-0.3cm]

(ElR6) \quad $\eval{ElR \_ \_ \_, Then \_ \_}{\mu}$ :=
 \texttt{False}\\[-0.3cm]

The cases (ElR2), (ElR3) and (ElR4) are obvious and do not require further explanations.\\[-0.3cm]

(ElR2) \/ $\eval{L, R}{\mu}$ := \\
\mbox{\qquad \qquad \parbox[t]{6.3cm}{
  \texttt{L=ElR \var{name1} \var{atts1} \var{r1}},\\
  \texttt{R=Then (ElR \var{name2} \var{atts2} \var{r2}) \var{s}},\\
 \texttt{$\eval{L, ElR \var{name2} \var{atts2} \var{r2}}{\mu}\\
 \wedge \eval{Epsilon, \var{s}}{\mu}$}}
}\\
          
(ElR3) \/ $\eval{L, R}{\mu}$ := \\
\mbox{\qquad \qquad \parbox[t]{7cm}{
  \texttt{L=ElR \var{name1} \var{atts1} \var{r1}},\\
  \texttt{R=Then (Or \var{s1} \var{s2}) \var{s}},\\
 \texttt{$(\eval{ElR \var{name1} \var{atts1} \var{r1}, Or \var{s1} \var{s2}}{\mu}\\
  \wedge \eval{Epsilon, \var{s}}{\mu})\\
  \vee (\eval{Epsilon, Or \var{s1} \var{s2}}{\mu}\\
  \wedge \eval{ElR \var{name1} \var{atts1} \var{r1}, \var{s}}{\mu})$}} \\
}\\

(ElR4) \/ $\eval{L, R}{\mu}$ := \\
\mbox{\qquad \qquad \parbox[t]{5.5cm}{
  \texttt{L=ElR \var{name1} \var{atts1} \var{r1}},\\
  \texttt{R=Then (Star \var{s1}) \var{s}},\\
 \texttt{$(\eval{ElR \var{name1} \var{atts1} \var{r1}, \var{s1}}{\mu}\\
   \wedge \eval{Epsilon, \var{s}}{\mu})\\
   \vee \eval{ElR \var{name1} \var{atts1} \var{r1}, \var{s}}{\mu}$}}
}\\

In the case of (ElR5), it is checked if the element node from the instance is in the macro body.
-- If so, the following hedge \var{s} has to be derivable to $\varepsilon$.
Alternatively, the element node is in \var{s}.
Then the macro body must be derivable to $\varepsilon$.\\[-0.3cm]

(ElR5) \/ $\eval{L, R}{\mu}$ := \\ 
\mbox{\qquad \qquad \parbox[t]{7cm}{
 \texttt{L=ElR \var{name1} \var{atts1} \var{r1}},\\
 \texttt{R=Then (MacroR \var{m}) \var{s}},\\
 \texttt{$(\eval{ElR \var{name1} \var{atts1} \var{r1}, MacroR \var{m}}{\mu}\\
  \wedge \eval{Epsilon, \var{s}}{\mu})\\
  \vee (\eval{Epsilon, MacroR \var{m}}{\mu}\\
  \wedge \eval{ElR \var{name1} \var{atts1} \var{r1}, \var{s}}{\mu})$}}
}\\

Both rules ($\Phi$) and ($\Omega$) are universal w.r.t. instance nodes.
A macro call in the schema has -- independent from the actual instance node -- an unfolding or substitution by the macro body in consequence, which is represented by exactly one OBDD-expression.\\[-0.3cm]

($\Phi$) \/ $\eval{\var{inst}, MacroR \var{mname}}{\mu}$ :=\\
\mbox{\qquad \qquad \parbox[t]{5.5cm}{
 \texttt{$\eval{\var{inst},\var{word}}{\mu}$} \\
 \for \texttt{\var{word} = $\func{getMacro}{2}$ mname $\mu$}}}\\

The same is with \lit{\texttt{Or}} in the schema -- independent from the concrete instance a matching case \var{r1} or case \var{r2} is validated.\\[-0.3cm]

($\Omega$) \/ $\eval{\var{inst}, Or \var{r1} \var{r2}}{\mu}$ :=\\
\mbox{\qquad \qquad \parbox[t]{5.5cm}{
  \texttt{$\eval{\var{inst}, \var{r1}}{\mu} \vee \eval{\var{inst}, \var{r2}}{\mu}$}}}\\[0.3cm]

\textbf{Example}

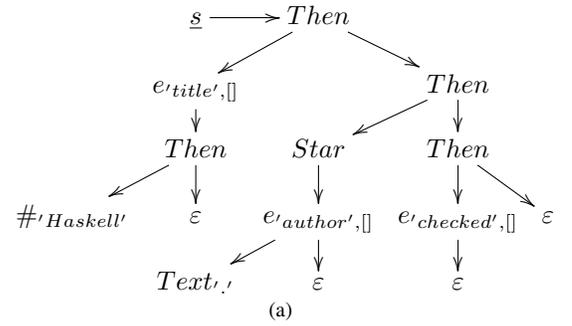
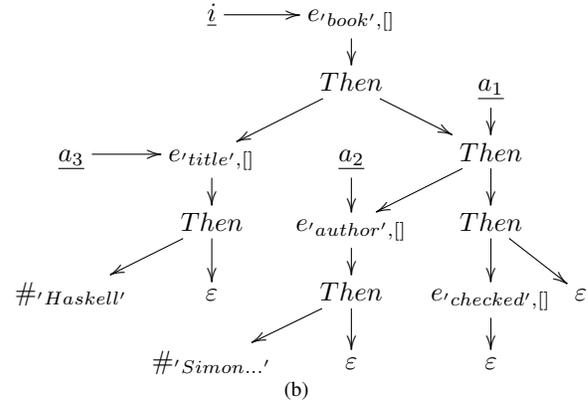
\begin{figure}[h]
\begin{center}
  \subfigure[]{\label{figure:DesignExamplesRegExFirst} 
\begin{minipage}{8cm}
\xymatrixrowsep{10pt}
\xymatrixcolsep{3pt}
 \xymatrix{
     & \underline{s} \ar[r]  & Then \ar[dl] \ar[dr] &  & \\
     & e_{'title',[]} \ar[d] &  & Then \ar[dl] \ar[d] & \\
     & Then \ar[dl] \ar[d] & Star \ar[d] & Then \ar[d] \ar[dr] & \\
    \#_{'Haskell'} & \varepsilon & e_{'author',[]} \ar[dl] \ar[d] & e_{'checked',[]} \ar[d] & \varepsilon\\
     & Text_{'.'} & \varepsilon & \varepsilon & 
 }
\end{minipage}
  }
\end{center}

\begin{center}
  \subfigure[]{\label{figure:DesignExamplesRegExSecond}
\begin{minipage}{8cm}
\xymatrixrowsep{10pt}
\xymatrixcolsep{3pt}
 \xymatrix{
     & \underline{i} \ar[r] & e_{'book',[]} \ar[d] &&\\
     & & Then \ar[dl] \ar[dr] & \underline{a_1} \ar[d] & \\
    \underline{a_3} \ar[r] & e_{'title',[]} \ar[d] & \underline{a_2} \ar[d] & Then \ar[dl] \ar[d] & \\
     & Then \ar[dl] \ar[d] & e_{'author',[]} \ar[d] & Then \ar[d] \ar[dr] & \\
    \#_{'Haskell'} & \varepsilon & Then \ar[dl] \ar[d] & e_{'checked',[]} \ar[d] & \varepsilon\\
     & \#_{'Simon...'} & \varepsilon & \varepsilon & 
 }
\end{minipage}
}
\end{center}
  \caption{Examples for Regular Expressions}
  \label{figure:DesignExamplesRegEx}
\end{figure}

The following example shows the validation of a simple document.
Let the schema \var{s} be given from fig.\ref{figure:DesignExamplesRegExFirst} and a corresponding instance document \var{i} from fig.\ref{figure:DesignExamplesRegExSecond}.

The textual notations are:\\

\begin{tabular}{rcl}
	\var{s}	  & = & \parbox[t]{6.5cm}{\texttt{ElR "book" \ [] Then (ElR "title" \ Then (TxtR "Haskell") Epsilon) Then (Star (ElR "\/author" \ [] (Then (TextR ".") Epsilon))) Then (ElR "checked" \ [] Epsilon) Epsilon}}\\\\
	\var{i}	  & = & \parbox[t]{6.5cm}{\texttt{ElR "book" \ [] Then (ElR "title" \ Then (TxtR "Haskell") Epsilon) Then (ElR "\/author" \ [] Then (TxtR "\/Simon...") Epsilon) Then (ElR "checked" \ [] Epsilon) Epsilon}}\\\\
	\var{a1}	& =	& \parbox[t]{6.5cm}{\texttt{Then (ElR "\/author" \/ [] Then (TxtR "\/Simon...") Epsilon) Epsilon}}\\\\
	\var{a2}	& = & \parbox[t]{6.5cm}{\texttt{ElR "\/author" \/ [] Then (TxtR "\/Simon...") Epsilon}}\\\\
	\var{a3} 	& = & \parbox[t]{6.5cm}{\texttt{ElR "title" \/ [] Then (TxtR "Haskell") Epsilon}}
\end{tabular}

The derivation looks as following:

\begin{tabular}{ll}
\multicolumn{2}{l}{$\eval[]{\var{i},\var{s}}{}$}\\

= & \mbox{\parbox[t]{5cm}{$\eval[]{$L$, ElR "book" \ [] ...}{}$\\
  $L$=\texttt{ElR "book" \ [] ...}}}\\

\parbox[t]{0.5cm}{=\\[-0.2cm]\begin{tiny}(ELR1)\end{tiny}}&
\mbox{\parbox[t]{7.5cm}{
  $($\texttt{"book"\/=="book"}$)\land \func{(qSort}{1}$\texttt{[]\/==\var{atts3}}$)\\
  \land$ $\eval[]{Then (ElR "title" \ [] ..., \var{r3})}{}$\\
  \for $($\texttt{ElR \_ \/ \var{atts3} \var{r3}}$)=\func{extractAttributes}{1}$\var{s}}}\\

= & \mbox{\parbox[t]{7cm}{
  $($\texttt{[]\/==\var{atts3}}$)\\
  \land \eval[]{Then (ElR "title" \ [] ..., \var{r3}}{}$\\
  \for $($\texttt{ElR \_ \/ \var{atts3} \var{r3}}$)=$ \var{s}}}\\

= & \mbox{\parbox[t]{7cm}{$\eval[]{Then (ElR "title" ..., Then ($R$}{}$\\
          \texttt{$R$=ElR "title"[] ...}}}\\

\mbox{\parbox[t]{0.5cm}{=\\[-0.2cm]\begin{tiny}(Then7)\end{tiny}}} &
\mbox{\parbox[t]{5.5cm}{
  $\lor$\texttt{[\ True}$\mid$\\
   \parbox[t]{13cm}{$(\var{t1},\var{t2}) \leftarrow \func{splits}{1} ($\texttt{Then $L$ $R$}$)$\\
   $L$=\texttt{(ElR "title" ...)},\\
   $R$=\texttt{Then (ElR "\/author" ...)},\\
$\ \land \ \eval[]{\var{t1}, ElR "title" ...}{}\\
   \land \ \eval[]{\var{t2}, Then (Star ...}{}$\texttt{\ ]}}}}\\

\mbox{\parbox[t]{0.5cm}{=\\[-0.2cm]\begin{tiny}(nd)\end{tiny}}} &
\mbox{\parbox[t]{6cm}{$\lor$\texttt{[\ True}$\mid$
$\eval[]{$L$, $R$}{}$,\\
 $L$=\texttt{ElR "title" \ [] ...},\\
 $R$=\texttt{ElR "title" ...},\\
$\ \land \ \eval[]{$L2$, $R2$}{},$\\
  $L2$=\texttt{Then (ElR "\/author" [] ..},\\
  $R2$=\texttt{Then (Star ...} \texttt{\ ]}}}
\end{tabular}
\begin{tabular}{ll}

= & \mbox{\parbox[t]{6.3cm}{
  $\eval[]{$L$, $R$}{}\\
  $L$=\texttt{ElR "title" [] ...},\\
  $R$=\texttt{ElR "title" ...},\\
  \land \eval[]{$L2$, Then (Star ...}{},\\
  $L2$=\texttt{Then (ElR "\/author" \/ [] ...}$}}\\

= & ...
\end{tabular}
\begin{tabular}{ll}
= & \mbox{\parbox[t]{7cm}{$\eval[]{Then (ElR "\/author" \/ [] ..., $R$}{}\\
 $R$=\texttt{Then (Star ...}$}}\\

\mbox{\parbox[t]{0.5cm}{=\\[-0.2cm]\begin{tiny}(Then7)\end{tiny}}} &
\mbox{\parbox[t]{7cm}{
  \parbox[t]{7cm}{$\lor$\texttt{[True}$\mid \\
  (\var{t1},\var{t2}) \leftarrow \func{splits}{1} ($\texttt{Then $L$ Then ...}$)$,\\
  $L$=\texttt{(ElR "\/author" \/ ...)},\\
$\land \ \eval[]{\var{t1}, Star (ElR ...)}{}$\\$\land \ \eval[]{\var{t2}, Then $R$ Epsilon}{}$\\
$R$= \texttt{(ElR "checked" \/ [] Epsilon)} \texttt{\ ]}}}}\\

= & \parbox[t]{6cm}{$\lor$\texttt{[\ True}$\mid$
  $\eval[]{\var{a1}, $S$}{}$,\\
  $S$=\texttt{Star (ElR "\/author" \/ [] $R$)},\\
  $R$=\texttt{Then (TextR ".") Epsilon},\\
  $\land \ \eval[]{$L2$, $R2$}{},\\
  $L2$=\texttt{Then (ElR ...) Epsilon},\\
  $R2$=\texttt{Then (ElR ...) Epsilon} $\texttt{\ ]}}\\

= & \mbox{\parbox[t]{7cm}{
  $\eval[]{\var{a1}, Star (ElR "\/author" \/ [] $R$)}{}\\
  $R$=\texttt{Then (TextR ".") Epsilon}\\
  \land$ $\eval[]{Then (ElR ...) Epsilon, $R2$}{}$\\
  $R2$=\texttt{Then (ElR ...) Epsilon}}}\\

\mbox{\parbox[t]{0.5cm}{=\\[-0.2cm]\begin{tiny}(Then6)\end{tiny}}} &
\mbox{\parbox[t]{7cm}{
  $\eval[]{\var{a1}, Star (ElR "\/author" \/ [] $R$)}{}\\
  $R$=\texttt{Then (TextR ".") Epsilon}\\
  \land$ $\eval[]{ElR "checked" \/ [] Epsilon, $R2$}{}$\\
  $R2$=\texttt{Then (ElR "checked" \/ [] Epsilon) Epsilon}}}\\

= & ...
\end{tabular}
\begin{tabular}{ll}
= & \mbox{\parbox[t]{7cm}{$\eval[]{\var{a1}, Star (ElR "\/author" \/ [] $R$)}{}$\\
  $R$=\texttt{Then (TextR ".") Epsilon}}}\\

\mbox{\parbox[t]{0.5cm}{=\\[-0.2cm]\begin{tiny}(Then6)\end{tiny}}} &
\mbox{\parbox[t]{7cm}{$\eval[]{\var{a2}, Star (ElR "\/author" \/ [] $R$)}{}$\\
  $R$=\texttt{Then (TextR ".") Epsilon}}}\\

\mbox{\parbox[t]{0.5cm}{=\\[-0.2cm]\begin{tiny}(Then7)\end{tiny}}} &
\mbox{\parbox[t]{6cm}{$\eval[]{\var{a2}, ElR "\/author" \ [] $R$}{}$\\
  $R$=\texttt{Then (TextR ".") Epsilon}}}\\

\mbox{\parbox[t]{0.5cm}{=\\[-0.2cm]\begin{tiny}(ElR1)\end{tiny}}} &
\mbox{\parbox[t]{7.3cm}{
  $($\texttt{"\/author" \/=="\/author"}$)\\
  \land (\func{qSort}{1}$\texttt{[]\/==\var{atts3}}$)\\
  \land$ $\eval[]{Then (TxtR "\/Simon...") Eps, \var{r3}}{}$\\
\for $($\texttt{ElR \_ \/ \var{atts3} \var{r3}}$)$ =\\
  $\func{extractAttributes}{1} ($\texttt{ElR "\/author" \/ [] Then (TextR ".") Epsilon}$)$}}\\

= & \mbox{\parbox[t]{7.1cm}{
  $(\func{qSort}{1}$\texttt{[]\/==\var{atts3}}$)\\
  \land$ $\eval[]{Then (TxtR "\/Simon...") Eps, \var{r3}}{}$\\
  \for $($\texttt{ElR \_ \/ \var{atts3} \var{r3}}$)$ = \texttt{ElR "\/author" \/ [] Then (TextR ".") Epsilon}}}\\

= & \mbox{\parbox[t]{7cm}{$\eval[]{Then (TxtR "\/Simon...") Eps, $R$}{}$\\
  $R$=\texttt{Then (TextR ".") Eps}}}\\

\mbox{\parbox[t]{0.5cm}{=\\[-0.2cm]\begin{tiny}(Then6)\end{tiny}}} &
\mbox{\parbox[t]{5cm}{
  $\eval[]{TxtR "\/Simon...", $R$}{}$\\
  $R$=\texttt{Then (TextR ".") Eps}}}\\

\end{tabular}


\begin{tabular}{ll}

\mbox{\parbox[t]{0.5cm}{=\\[-0.2cm]\begin{tiny}(\#1)\end{tiny}}} &
\mbox{\parbox[t]{7.3cm}{
  $\lor$ \texttt{[True}$\mid$\\\parbox[t]{5.5cm}{$(\var{s1},\var{s2})\leftarrow \func{splitText}{1} $ \texttt{"\/Simon..."}\\
  $\land \ \eval[]{TxtR \var{s1}, TextR "."}{}{}$\\
  $\land \ \eval[]{TxtR \var{s2}, Epsilon}{}{}$\texttt{]}}}}\\

= & \mbox{\parbox[t]{6cm}{
  $\eval[]{TxtR "\/Simon...", TextR "."}{}$\\
  $\land \ \eval[]{TxtR "\/", Epsilon}{}{}$}}\\

\mbox{\parbox[t]{0.5cm}{=\\[-0.2cm]\begin{tiny}(\#6)\end{tiny}}} &
\mbox{$\eval[]{TxtR "\/", Epsilon}{}$}\\

\mbox{\parbox[t]{0.5cm}{=\\[-0.2cm]\begin{tiny}(\#2)\end{tiny}}} &
\mbox{\texttt{True}  $\/_{\blacksquare}$}
\end{tabular}\\[0.3cm]


\begin{tabular}{ll}
\multicolumn{2}{l}{$\eval[]{\var{a3},\var{a3}}{}$}\\

\mbox{\parbox[t]{0.5cm}{=\\[-0.2cm]\begin{tiny}(ElR1)\end{tiny}}} &
\mbox{\parbox[t]{7cm}{
  $\eval[]{Then (TxtR "Haskell") Eps, \var{r3}}{}$\\
  \for $(\texttt{ElR \_ \/ \var{atts3} \var{r3}}) =\\
   \func{extractAttributes}{1}
   ($\texttt{ElR "title" \/ [] Then}\\
   \texttt{(TxtR "Haskell") Epsilon}$)$}}\\

= & \mbox{\parbox[t]{7cm}{
  $(\func{qSort}{1}$\texttt{[]\/==\var{atts3}})$\\
  \land \ \eval[]{Then (TxtR "Haskell") Eps, \var{r3}}{}$\\
  \for $($\texttt{ElR \_ \/ \var{atts3} \var{r3}}$)=$\\
  \texttt{ElR "title" \/ [] $R$}\\
  $R$=\texttt{Then (TxtR "Haskell") Epsilon}}}\\

= & \mbox{\parbox[t]{6.5cm}{
   $\eval[]{$L$, Then (TxtR "Haskell") Eps}{}$\\
   $L$=\texttt{Then (TxtR "Haskell") Eps}}}\\

\mbox{\parbox[t]{0.5cm}{=\\[-0.2cm]\begin{tiny}(Then6)\end{tiny}}} &
\mbox{\parbox[t]{7cm}{
  $\eval[]{TxtR "Haskell", $R$}{}$\\
  $R$=\texttt{Then (TxtR "Haskell") Epsilon}}}\\

\mbox{\parbox[t]{0.5cm}{=\\[-0.2cm]\begin{tiny}(\#1)\end{tiny}}} &
\mbox{\parbox[t]{7.1cm}{
  $\lor$ \texttt{[True}$\mid$\parbox[t]{5.7cm}{
   $(\var{s1},\var{s2})\leftarrow \func{splitText}{1}$ \texttt{"\/Haskell.."}\\
   $\land \eval[]{TxtR \var{s1}, TxtR "Haskell"}{}$\\
   $\land \eval[]{TxtR \var{s2}, Epsilon}{}{}$ \texttt{\ ]}
}}}\\

= & \mbox{\parbox[t]{7.1cm}{
    $\lor$ \texttt{[True}$\mid\\
       \eval[]{TxtR "Haskell", TxtR "Haskell"}{}\\
       \land \eval[]{TxtR "\/", Epsilon}{}{}$\texttt{]}}}\\

= & \mbox{\parbox[t]{7.1cm}{
    $\eval[]{TxtR "Haskell", TxtR "Haskell"}{}\\
    \land \eval[]{TxtR "\/", Epsilon}{}{}$}}\\

\mbox{\parbox[t]{0.5cm}{=\\[-0.2cm]\begin{tiny}(\#7)\end{tiny}}} &
\mbox{\parbox[t]{7.1cm}{
  $($\texttt{"Haskell"\/=="Haskell"}$)\\
  \land \eval[]{TxtR "\/", Epsilon}{}$}}\\

\mbox{\parbox[t]{0.5cm}{=\\[-0.2cm]\begin{tiny}(\#2)\end{tiny}}} &
\texttt{True}$\ _{\blacksquare}$
\end{tabular}\\[0.3cm]


\begin{tabular}{ll}
\multicolumn{2}{l}{
  \parbox[t]{7.5cm}{
  $\eval[]{ElR "checked" [] Eps, $R$}{}$\\
  $R$=\texttt{Then (ElR "checked" \/ [] Eps) Eps}
}}\\

\mbox{\parbox[t]{0.5cm}{=\\[-0.2cm]\begin{tiny}(ElR2)\end{tiny}}} &
\mbox{\parbox[t]{6cm}{
  $\eval[]{$L$, $L$}{}$\\
  $L$=\texttt{ElR "checked" \ [] Epsilon}\\
  $\land \eval[]{Epsilon, Epsilon}{}$}}\\

\mbox{\parbox[t]{0.5cm}{=\\[-0.2cm]\begin{tiny}(ElR1)\end{tiny}}} &
\mbox{\parbox[t]{7.3cm}{
  $($\texttt{"checked"\/=="checked"}$)$\\
  $\land (\func{qSort}{1}$\texttt{[]==}$\var{atts3})$\\
  $\land \eval[]{Epsilon, \var{r3}}{}$\\
  \for \texttt{(ElR \_ \/ \var{atts3} \var{r3})} $=\\
  \func{extractAttributes}{1} ($\texttt{ElR "checked" [] Eps}$)$}}\\

= & \mbox{\parbox[t]{7cm}{
   $(\func{qSort}{1}$\texttt{[]==}$\var{atts3}) \land \eval[]{Epsilon, \var{r3}}{}$\\
   \for \texttt{(ElR \_ \/ \var{atts3} \var{r3})} $=$\\
   \texttt{ElR "checked" \/ [] Epsilon}}}\\

= & \mbox{\texttt{True}}$\ _{\blacksquare}$
\end{tabular}

\texttt{ }\\\\\\

In the derivation below the equality sign the applied validation rules are provided.
Rules labelled with \lit{(nd)} indicate non-deterministic selection is made.
When searching for a solution, the program requires numerous executions, which have to be refined each time.
The rule \lit{(nd)}, therefore, is only of didactic help.
The continuations \lit{...} are shortened regular expressions --- those complete preceding rules.

Although the selected instance document is relatively small, the derivation shows that even a few non-deterministic cases can lead to extensive search.

\newpage

\section{Implementation}
\label{section:Implementation}

\subsection{Overview}

Haskell is recommended for implementing denotational semantics because of the lack of side-effects and its functional paradigm.
Haskell's features are very close to the syntax and semantic denotational semantics.
It includes, for instance, higher-order functions, strict static typing, data encapsulation, lazy evaluation and generic polymorphism \cite{Wadler:00}, \cite{Tho:1999}.
However, some functional programming languages, like LISP or Miranda do not have at least one of the mentioned advantages.
That is why Haskell is chosen.\\

The Haskell XML-Toolbox (HXT) \cite{HXT:04} provides a huge library of functions processing XML.
In version 7.0 HXT contains over 114 non-empty sub-packages.
It includes an XML data model, XML parser and serialiser, a validator for RelaxNG and DTD schemas and processors for XPath and XSLT.
Apart from that, HXT has numerous navigation function and constructors.
The function \lit{\texttt{readDocument}} does parsing.
Serialisation is done by \lit{\texttt{writeDocument}}.
Both functions use stateful arrows (cf. sect.\ref{section:Analysis}), which guarantees referential transparency to the outside. Both functions obey a strict sequential evaluation ordering.
Filters may be used as a substitution for arrows.
Filters can be used without the binary sequential operator \lit{\texttt{>>>}} (comparable to "`;"' in C or Pascal).
It allows lazy evaluation, and not needed calculations may be dropped.
HXT is a toolbox. HXT does not come as a framework because control always sticks to the application programmer and never changes.
Just a few filters allow semi-automatic processing of XML documents by using user-defined helper functions.

Localisation of files works with \textit{URI}s-addressing.
Thus XML sources are independently addressable from the underlying system.
Unfortunately, the recent HTTP module does not support relative addressing to the full extend.

In contrast, \textit{HaXML} \cite{HAXML:04} includes 25 non-empty sub-packages.
HaXML is reduced to essential XML-operations, a pretty-printer and HTML-processing.
It uses filters as combinators.
Arrows are currently not foreseen.
Despite that, the integrated HaXML-parser is based on memoisation.
So, needed previously calculated subexpressions are calculated only once.

The selection for the right toolkit wins HXT due to its huge amount of supported features.

\subsection{Architecture}

This section introduces the main functions and modules written in Haskell.
The implementation of instantiator and validator are described briefly.
Apart from that, tests are shortly demonstrated in order to assure proper implementations.
Introduced models are checked visually and briefly.

\subsubsection{Function Dependency Graph}

The function dependency graph for each instantiator and validator is illustrated in fig.\ref{figure:DesignFunctionDependencyGraph}.
Essentially, the implementation consists of three modules: \texttt{Main}, \texttt{Instantiator} and \texttt{Validator}.
Helper functions as \lit{\texttt{getXPathSubTrees}} are skipped.

\begin{figure}[h]
\begin{center}
\mbox{
\begin{xy}
 (-40,-12) = "main" *+!RD{main}*\frm{-},
 (+20,  0)  = "inst"  *+!RD{instantiateXTLArr}*\frm{-},
 (+20, -6)  = "get"  *+!RD{getXmlDocument}*\frm{-},
 (+20,-12)  = "read"  *+!RD{readDocument}*\frm{-},
 (+20,-18)  = "write"  *+!RD{writeDocument}*\frm{-},
 (+20,-24)  = "val"  *+!RD{validateDocumentArr}*\frm{-},
 {\ar @/^1pc/ (-40,-10);(-14,2)},
 {\ar @/^1pc/ (-40,-10);(-11,-4)},
 {\ar @/^1pc/ (-40,-10);(-6,-10)},
 {\ar @/^1pc/ (-40,-10);(-7,-16)},
 {\ar @/^1pc/ (-40,-10);(-17,-20)},
\end{xy}
}\\[0.7cm]
\end{center}

\begin{center}
\mbox{
\begin{xy}
 (  0,  0)  = "instDoc" *+!RD{instantiateDocument}*\frm{-},
 (  0, -12)  = "instXTL"  *+!RD{instantiateXTL}*\frm{-},
 (-20,-30)  = "inst"  *+!RD{instantiate}*\frm{-},
 (-35,-40)  = "inst2"  *+!RD{instantiate2}*\frm{-},
 (-20,-48)  = "matchXTLM"  *+!RD{matchXTLMacro}*\frm{-},
 (+20,-30)  = "redXTLAtts"  *+!RD{reduceXTLAttributes}*\frm{-},
 (+20,-40)  = "matchXTLAtts"  *+!RD{matchXTLAttribute}*\frm{-},
 {\ar (-15,0);(-15,-7)},
 {\ar (-15,-12);(-30,-25)},
 {\ar (-15,-12);(+5,-25)},
 {\ar (-15,-12);(+5,-25)},
 {\ar (-30,-30);(-45,-35)},
 {\ar (-30,-30);(-30,-43)},
 {\ar (0,-30);(0,-35)},
\end{xy}
}\\[0.7cm]
\end{center}

\begin{center}
\mbox{
\begin{xy}
 (  0, +12)  = "valDoc" *+!RD{validateDocument}*\frm{-},
 (-10,   0)  = "val"  *+!RD{validate}*\frm{-},
 ( 20,   0)  = "val"  *+!RD{extractMacros}*\frm{-},
 (-10, -12)  = "val"  *+!RD{matches}*\frm{-},
 (-35,-30)  = "inst"  *+!RD{splits}*\frm{-},
 (-35,-40)  = "inst2"  *+!RD{getMacro}*\frm{-},
 (+20,-30)  = "redXTLAtts"  *+!RD{splitText}*\frm{-},
 (+20,-40)  = "matchXTLAtts"  *+!RD{frontSplitText}*\frm{-},
 (+20,-12)  = "matchXTLAtts"  *+!RD{frontSplits}*\frm{-},
 {\ar (-15,12);(-15,5)},
 {\ar (-10,2);(-6,2)},
 {\ar (-15,0);(-15,-7)},
 {\ar (-15,-12);(-35,-25)},
 {\ar (-15,-12);(+5,-24)},
 {\ar (-15,-12);(+5,-34)},
 {\ar (-10,-10);(-1,-10)},
 {\ar (-15,-12);(-35,-34)},
\end{xy}
}
\end{center}

  \caption{Function Dependency Graph for \texttt{Instantiator} and \texttt{Validator}}
  \label{figure:DesignFunctionDependencyGraph}
\end{figure}

Functions for instantiation and validation are located in module \texttt{Main}.
Here the functions that appear filled are used by both programs.
In contrast to function \lit{\texttt{readDocument}}, \lit{\texttt{getXmlDocument}} reads XML-documents by avoiding arrows, so further XML-documents in \lit{\texttt{main}} can be processed.

The function \lit{\texttt{instantiateXTL}} performs an instantiation using a template and a source for instantiation data.
\lit{\texttt{validateDocument}} performs validation of an instance document against an XTL-schema.
Both functions are implemented as arrows. These are in \lit{\texttt{main}} together with input and output function in a \lit{\texttt{do}}-environment.
Because \lit{\texttt{instantiateDocument}} and \lit{\texttt{validateDocument}} require a second document, the effective function is implemented as partially defined arrows.

\lit{\texttt{instantiateDocument}} transforms multiple \lit{\texttt{XmlTree}} into the data model \lit{\texttt{XTL}}.

It corresponds to $\eval[Start]{}{}$ in terms of denotational semantics.
An instantiation is triggered by \lit{\texttt{instantiateXTL}}.
A pre-calculation step within the top-level node of the template XTL-attributes is united with preexisting attributes from element nodes.
That corresponds to $\eval[r]{}{}$.
Then instantiation resumes recursively.
For an element node, this is described by the function
\lit{\texttt{instantiate}} (namely $\eval[]{}{}$) and \lit{\texttt{instantiate2}} ($\eval[\alpha]{}{}$) describes it for hedges.
\lit{\texttt{matchXTLMacro}} distinguishes on the top-level for an element node, if it is a macro or not.

Validation is triggered by \lit{\texttt{validate}} as soon as a given \lit{\texttt{XTL}} is transferred to the regular data model \lit{\texttt{Reg}}.
Validation equals $\eval[Start]{}{}$ in terms of denotational semantics.
A call to \lit{\texttt{extractMacros}} filters at the beginning of all macros.
It corresponds to $\eval[r]{}{}$.
Validation continues recursively with \lit{\texttt{matches}} with regular expressions for both instance and schema.
The function \lit{\texttt{matches}} corresponds to $\eval[]{}{}$.
Macro calls are implemented by \lit{\texttt{getMacro}}.
The non-deterministic splitting of interleavings is done by \lit{\texttt{frontSplitText}}, \lit{\texttt{splitText}}, \lit{\texttt{splits}} and \lit{\texttt{frontSplits}} (cf. sect.\ref{section:AnalysisValidation}) and depends on the node type for each element node.\\

Both implementations of both processes strictly follow the denotational semantics from the appendix.
So, on validation, match-rule (E1) is noted in Haskell as following (see sect.\ref{section:DesignMain}):\\[-0.3cm]

\switchHaskell
\begin{tabular}{l}
\qquad
\begin{lstlisting}
matches Epsilon (TxtR "") macros = True.
\end{lstlisting}
\end{tabular}\\[-0.3cm]

The same holds for instantiation (cf. app.\ref{appendix:DenotationalSemantics}).\\

Not well-formed documents cannot be instantiated.
Since \lit{\texttt{readDocument}} parses lazily and traverses XML documents in pre-order, errors are issued with a position.
Element nodes, which do not obey conventions from sect.\ref{section:mainXTL}, are interpreted as usual element nodes (cf. \cite{XTLSpec:2007}).

As presented in sect.\ref{section:DesignInstantiationSemantic}, instantiation data shall be passed in a context in \lit{\texttt{xtl:for-each}}.
Binding instantiation data implicitly to the PHP function is disadvantageous --- children of \lit{\texttt{xtl:for-each}} access portions of instantiation data.\\

Errors are issued locally.
However, this does not mean a thrown error is the reason for a failure during the validation (cf. sect.\ref{section:DesignMain}).
The pre-order processed schema is serialised pre-order to meet the requirement of error localisation --- this allows a faster localisation by the user. 
An error stack would simplify bug tracking on the one side. On the other side, it would, however, rapidly increase execution time.
The serialised document contains the last valid node because validation operates lazy.

\subsubsection{Checking Validity}

The proof for a correct model transformation was done in sect.\ref{section:DesignDataModels}.
The proof for completeness of the data models is already done in sect.\ref{section:DesignMain}.
Here, all XML-document composed of text and element nodes can be expressed by \lit{\texttt{XTL}} and \lit{\texttt{Reg}}.
Properties of instantiation and validation are discussed in sect.\ref{section:Analysis}.

Correctness of the implementations is guaranteed by a precise translation of the denotational semantic and Haskell's referential transparency (cf. sect.\ref{section:DesignInstantiationSemantic}, \ref{section:DesignValidationSemantic}).
Implementation validity is assured by HUnit-tests \cite{HUnit:2003}.
During instantiation, XPath is prototypically used as a command language.
Tests cover the rules of the denotational semantics and also in combination with other rules.
Tests check for valid and invalid input.
The validator has currently 818 tests, where the instantiator has 62 tests.
The big discrepancy in the exponential rise of complexity is due to the additional transitions by the validator.
The instantiator is determined and much simpler in comparison to the validator.

Documentation for the Haskell functions is given by a Haddock-helpfile \cite{Haddock:2001}.\\

As suggested in sect.\ref{section:DesignMain}, test cases should be enriched by infinite documents.
Infinite OBDDs can be passed to \lit{\texttt{validate}}-functions.
The termination behaviour may practically be restricted, since in case of a left-recursion, it is for sure after a finite number of steps a result eventually is available, or a non-termination \lit{$\perp$} is the "`result"'.
The following example shows infinite data structures as an extension to existing test cases:

\switchHaskell
\begin{center}
\begin{tabular}{l}
\begin{lstlisting}
genTree,genTree2,genTree3::Tree String
genTree = Node "a" (Node "b" genTree Eps) 
                   (Node "c" genTree Eps)
genTree2 = Node "a" (Node "b" Eps genTree2)
                    (Node "c" genTree2 Eps)
genTree3 = Node "a" Eps 
            (Node "b" genTree3 Eps)

test Eps _ = True
test (Node _ l r) c = 
	if (c==10) True
	else or [(test l (c+1)), (test r (c+1))]
\end{lstlisting}
\end{tabular}
\end{center}

For the sake of a clear explanation, a binary tree \lit{\texttt{Tree}} is used here.
This tree has a similar structure as regular expressions \lit{\texttt{Then}} and \lit{\texttt{Or}} in \lit{\texttt{Reg}}  (cf. sect.\ref{section:DesignReg}):

The underlying data model \texttt{Tree} is defined as

\begin{center}
\begin{tabular}{c}
\begin{lstlisting}
data Tree a = Node a (Tree a) (Tree a) | Eps.
\end{lstlisting}
\end{tabular}
\end{center}

The test function \lit{\texttt{test}} has the type \lit{\texttt{Tree a$\rightarrow$Integer$\rightarrow$Bool}}.
The starting value \texttt{c==0} iterates a polymorphic infinite \lit{\texttt{Tree}} until a node of depth ten is visited for the first time.
The equivalence partition of positive test cases encloses all (\texttt{Tree Integer}, $c$), where \lit{$c$} is an integer less equals 10.
For this partition, the function terminates with the result \lit{\texttt{True}}.
For \texttt{c>10}, it results in \lit{$\perp$}, because the data structure is infinite and the base case "`\texttt{test Eps _ = True}"' is unreachable.
This example demonstrates the meaning of infinite OBDDs as a test case.
If the test function were passed a negative finite input, it would always return a result not equal to \lit{$\perp$}.
For an invalid input, the test function returns \lit{$\perp$}.
Match-rules of the validator returning a result for \lit{$\perp$}, and \lit{$\perp$} otherwise have similar behaviour.
That is why test functions should consider selected unlimited test data or data structures as \lit{\texttt{genTree}}, \lit{\texttt{genTree2}} and \lit{\texttt{genTree3}}.

\subsection{Optimisations}

Recommendations for improvement on two examples in Haskell are shown.

\subsubsection{Multiple Iterations}

Whilst instantiation hedges are iterated several times in order to filter nodes with a particular property.
The rules (E) and (I3) are considered patterns from app.\ref{appendix:DenotationalSemanticInstantiation}:\\[-0.3cm]

 \<let> \parbox[t]{10cm}{\var{attDefs} = $\func{filter}{2} ( \lambda \var{child}. \eval[MA]{\var{child}}{}) \var{c}$, \\
        \var{nodes} = $\func{filter}{2} ( \lambda \var{child}. \func{not}{1} \; \eval[MA]{\var{child}}{}) \var{c}$ .}\\

It would be more efficient to comprehend both resulting variables, which would be a tuple, and accumulate one of both tuple-variants.
The following fragment shows a corresponding improvement towards reducing the number of iterations:\\[-0.7cm]

\begin{center}
\begin{tabular}{c}
\begin{tabular}{l}
 \<let> (\var{attDefs}, \/ \var{nodes}) = \\
\parbox[t]{14cm}{$\func{foldl}{3} ( \lambda (\var{m},\var{n}) \var{c}. \<if> \ \eval[MA]{\var{c}}{} \ \<then> \ (\var{m} \pplus [\var{c}], \var{n})$\\
\<else> $(\var{m},\var{n} \pplus [\var{c}]))$ \texttt{([],[]) \var{c}} .
}
\end{tabular}
\end{tabular}
\end{center}

By bundling, child nodes are processed once.
Lazy evaluation still holds, but in every case, \lit{\texttt{let}} is evaluated before the function body.
The performance bargain looks reasonable initially because the new variant has an alternative with a condition and alternative.
A rather complex program buys the supposed advantage in performance since the new variant has an alternative condition.
The rows, which initially used to be relatively short, still hardly differ.
The gained improvement has an unsymmetrical behaviour.

The separation between program and optimisation rules can be achieved in GHC by introducing an additional Haskell program comment.
This comment is checked while interpretation and matching optimisation rules are applied to defined functions.

Concerning sect.\ref{section:ImplementationHaskell2Java}, this simplification has little meaning because object-oriented modelling interprets operations over aggregated child nodes undoubtedly different.

\subsubsection{Lazy Evaluation}
Left-associative folds have a performance advantage over right-associative folds because of the lazy evaluation.
Because of the homomorphism-condition holding during validation (see sect.\ref{section:AnalysisProperties}), regular subexpression may be evaluated in arbitrary ordering.
Subexpressions, however, may be skipped.
So, list comprehensions, which are left-associative, lead to \lit{\texttt{frontSplits}} calculates all partitions are ascending by the length of a regular expression.
This calculation skips partitions quicker.

Hash-tables should be used to access attribute entries during validation because attributes are accessed quite often and those often not canonised in applications.
The algorithm from app.\ref{appendix:DenotationalSemantics} uses a lazy evaluation strategy, speeding up further using hash-tables.
The table size should grow in proportion to the derived regular expression length (see sect.\ref{section:DesignMain}).

\subsection{Implementation in Java}
\label{section:ImplementationHaskell2Java}

Before the Haskell implementation is translated into Java, several main caveats need to be considered.
Xerces \cite{Xerc} may be used for the input and output of XML documents. JXPath \cite{JXP:04} may be used for XPath-queries.

Data models \lit{\texttt{XTL}} and \lit{\texttt{Reg}}, which are given as algebraic data types, must first be modelled as parametrised classes.
So, for \lit{\texttt{Reg}}, an abstract class \lit{\texttt{RegEx}} is defined. \lit{\texttt{RegEx}} has subclasses, for instance, \lit{\texttt{Then}} and \lit{\texttt{Or}}.

Haskell's \textit{generic polymorphism} is restricted further by \textit{ad-hoc} polymorphism \cite{Card:1999} and is used, for instance, by sorting and when processing instantiation data.
Java 5.0 provides the possibility to replace generic polymorphism with templates that are determined on runtime.

Unfortunately, lazy evaluation cannot directly be simulated in Java.
By explicit checks, the evaluation order needs to be influenced, s.t. many cases are eliminated. Alternatively, the most likely cases need to be checked.
For the encapsulation of lazy methods, the STRATEGY pattern is promising.

Higher-order function needs to be mimicked by polymorph classes.
Here, \textit{HOF}s are implemented as concrete classes, which calls polymorphic class methods.
Partial functions can be mimicked in Java without sending messages.
Partial arguments can be determined by queries to \lit{\texttt{get}}-methods on the called object.
Interfaces can describe static typing of the PHP function.

Matching rules of the validation algorithm can either be adopted as is.
In this case, the validator operates recursively descendant and consumes many resources.
The alternative is to consider a pushdown-automaton.
However, the overall functional paradigm will severely change, and so will the denotational semantics.
In this case, it may be better to refer to operational semantics instead.
However, this is not the aim of this work (see sect.\ref{section:AnalysisStateOfArt}).

\subsubsection{Proposition of Class Diagram}

\begin{figure}[h]
\begin{center}
\begin{rotate}{-90}
\mbox{
\begin{minipage}{21cm}
\includegraphics[width=20cm]{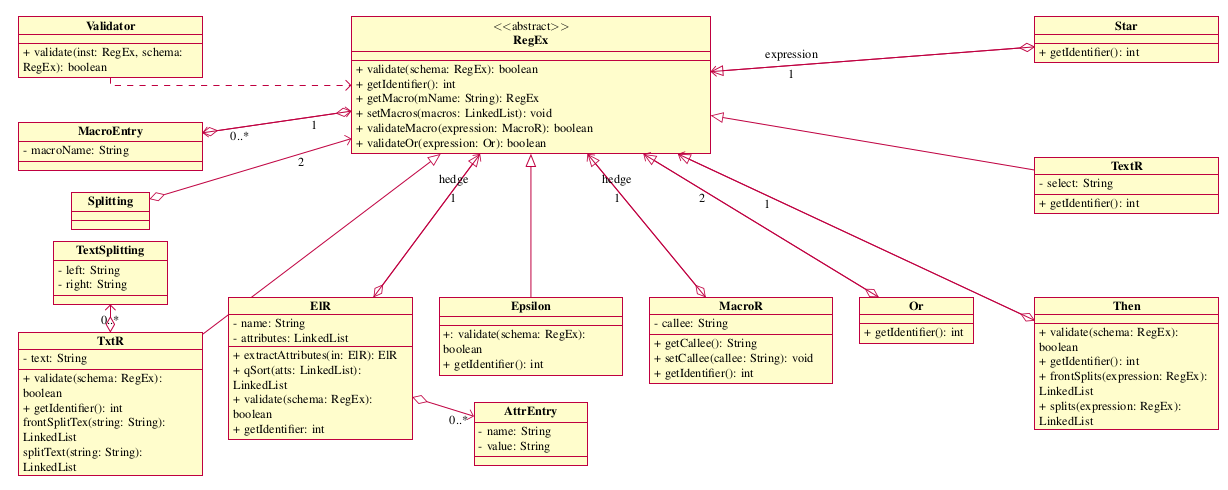}\\
  \caption{Validator Design for Java}
\end{minipage}}
\end{rotate}
\end{center}
  \label{figure:ImplementationClasses}
\end{figure}

Fig.\ref{figure:ImplementationClasses} shows a possible design for validation.
The class \texttt{Validator} represents the caller, which initiates an instance document against a given schema document.
The class \texttt{RegEx} is abstract, represents regular expressions, and currently has eight subclasses.
Subclasses containing one or two \texttt{RegEx} allow variable child nodes.
The abstract methods \texttt{validate} and \texttt{getIdentifier} are implemented in the subclasses.
The method \texttt{getIdentifier} returns the identifier of a subclass.
This method breaks encapsulation.
However, it may be tolerated if all regular expressions are fully covered.
By object-centric identifiers, every \texttt{validate} method can be implemented by \textit{switch}-constructs.
Implementation follows the rules of the corresponding denotational semantics closely herewith.

Since the methods \texttt{validateOr} and \texttt{validateMacro} are universal (see sect.\ref{section:DesignValidationSemantic}), \texttt{RegEx} can implement those.
The class \texttt{MacroEntry} represents an entry of the macro-environment $\mu$ (see sect.\ref{section:mainXTL}).\\

In this class diagram, several design and architectural patterns are hidden (after Fowler \cite{fowl:2000} and Kernievsky \cite{ker:2006}).

Initially, regular expressions are constructed by the counted subclasses by using the BUILDER pattern.
Here \texttt{Validator} acts as \textit{Director} and \texttt{RegEx} as \textit{AbstractBuilder}.
The standard method is \texttt{create}.
The DECORATOR pattern allows some \texttt{RegEx}-classes the reuse of previously implemented code.
The classes \texttt{TxtR}, \texttt{Epsilon} and \texttt{TextR}
act as \textit{ConcreteComponent}, \texttt{RegEx} as \textit{Component} and \texttt{ElR}, \texttt{Star}, \texttt{MacroR}, \texttt{Or}, \texttt{Then} as \textit{ConcreteDecorator}.

The INTERPRETER pattern describes a regular language.
The client \texttt{Validator} triggers validation by calling the method \texttt{validate}.
The IMPLICIT-LANGUAGE pattern consists of terminals and non-terminals, which generate the language.
Those subclasses that do not contain composition act as \textit{Terminal}s
Classes, which have a simple composition \texttt{RegEx}, act as \textit{Non-Terminal}.
Both symbols are \textit{interpretable expressions}.

On validation, each node type is matched with an instance node.
Invalid combinators are skipped and validated against \lit{\texttt{False}}.
The classes do not implement \texttt{validate} for every \texttt{RegEx}-subclass.
That is also because the structure of regular expression for schemas does not extend in a meaningful way.
Hence, each class type during evaluation returns identifiers or type information.
The function \texttt{getIdentifier} can obtain this information.
That is why a separation between the data model and tree traversal does not make sense.

Hence, an implementation by the VISITOR pattern is not appropriate here.
Besides, the universal DESCRIPTOR pattern occurs whose \textit{INSTANCE} is the class \texttt{TextSplitting} and whose \textit{DESCRIPTION} is \texttt{TxtR}.
In analogy, \texttt{ElR} and \texttt{AttrEntry} also match the DESCRIPTOR pattern.

Apart from this, the TEMPLATE-METHOD pattern can be found.
The abstract class here is \texttt{RegEx}, where \texttt{validateMacro} and
\texttt{validateOr} act as \textit{Template}.
The subclasses listing the same methods are \textit{Hooks}, which are referring to templates.\\\\

In the next step, the data model \lit{\texttt{XTL}} and polymorphic instantiation data and the instantiator can be modelled.
The proposed test suite can be used in addition to the module test.
\newpage

\newpage

\section{Comparison}
\label{section:ComparisonMain}

In this section, dedicated schema languages are compared according to previously selected criteria.
Here, language features are more of importance than implementations.
The following questions will be taken into consideration:

\begin{enumerate}
 \item What are the positive factors towards unification?

Here, semantics, syntax and algorithms for both instantiation and validation shall be considered.
The criteria weights shall be considered thoroughly.
Criteria in favour and against unification shall be clarified.

 \item Which linguistic features are in favour and which ones are against unification?

Is it possible to adapt and adequately represent features from other schema languages?
A metric shall be provided if possible.
If the other XML schema languages show up weakness, investigate if the weakness may be resolved.

 \item Which consequences do unification have in practice?

What is the practical benefit of unification?
\end{enumerate}

Furthermore, differences and mutual grounds of features towards a unification from previous sections shall thoroughly be analysed.

\subsection{Considered Languages}

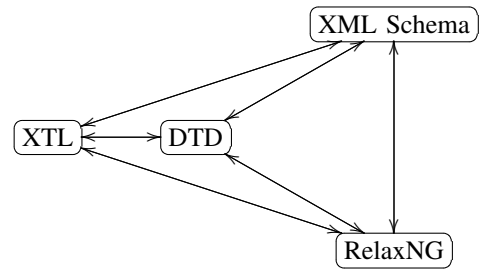
\begin{figure}[h]
\begin{center}
 \begin{minipage}{2cm}
  \xymatrixrowsep{30pt}
  \xymatrixcolsep{30pt}
  \xymatrix{
    && *+[F-:<3pt>]\txt{XML\ Schema} \ar[lld] \ar[ld] \ar[dd] \\
    *+[F-:<3pt>]\txt{XTL} \ar[r] \ar[rru] \ar[rrd] & *+[F-:<3pt>]\txt{DTD} \ar[l] \ar[ru] \ar[rd] &\\
    && *+[F-:<3pt>]\txt{RelaxNG} \ar[uu] \ar[lu] \ar[llu]
  }
  \end{minipage}
\end{center}
  \caption{Considered Schema Languages}
  \label{figure:ConsideredLanguages}
\end{figure}

The considered languages are illustrated in fig.\ref{figure:ConsideredLanguages}.
It is XTL \cite{XTLSpec:2007} which may be considered as one fraction, and rather popular schema languages like XSD \cite{Fal:2004}, \cite{Tho:2004}, \cite{Bir:2004}, RelaxNG \cite{Cla:2001} and DTD \cite{DTDGen:07}.
Since DTD may not be described within XML, DTD will only be referred to at some positions only where appropriate for comparison purposes.

XSLT is considered for the sake of comparison between template and schema languages \cite{Cla:1999}.

\subsection{Criteria}
\label{section:ComparisonCriteria}

The criteria are applied to XTL as well as to selected schema languages.
Criteria are supposed to be as common as possible and are focused on unification.
In conclusion, comparisons are mostly qualitatively.

A unification of template expansion and schema validation is achieved by involved documents and referred schema languages (cf. sect.\ref{section:Analysis}).
Orthogonality and distributivity are properties that have little meaning here.
Both properties may already be covered by modularity.\\

\textbf{Similarity}\\[-0.3cm]

Similarity refers to template and instance documents on the one side and template and schema documents on the other side.
If the template and schema document are the same, then unification is achieved.
Otherwise, there are commands in the template language which do not correspond to commands in the schema language or vice versa.
That is distinguishing instantiation from validation.

Moreover, similarity means an adequate representation of a template document w.r.t. an instance document.
If both are too different, then the similarity is too little.
It means template markups can be transferred into instance markups by only a considerable amount of changes.\\

If template and schema document or template and instance document were only close enough, then a general unification is achievable.\\

\textbf{Expressibility}\\[-0.3cm]

Expressibility asks if features may express a template language from the schema language and vice versa.

Regularity properties and type safety are researched for unification.
An essential question to find here is whether functions and symbols can be validated as expression.

Even the representation of rules themselves may influence the unification.
For instance, one question emerges if rule representation may improve unification if rules are based purely on filters and pattern-matching.
It was observed that pattern-matching often is better for more compact notation, if possible at all, than filters, for instance.

\subsubsection{Classification}

The taxonomies are shown in tab.\ref{table:ComparisonTaxonomyCriteria}, \ref{table:ComparisonTaxonomyCriteriaBC} may be derived from sect.\ref{section:Introduction}.
The three views are mostly related to schema languages.

\begin{table}[h]
\begin{center}
\begin{tabular}{|c|c|c|}
  \hline
  Syntax & Semantic & Complexity\\
  \hline
  \hline
  Schema-Style			&	Typing			 & Time- / Memory usage \\
  Ordering			&	Functions		 & Automata Class \\
  Syntactic Sugar	& 						Constraints		 &	Evaluation Ordering \\
  Pattern-Matching		&	Error Handling & \\
  Regularity								 &	& \\
  Symbols				&	 & \\
  Control Flow	&	 & \\
  \hline
\end{tabular}
\end{center}
  \caption{Taxonomy from the perspective of a Schema Language}
  \label{table:ComparisonTaxonomyCriteria}
\end{table}

\begin{table}
\begin{center}
\begin{tabular}{c}
\subfigure[from the perspective of a developer]{

\begin{tabular}{|c|c|c|}
	\hline
	Openness	&	Extensibility	&	Variability \\
	\hline \hline
	Rules set	&	new Command Tags & Command Language \\
	Constraints 	& 	Modules/Namespaces	& Instantiation Data \\
	\hline
\end{tabular}}\\\\

\subfigure[according to the degree of objective consideration]{
\begin{tabular}{|c|c|c|}
	\hline
	Objective		&	Subjective \\
	\hline \hline
	Language Type	&	Tool support \\
	Complexity	&	Usability \\
	& 	Self-explanation \\
	\hline
\end{tabular}}
\end{tabular}
\end{center}

\caption{Taxonomies of the Comparison Criteria}
\label{table:ComparisonTaxonomyCriteriaBC}
\end{table}

Tab.\ref{table:ComparisonTaxonomyCriteria} differentiates between schema features' syntax, semantic validation concepts, and corresponding algorithms' footprints.
The column "`Syntax"' denotes "`Symbols"' hedges and attribute unions.
Since hedges and attributes of templates and schemas do not alter, the corresponding identifiers may be considered symbols.
The remaining columns are not ambiguous (see sect.\ref{section:Basics}).
The existing column "'Typing"' ask whether generated elements always have a type and if validation may result in an intersection of command-tags.
"`Functions"' expresses representation and interpretation of functions which have an arity of one or more.
Constraints are a possibility to restrict nodes and texts even more.
Granularity plays a key role here.
"`Syntactic Sugar"' means if, for instance, parametrised loops and other recurring idioms may be replaced by a more elegant syntactic notation.

Automata class is determined by the level of determinism and evaluation order. The class is a measure of how complex a concrete validation implementation may be.

The taxonomy (a) from tab.\ref{table:ComparisonTaxonomyCriteriaBC} splits the comparison criteria from (an application) developer's view.
The level of extensibility and variability determines the partition.
Modules are on possibility in order to bind arbitrary schemas by namespaces and symbols.
Variability means, for example, how much a command language, instantiation data and the evaluation order might be influenced.
"`Evaluation Order"' means if expressions are evaluated one after another in the order they appear in the incoming document or if they are only used when needed.

Taxonomy (b) compares subjective and objective criteria.
Self-explanatory means similarity to a document.
However, it may also mean a concise description of a schema node.

\subsubsection{Assessment}

In this section, comparisons are based on tab.\ref{table:ComparisonTaxonomyCriteria}.
Taxonomies (a) and (b) of tab.\ref{table:ComparisonTaxonomyCriteriaBC} are compared with the first table.
The criteria from tab.\ref{table:ComparisonTaxonomyCriteria} are loosely coupled.
Constraints and typing are of overall practical meaning.
"`Error handling"' is a secondary requirement to usability.
Functions are of utmost importance to the expressibility of template languages.

Syntax summarises the most important criteria in comparison to semantic and complexity.
"`Ordering"' and "`Symbols"' may be considered as syntactic sugar.
Symbols replace well-defined nodes and hedges. Therefore, they improve reuse.
Control flow in the validation process is of crucial importance.

From the application’s perspective, the schema style is essential.
If a schema language is pattern-styled, then even complex schemas with a difficult syntax description may easily be expressed.
That relates to the regularity of schema languages.

W.r.t. to complexity, time and memory consumption play a key role.
However, because of heterogeneous program systems, this criterion is often totally underestimated.
Because of the prototypical Haskell implementation, a dynamic profile does not make sense.
Both evaluation order, as well as eliminated bottlenecks, directly affect the efficiency of validators.\\

In addition to openness,structured rules may improve the comparison towards unification.
Excluding constraints may cause contradiction towards unification and make the definition of new predicates error-prone and bloated.
That is why an investigation of including and excluding specifications shall be done.
The following points seem most promising regarding modularisation of template and schema: integration into namespaces, node inclusion, and exchangeable command languages.
The taxonomies from tab.\ref{table:ComparisonTaxonomyCriteria} (a) already cover the criteria from tab.\ref{table:ComparisonTaxonomyCriteriaBC} (b).

\subsection{Semantic}

This section introduces features of the considered schema languages and one template language.
Typing is here of outstanding interest in schema and command languages and the representation of functions, the use of constraint and error handling.

\subsubsection{Typing}
\label{section:ComparisonTyping}

Each template node must be well-typed.
It is a precondition for instantiation and validation.
On the one hand, there are element and text nodes.
On the other hand, there are command tags. Depending on the concrete command, those tags also return either an element node, a text node or may return a hedge containing both as an instance (cf. \cite{Ata:2004}).
If this condition is fulfilled, then implicitly well-formedness of the instance is guaranteed.
Access to attributes can be replaced as access to element nodes, as discussed earlier, and therefore does not require further investigation.
Depending on a given schema language, typing may also include referential integrity.

Instantiation uses PHP functions, where validation does not.
Validation is considered when validating the associated command tag type, so descriptions are reused in a unified approach.

Unification does not matter about command tags.
The fewer possibilities exist to instantiate a template node; the faster validation can perform (see sect.\ref{section:ComparisonResourceConsumption}).\\

In XSLT, type safety is guaranteed in XPath-expressions for command tags.
The result is either a hedge, an element node or a boolean value (see \cite{DeR:1999}, \cite{Wad:2000}).
Numbers and dates are included in hedges whose children nodes have exactly one node and a list of \texttt{<book/>}-nodes as a hedge with a certain amount of element nodes.
Similarly to XTL, singular types are used to include and boolean types for controlling the instantiation.
The consideration of each command tag's typing is essential to validation because, for each command tag, a decision needs to be made with how many instance nodes it corresponds.
In other words, this means the type of the inferred template node is compatible with the type of the instance node.

That raises whether a command retrieving text could not accidentally return a singular node or a hedge.
However, the text's interpretation as element nodes would insist tag-brackets are generated at least in the output.
However, this is not possible in XTL and XSLT's default configuration because special characters are treated as \textit{XML entities}.
Type safety on text output can be influenced in XSLT by the attribute \lit{\texttt{disable-output-escaping}}.
Even well-formedness is not guaranteed in XSLT because an upper-most root node does not need to exist.
Because of this, XSLT is not appropriate for the unification of instantiation and validation.
Despite this, XTL does guarantee type safety.\\

Guaranteeing referential integrity is essential, especially in XML databases.
In XTL, most existing command tags can be adapted without prior configuration. However, unique values and foreign keys must be supported by the command language.

XSD has techniques to assure referential integrity.
So, for instance, DTD and RelaxNG do not support keys.
DTD supports only fundamental unique elements.
However, there is no modularisation. Therefore, no separation and no distinction by element names are possible.

Information about keys and uniqueness does not hinder unification.
However, they are only relevant to validation, but not to instantiation.
Another approach to separate a schema is to hoist relationships over keys.
In order to do that, schema nodes need to be referenced.
That is done by XPath-expressions --- in analogy to XSD.
The key-referenced relationship remains free of redundant nodes, but it may also invalidate schema relations if path expressions become invalid.

From the standpoint of standard rules (see sect.\ref{section:ComparisonRuleset}), the support for referential integrity is a violation of the unification of instantiation and validation.
Hence, instantiation primary runs for a given document only once.
However, validation runs the first time in order to localise identifiers and unique elements and attributes. The first run is then still needed in order to check uniqueness and keys.

\subsubsection{Functions}
\label{section:ComparisonFunctions}

Functions are a vital acceptance criterion for template languages.
In JSP and ASP, small calculations can be performed by functions.
A local and remote function call may be triggered.
Template languages like XSLT define functions with arity greater equals zero by named templates (cf. \cite{Hab:2006}).
The most popular schema languages do not have corresponding mechanisms.
Functions require that among its values, there is a possibility to check each value has a correspondence to at least one instantiation data field.
If this is the case, then this value is valid according to the function.
Otherwise, the function value is not valid according to a schema node.
In other words, functions must be invertible, and the instantiation of a template must be an isomorphic mapping \cite{Hab:2006}.

Since this harsh restriction is too hard in practice, functions may not be validated in general.
It means generalised functions prevent the unification. Restrictions must be imposed.\\

Functions may be not defined in XTL, which would not even be meaningful for the mentioned reason.
Referential transparency disallows it.
Only macros may be stored locally in the global symbol environment.
That is why only functions within the command language may be used.

\subsubsection{Constraints}
\label{section:ComparisonConstraints}

Constraints may be required in both processes.
During validation, constraints restrain nodes and text.
So, specifications may be including or excluding.
XTL, RelaxNG and DTD belong to schema languages with including specifications.
XSD is a hybrid. XSD may allow contradicting specifications for which it is impossible to define a valid document. For instance:

\switchXSD
\lstset{tabsize=2}
\lstset{basicstyle=\ttfamily\small}
\lstset{numbers=none}
\lstset{numberstyle=\tiny}
\lstset{showstringspaces=false}
\lstset{captionpos=b}

\begin{center}
\begin{tabular}{c}
\begin{lstlisting}
<xsd:element name="top"> 
 <xsd:complexType>
   <xsd:sequence>
    <xsd:element name="second" type="BBB2"
     minOccurs="0"/> 
  </xsd:sequence>
 </xsd:complexType>
</xsd:element>

<xsd:complexType name="AAA">
  <xsd:sequence> 
  <xsd:element name="x" type="xsd:string"
     minOccurs="3"maxOccurs="3"/> 
  </xsd:sequence>
</xsd:complexType>

<xsd:complexType name="BBB2">
  <xsd:complexContent>
    <xsd:restriction base="AAA">
      <xsd:sequence maxOccurs="1"> 
          <xsd:element name="x"
           type="xsd:string"
           minOccurs="3"
           maxOccurs="3"/>
      </xsd:sequence>
    </xsd:restriction>
  </xsd:complexContent>
</xsd:complexType>
\end{lstlisting}
\end{tabular}
\end{center}

\texttt{<second/>} is not a valid instance, because \lit{\texttt{BBB2}} requires an empty \textit{content-model} and requires an element node as child, so "`\texttt{<second><x>1</x></second>}"' is also invalid.\\


An example of constraints is the explicit null.
\texttt{AAA} may not have child nodes within DTD in order to specify an element node. 
Children shall be specified \texttt{EMPTY}.
The full element node corresponding will be  \texttt{<!ELEMENT AAA EMPTY>}.
In RelaxNG \texttt{<empty/>} has the same effect.
In XSD the same is achieved by an empty sequence.

Constraints on elements are helpful to restrict.
However, partially specified names can be inconvenient if names do no more differ (enough).
In contrast to that, the restriction of child nodes is important. So, for example, a schema containing \texttt{<a/>} with a text node, followed by \texttt{<b/>} is specified in XTL as \texttt{<a><xtl:text select="/a"/><b/></a>}.
Similar looks the specification for RelaxNG and DTD.
The XSD is relatively long (see fig.\ref{figure:ComparisonXSDSeqTextNodes}).
Here, \texttt{<a>\#<b/></a>} is not exactly expressed, where \lit{\#} denotes an arbitrary text node.
The weakness of content models in XSD is their imprecise position for text and element nodes.
So, \texttt{<a><b/>\#</a>} and \texttt{<a>\#<b/>\#</a>} are accepted, even so this was not originally intended.

That is why including schemas is easier to check than excluding because validation has only to check whether both schema and instance nodes fulfil the same predicates.
When excluding instance nodes in a schema, all listed specifications must be checked.
Excluding schemas make unification more difficult because a fixed amount of excluding predicates needs to be considered.

On unification, attribute constraints may not specify attribute values any closer.
For this reason, attribute names must be given in a complete form -- the same as it is with element nodes.
In contrast to elements, the exclusion of attributes in schema languages is weaker and has to obey weaker restrictions.
So, in RelaxNG, within an \lit{\texttt{except}}-node \lit{\texttt{nsName}} may exclude a particular namespace.
In XSD, only a specific namespace may either be included or excluded.
On the contrary, arbitrary attributes without namespaces can be added by the \lit{\texttt{anyAttribute}}.
It reminds the mixed content model, but is still different.

In XTL Tag \lit{\texttt{attribute}} allows specifying attributes arbitrarily.
From this perspective, XTL offers the best usability.\\

\begin{figure}[h]
\switchXSD
\lstset{tabsize=2}
\lstset{basicstyle=\ttfamily\small}
\lstset{numbers=none}
\lstset{numberstyle=\tiny}
\lstset{showstringspaces=false}
\lstset{captionpos=b}
\begin{center}
\begin{tabular}{c}
\begin{lstlisting}
<?xml version="1.0"?>
<xsd:schema xmlns:xsd="http://www.w3.org/..">
 <xsd:element name="a">
  <xsd:complexType mixed="true">
   <xsd:sequence>
    <xsd:element name="b"/>
   </xsd:sequence>
  </xsd:complexType>
 </xsd:element>
</xsd:schema>
\end{lstlisting}
\end{tabular}
\end{center}
\caption{XSD-Schema for a mixed Content-Model}
\label{figure:ComparisonXSDSeqTextNodes}
\end{figure}

Multiplicities are another possibility to restrict hedges.
In XSD, any non-negative multiplicities are expressible by attributes.
RelaxNG only allows the multiplicities \lit{0..*} and \lit{1..*}.

Multiplicities cannot be expressed dynamically but only by predetermined values and variables.
It ensures that, for instance, in XSD, regular expressions can be composed (see sect.\ref{section:ComparisonAutomatonClass}), and cycles do not depend on variables.
In XTL, there is currently no option to repeat element nodes at any time.
A fixed number of repetitions is allowed after all.
That is why \texttt{<a/><a/><a/>} cannot be described by $L(a^{3})$. Multiplicities in XTL are recommended for the same.
Kleene's star operator \lit{unbound} may remain unnoticed on a definition because it is already covered by \lit{xtl:for-each}.
Since multiplicities remain the properties on regularity untouched, unification gets another valid construct to be used.

\subsubsection{Error Handling}

Error handling strongly depends on the concrete automaton (see sect.\ref{section:ComparisonAutomatonClass}).
The handling of errors in XTL differs from the behaviour in XSD and RelaxNG.
That is mainly due to the prototypical stage of XTL.
XSD and RelaxNG issue an error message containing the error position. The schema is well-formed and suffices the syntactic requirements of the schema language.
Although RelaxNG may also be simulated by a non-deterministic automaton (see sect.\ref{section:ComparisonAutomatonClass}), the output of all previous locally occurred errors is avoided by implementing an error stack.
The automaton is a big help to schema developers but requires further resources.
In XTL, there shall be an error stack too.

\subsection{Syntax}

In this section, syntactic features of schema languages are considered.
The more features a schema language has, the more powerful it is -- this assumption is false in general.
Besides unification criteria, learnability is also being considered.
The goal of this section is to find answers to the following questions:

\begin{enumerate}
 \item \textbf{Do linguistic features improve unification?}

What can be done, if possible at all, in order to increase unification?

 \item \textbf{Can lingual features of other languages be expressed within XTL?}

Apart from the question if it is possible in general, the question of complexity arises second and of type safety third (see sect.\ref{section:ComparisonTyping}).
Does the other way round also function?
	
 \item \textbf{How significant are those lingual features in detail?}

 \item \textbf{Is XTL minimal w.r.t. lingual feature amount?}	
Are some template and schema features missing?
\end{enumerate}

XTL combines the advantages of XML-schema languages, especially the vast amount of available tools and APIs.
One advantage, however, is still missing in RelaxNG and XTL: it is the fixed association between the document instance and schema.
It should explicitly be placed in an XML processing instruction or the top element node, so the relationship does not get lost, primarily when several schemas are used.
The information regarding embedded command languages should also be explicit.
The \lit{\texttt{realm}}-attribute achieves this.

\subsubsection{Symbols}
\label{section:ComparisonSymbols}

In schemas, it is often helpful to summarise hedges and reuse them at different locations.
The insertion of a hedge is comparable to constant symbols because it is assigned only once and remains invariant afterwards.
The invariance touches template and schema nodes.
That is why invariant hedges are sometimes called "'\textit{symbols}"' which can be expressed by constants.
Symbols are visible everywhere within a given schema for all considered schema languages.
Some template languages like XSLT version 2.0 and Prolog (cf. \cite{Hab:2006}) offer variable agreements, which may be used for memoisation and as a placeholder.
These variables are not considered symbols because they are dynamic and therefore violate referential transparency.

Symbols are so-called \textit{Substitution groups} in XSD, \textit{Definitions} in RelaxNG, and \textit{Macros} in XTL.
Symbols in XSD may restrict or extend and always refer to element nodes.
Symbols introduced in RelaxNG behave isomorphic to \lit{\texttt{xtl:call-macro}} and \lit{\texttt{xtl:macro}}.

In contrast to XSD in XTL and RelaxNG, text and element nodes may be appended before and after a macro call, where the ordering does not change (see sect.\ref{section:ComparisonOrdering}).
Symbols do not violate referential transparency due to determinism, so these are well-defined before instantiation and validation.
That is why they harmonise both processes, instantiation and validation.
Variables with mutable values on the other side hinder unification because these immediately influence validation (see sect.\ref{section:ComparisonFunctions}).\\

The following coarse-grained structure would be namespaces.
In XTL, namespaces are present, but because of their prototypical characteristics, they are not supported by XTL-implementations.
In other schema languages, namespaces can either be within the top root node or RelaxNG or be local in RelaxNG in all children nodes.
Apart from that, namespaces in RelaxNG and XTL can also be applied to attributes.
The attribute \lit{\texttt{ns}} in element nodes does just that.
Namespaces are modules too.
By doing so, namespaces contribute to an ongoing unification of template language and schema languages.

XSD has numerous simple and complex data types.
In contrast to this, RelaxNG has only 2, and \textit{vocabularies} of other schema languages are imported.
XTL has only the data types from the previous section.

\subsubsection{Control Flow}

XML tags determine the control flow in schemas.
The flow controls validation.
Upon closer examination, \lit{\texttt{xtl:if}} and \lit{\texttt{xtl:for-each}} can be interpreted as regular expressions.
Simple conditions are represented by "`$r\mid \varepsilon$"', nested conditions with optional alternatives, and by a selection of "`$r_{1} \mid r_{2} \mid ... \mid \varepsilon$"', where \lit{$r_{j}$} denote hedges of corresponding consequences.

\lit{\texttt{xtl:if}}, \lit{\texttt{xtl:for-each}} and \lit{\texttt{xtl:call-macro}} determine the control flow in XTL, where the latter correspond to hedge substitutions (cf. sect.\ref{section:ComparisonConstraints}).
A matching \lit{\texttt{xtl:for-each}} can substitute every \lit{\texttt{xtl:if}}, but not vice versa.
Hence, \lit{\texttt{xtl:for-each}} is a universal element.
In XSD, conditions may also be expressed by multiplicities with lower bound \lit{0} and upper bound \lit{1}.
\lit{\texttt{xtl:if}}-conditions can express multiple selections and nested conditions.
These are the same from the standpoint of validation if the proposed right-associated OBDD is chosen (cf. sect.\ref{section:DesignMain}).
So, a simple and a nested condition as well as multiple selections are preferred for unification.
When it comes to instantiation, refined conditions are beneficial since unconsidered cases could be dropped, and the overall instantiation increased.

\subsubsection{Ordering}
\label{section:ComparisonOrdering}

The ordering is of utmost importance to validation.
Without an ordering, the increase of complexity for any validation algorithm could be tremendous (cf. sect.\ref{section:ComparisonResourceConsumption}).
Hence, an unspecified ordering requires testing for all possible \textit{permutations} of all possible hedges.
The amount of permutations is $n!$, where $n$ is the number of children for a given hedge.

In XSD, the ordering can optionally be specified.
By doing so, readability suffers and redundancy increases.
A specification would be more efficient in terms of adequacy if only the provided ordering were allowed.
The concrete ordering should always be taken out by expressions of a given schema language and should be done explicitly. Often certain permutations are not desired, and systematic exclusion of all unwanted would critically increase complexity too.
For example, in XSD, there exist different representations of "`\texttt{<a/><b/>}"'.
This hedge may be represented by the unordered regular expression $\texttt{(a|b)}^{*}$. This expression is interpreted under a specific ordering mode and is expressed by multiplicities.
So, 

\begin{center}
\texttt{<xsd:choice minOccurs="0" \/maxOccurs="\/unbounded"\/>}$\var{r}$\texttt{</xsd:choice>}
\end{center}

describes the same tree language as:

\begin{center}
\texttt{<xsd:sequence minOccurs="0" \/ maxOccurs="\/unbounded"\/>}$\var{r}$\texttt{</xsd:sequence>}
\end{center}

The ordering would be restricted in RelaxNG by the tag \lit{notallowed} to empty hedges only.

In XTL and DTD, only those orderings are valid, matching the ordering specified in the schema.
The implicit permutation of hedges is disallowed.
In general, a permutable hedge \lit{$e_{0}e_{1}...e_{n}$} can be expressed as a regular expression

$$\texttt{OR} \ (e_{0} \cdot \func{permutate}{1}(e_{1}...e_{n})) \ \ldots (e_{n} \cdot \func{permutate}{1}(e_{1}...e_{n-1}))$$

(see sect.\ref{section:DesignMain}).

This representation of $\func{permutate}{1}$ is syntactic sugar and is believed to be safe due to its closeness.
However, permutation would need to have a meaningful counterpart on instantiation when it comes to unification.
If instantiated hedges ought to be permuted, then it can be stated that permutation increases unification w.r.t. instantiation and validation.
A \lit{\texttt{permutate}}-tag shall be introduced to XTL for the explicitly specified permutation. 

Because of the described dramatic drop in performance on validation in general, unknown hedges and suffixes of known nodes are not considered currently in XTL and therefore are not implemented yet.
Unknown prefixes can be expressed by \lit{include}-tags and are guarded by \lit{xtl:for-each}.
The only severe disadvantage is an extensive search right at the beginning of the unknown prefix.
It is a practical advantage to place known nodes at the start of a hedge and as close as possible to the document's beginning.

\subsubsection{Patterns}

XSD is the most pattern-styled language among all considered schema languages (cf. \cite{Mur:2001}).
In contrary, XTL, RelaxNG and DTD have a grammar-styled representation which is too less pattern-styled (see \cite{XTLSpec:2007}, \cite{Cla:2007}).

Besides the style of a schema language, syntactic notation, e.g., using regular operator, greatly influences usability.
In RelaxNG, the tag \lit{\texttt{oneOrMore}} is used as a replacement for the plus-operator.
Operators of the command language do not obey regularity criteria in XSD because restrictions on symbols (see sect.\ref{section:ComparisonSymbols}) violate those.
Expressions of the command language do not have any significance because they are ignored on validation.

\subsubsection{Usability}

Usability is worsened in XSD by redundant \lit{\texttt{complexType}} definitions.
Both XSD and RelaxNG lack adequate representations of element nodes.
So, \texttt{<xsd:element name="..."\/>} or \texttt{<element> <name .../>...} must be used as node constructors.
Node definitions in DTD are not adequate. However, this cannot be resolved by any other means because of its non-XML notation.
In contrast to this, XTL can take schema nodes exactly as they are.
Specifications take nodes exactly as they are and simplify unification because template, schema, and instance nodes are all congruent.

\subsubsection{Syntactic Sugar}

Syntactic sugar in this work's scope means command tags that can be removed from a given language, s.t. expressibility neither regarding schema language nor template language diminishes.
The more features are shared among schema and template languages, the more unification increases because expressibility increases relatively.

Idioms and syntactic sugar have the following characteristics:

\begin{itemize}
 \item \textbf{Openness:} Idioms must be expressible just by some core functions.

Herewith, neither referential transparency may be violated, nor additional assumptions about instantiation data are allowed.

 \item \textbf{Extension:} Idioms may not extend the expressibility of a schema language.

For example, this means that regular schema languages may naturally recognise ranked-competing fragments due to a weaker tag.
Nevertheless, ranked-competing schema languages may not introduce tags, which would enable regular schema language recognition.
\end{itemize}

Openness causes both processes, validation and instantiation, are proceeded by another step is turning its product, the document, into a form free of sugared tags.
Simple forms lead validators and schema simplifiers (see sect.\ref{section:ComparisonRuleset}) to heavily reduced rules sets.
For instance, in RelaxNG, this is called "'simple syntax"' \cite{Cla:2001}.\\

An extension of a ranked-competing schema language to a regular schema language leads to most rules of a validator must be dropped since an increase in non-determinism leads immediately to further matching cases.
Here, a validator is assumed to be fully described by a rule set.
The inversion makes a conflict visible now since non-deterministic matching rules need to be excluded by the syntax.

It is compulsory syntactic sugar has unique semantics in both the template language and the schema language.
If an idiom has only in one of both languages a non-empty semantic, then exclusions may only be hard to formulate and are less plausible.

\subsection{Complexity}
\label{section:ComparisonResourceConsumption}

A closer description of complexity is given in sect.\ref{section:DesignMain}.
The validation algorithm is not bound by polynomial complexity w.r.t. schema length (see \cite{Ata:2004})  herewith.
The complexity is caused by rules splitting string and \lit{\texttt{Then}}-nodes non-deterministically.
As a result of this, the potential search space for validation is drastically expanded.
Nonetheless, this approach ends with a higher expressibility than XSD or DTD.
The memory consumption is directed by runtime behaviour. When an alternative occurs, the last valid state before entering the recursion has to be stored.
The XTL validation requires only the memory needed according to the maximal recursion depth.
Solutions once dropped are not considered a second time.
Only open solution not yet considered must be stored on lazy evaluation.

The separate handling of attributes does not cause a significant rise in complexity.

The infinite OBDDs proposed in sect.\ref{section:Implementation} have non-polynomial complexity on lazy evaluation, but only if fixpoints exist in the schema.
From the practical perspective, infinite instance documents are disallowed because validation only considers enclosed documents.
Instantiation can accept infinite OBDDs as input. Because in contrast to validation, instantiation does not require the entire document is present.
Consecutive tags in a hedge may be instantiated independently (see sect.\ref{section:StateOfTheArtInstantiation}).
\textit{Exceptions} are macros having infinite bodies but whose overall structure is well-defined.
In the case of finite templates without left-recursion, instantiation always terminates with a well-formed resulting instance document.
The runtime complexity of instantiation is bound by a polynomial which degree is the maximum number of nested loops.
However, this does not hold in general for macro calls.

Macros lead to a non-polynomial runtime behaviour, which may be compensated by lazy evaluation.

\label{section:ComparisonAutomatonClass}

Depending on a tree language (see sect.\ref{section:Analysis}), a schema language automata may recognise different granularity levels.
DTD is expressed by a single-typed grammar, where a ranked-competing grammar expresses XSD.
The latter allows more freedom \cite{Mur:2001}.
RelaxNG and XTL are generated by regular grammars and allow maximal expressibility.
So, XTL allows defining arbitrary (but at most regular) sequences of nodes.

On the other side, there is the syntax to be considered.
XTL has a minimalistic syntax since no element is sugared (see sect.\ref{section:Analysis}, \ref{section:DesignMain}).
RelaxNG has a relatively tiny amount of features, for instance, in comparison to XSD. However, this does not affect expressibility.
XSD has poor expressibility but lots of sugar, which quickly leads to hard to read documents.

By missing macros or definitions, schema validation always terminates in XSD.
Only for XTL and RelaxNG closedness properties hold regarding intersection, union and complement.
That is why those languages are perfect for extensions.
The same expressibility class practically means transformations into each other are possible without any severe hinder.
There may also be transformations from XSD and DTD to XTL and RelaxNG. Beware the other direction is not possible in general.

Top-down deterministic validators recognise schemas in DTD.
Top-down non-deterministic validators recognise RelaxNG and XTL.
However, XSD is generated by a single-typed tree grammar \cite{Mur:2001}.
That is why a top-down deterministic validator recognises XSD.

\subsection{Unification}

This section unification is considered in general towards documents and rules sets.

\subsubsection{Documents}

In order to unify templates and schemas, the following restrictions are recommended:

\begin{enumerate}
  \item \textbf{Type Safety}.
Both template and instance document must be XML.
Existing command tags need to have a semantic, which does not depend on surrounding tags and is still distinct among other command tags.
In conclusion, this means validation should have relatively a few possibilities only to validate against an instance node.

 \item \textbf{Abstraction from the Command Language}.
Expressions to be calculated should be enclosed in the document structure.
It means, functions must be eliminated and command languages ignored during validation.
Instantiation data must be hidden during validation.
Instantiation data may not be in the schema, so access is granted by attribute entries in command tags.

 \item \textbf{Self-Similarity}.
The more significant similarities between schema and instance document are the more straightforward validation is to describe and the bigger the intersection of shared language features.
\end{enumerate}

\subsubsection{Rules Set}
\label{section:ComparisonRuleset}

The essential difference between instantiators and validators is the underlying algorithms.
A validator matches nodes, where an instantiator substitutes nodes (see \cite{Rahm:01}).
The instantiator may process nodes in parallel, where a validator processes an instance document sequentially.

Instantiators and validators are based on rules sets.
Rules may express denotational semantics for instantiation and validation if those are of the kind: $p\rightarrow_{a_{0},...,a_{n}}q$.
Here, \lit{$p$} denotes a premise. The premise of instantiation contains a template node, where the premise of validation contains an instance node.
\lit{$a_{\texttt{j}}$} represent constraints.
\lit{$q$} denotes the result of an instantiation or validation.
The evaluation ordering of instantiation and validation is controlled by constraints (cf. \cite{Hab:2006}).
Rule sets do not contain instantiation data.
However, this is an obstacle, because validation iterates the instance document in parallel to the schema document.
Instantiation does not allow an actual procedure.
First, this is because instantiation data may be arbitrary heterogeneous structures, which cannot be compared with the template document in general.
Second, any assumptions about the internal structure of instantiation data are strictly prohibited, as shown in sect.\ref{section:Analysis}.\\

An essential benefit of a rule-based semantic is the openness towards compilers and interpreters with a rigid structure (cf. \cite{Cza:2000}) -- as is the case of denotational semantics in sect.\ref{section:DesignMain} and implementation in sect.\ref{section:Implementation}.
So, new elements are comfortably inserted into or removed from existing rules.
So, a$m+n$ new cases are inserted to the rule set of a validator in case of insertion of a new tag, where the bipartite graph consists of at most $m \times n$ edges (cf. fig.\ref{figure:DesignInterleaving}).
It means the insertion of a new element has a linear complexity increase in conclusion (even the constant factor "`1"').
Despite this are bottom-up parsers, which often require a change in a considerable amount of rules.
It is worth mentioning that $m+n$ cases could have further distinction, for instance, for the concatenation of (possibly empty) command tags (cf. sect.\ref{section:DesignValidationSemantic}).

In contrast to this, an instantiator requires just one rule into the denotational semantics since there are no non-deterministic cases and a new command tag has the same behaviour as other tags.\\

The reuse of short and self-explanatory templates is simplified.
In order to effectively reuse documents of a schema language, simplifications with the document structure are recommended.
That is why rule-based simplifiers shall be developed, where weak regular schema languages are advantageous.
For example, the schema $a^{*}(ba^{*})^{*}$ shall be restructured to $(a|b)^{*}$  \cite{Du:2001}.\\\\

From a verbal perspective, commands hinder unification, which either supports instantiation or validation.
However, their implementation implies the assumption that the other function does not make assumptions about the origin.
On instantiation, e.g. XSLT-stylesheets are XSLT-templates, and regular text patterns on validation.

\section{Summary}

This work examined how much template instantiation can narrow schema validation for XML documents in a unification attempt.
First, instantiation and validation were formalised. Properties towards their practical meaning were probed, and implementation was developed.
Requirements for unification were elaborated, and a comparison was made based on these results.\\

The semantics were formulated in different ways.
On the one side, denotational semantics specified the programs' behaviour.
On the other side, rules demonstrated introduced data models used and transformed.
The tree automaton model was used for evaluation.
Optimisation techniques were discussed.
The formalisation made it more evident that instantiation is adequately represented as a term-rewriting system and validation as graph-matcher, also a rule and term-based system.

Both semantics showed that the rules set for both instantiation and validation could not entirely be unified.
However, the reuse of simplified code simplifies unification.\\

The implementation allowed the unification of both processes on the document level.
A comprehensive test suite guarantees the validity of all implementations.
A stack for error should be integrated in future with a Java implementation.
The regular data model was prototypically introduced in Java.

Analysis showed that XTL has regular grammar properties, except macros, which extend expressibility and violate specific closeness properties.
The extension requires further research towards practical means.

Moreover, it was shown termination does not hold and should not hold in general.
An explanation was given why filters and arrows are not best, especially when XTL will be variable and extensive.
Recommendations for improvement were given.

Instantiation showed XTL is not as universally applicable as, for instance, XSLT. For instance, there is no possibility to define arbitrary functions in a schema, which could be used later.
It was found the expressibility of XTL directly depends on the command language.
Instantiators work deterministic, where validators do not by default. Here, parallel validation should be considered further.\\

It was found that it is advantageous to restrict unification.
If a schema language assigns a type to each slot, then validation may be simplified quite considerable.
Regular properties imply syntactic sugar can be defined without any change in the schema language's expressibility.
In order to obtain the most flexibility, command languages require adaptations.
The introduced rules for instantiation and validation have, in consequence, that changes can be done quickly.
A practical possibility to beat non-determinism is the construction of automata as described and the restriction of a schema language's expressibility.
It was noted, however, that there can be drawbacks here.
First, Rabin-Scott's powerset construction may cost too high efforts for XML. Second, a restriction to single-typed or ranked-competitive is not a solution because closeness-properties are violated.\\

The comparison showed that potentially all generated instance documents significantly impacted the unification --- much more than the expressibility of encapsulated queries.
Comparison criteria were introduced regarding syntax and semantics.
Comparisons were taken out and marked accordingly.
Despite its colossal syntax definitions, XSD was found weaker than XTL or RelaxNG. XTL as template language is quite universal.
Because of its universality it is possible, for instance, to define keys for referential integrity.
Variable orderings of foreseen attributes shall be considered as syntactic sugar.
It is recommended to define a rule-based simplifier for XTL-schemas because it is estimated simple schemas, and tools will raise the acceptance of XTL.

\newpage
\section{Glossary}

\myglsentry{Abstraction}
\myglsdesc{denotes an anonymous function in the $\lambda$-calculus.}

\myglsentry{Ad-hoc polymorphism}
\myglsdesc{similar to $\nearrow$ generic polymorphism, abstracts from and restricts a data type, e.g. by subclassing.}

\myglsentry{Active Server Pages}
\myglsdesc{a certain $\nearrow$ template-language.}

\myglsentry{Ambiguity}
\myglsdesc{non-determinism in grammars caused by overlapping rules.}

\myglsentry{Application}
\myglsdesc{denotes in terms of $\lambda$-calculus the application of a given term to an $\nearrow$ abstraction. Is equal to $\beta$-reduction. Applications with term and abstraction being the same are self-applicative.}

\myglsentry{Arrow}
\myglsdesc{generalised $\nearrow$ monad having functions as input.}

\myglsentry{Arity}
\myglsdesc{The amount of parameters a function has. A function with arity zero is a constant.}

\myglsentry{Command Language}
\myglsdesc{A language embedded in some $\nearrow$ template-language for accessing $\nearrow$ instantiation data by placeholder plugins. XPath is the command language for $\nearrow$ XSLT, where JXPath is a reference implementation.}

\myglsentry{Ranked-Competing}
\myglsdesc{$\nearrow$ Regular tree grammar whose $\nearrow$ hedges are uniquely decomposable.}

\myglsentry{Bypass Attribute}
\myglsdesc{a $\nearrow$ XTL-attribute for the stepwise instantiation of a $\nearrow$ template.}

\myglsentry{Call-by-need evaluation}
\myglsdesc{a particular case of $\nearrow$ lazy evaluation on which intermediate results are not calculated twice.}

\myglsentry{Constraints}
\myglsdesc{denotes general restrictions. Constraints during $\nearrow$ instantiation and $\nearrow$ validation select and specify nodes with $\nearrow$ command tags. In $\nearrow$ XTL constraints are specified by "`select"'-expressions. Constraints may also be used to specify valid and invalid variable and function domains. In the context of databases constraints guarantee referential integrity.}

\myglsentry{Content-Model}
\myglsdesc{describes a $\nearrow$ hedge in $\nearrow$ XSD.}

\myglsentry{Definitions}
\myglsdesc{denote dedicated $\nearrow$ symbol nodes in $\nearrow$ RelaxNG.}

\myglsentry{Denotational Semantics}
\myglsdesc{\nocite{InfDud:1993} also known as functional semantics, denotes the functional behaviour for a given program. It uses an universal language for its syntax, e.g. set operators or an abstract programming language, and defines a relation between syntactical constructs and their meaning.
Denotational semantics abstract from a certain machine platform and focuses on calculating output for a given input.}

\myglsentry{Derivatives}
\myglsdesc{denote mappings from regular expressions to regular automata as proposed by Brzozowski and Glushkov.}

\myglsentry{Document reconstruction}
\myglsdesc{reconstructs unknown $\nearrow$ instantiation data.
It is in contrast to $\nearrow$ validation, where instantiation data is known.  Document reconstruction can be considered as an inverse operation to $\nearrow$ instantiation.}

\myglsentry{Document Object Model}
\myglsdesc{is a data model for a given XML document.}

\myglsentry{Regular (Tree Grammar)}
\myglsdesc{the most powerful of all considered $\nearrow$ regular tree grammars, which has no restriction.}

\myglsentry{Endomorphism}
\myglsdesc{mapping whose domain and codomain both denote the same set.}

\myglsentry{Exhaustive Search}
\myglsdesc{searches for a proper non-deterministic $\nearrow$validation.}

\myglsentry{Lazy Evaluation}
\myglsdesc{evaluation ordering which includes only those steps essential for obtaining the final result. Within the $\lambda$-calculus, it corresponds to the outermost term reduction.}

\myglsentry{Filter}
\myglsdesc{functions having a polymorphic $\nearrow$ type \texttt{a}$\rightarrow$\texttt{[a]}.}

\myglsentry{fixpoint}
\myglsdesc{in geometry denotes a point that is fixed for a given mapping. Similarly, a fixpoint in templates and schemas is a synonym for conditions that in loops do no further change. In $\lambda$-calculi, a fixpoint denotes a condition with an invariant (state) in recursions. $\nearrow$ left-recursions lead to unreachable fixpoints during $\nearrow$ instantiation and $\nearrow$ validation.}

\myglsentry{Functional}
\myglsdesc{$\nearrow$ higher-order functions.}

\myglsentry{Higher-order Functions}
\myglsdesc{functions consuming functions as input and output. Known list-$\nearrow$functionals include fold-left \lit{foldl}, mapping \lit{map} and filter \lit{filter}.}

\myglsentry{Generic Polymorphism}
\myglsdesc{abstraction of a certain data-type without further restrictions, see $\nearrow$ ad-hoc polymorphism.}

\myglsentry{Glushkov-Automaton}
\myglsdesc{finite determined automaton which recognises regular expressions.}

\myglsentry{Grammar Style}
\myglsdesc{schema language whose syntax can be expressed well by a grammar --- in contrast to $\nearrow$ pattern-like schema languages.}

\myglsentry{Graph-Matcher}
\myglsdesc{program, which checks if two graphs are identical.}

\myglsentry{HaXML}
\myglsdesc{Haskell-API for processing XML documents (also see $\nearrow$ HXT).}

\myglsentry{Hedge}
\myglsdesc{synonymous for a list of children nodes in an XML document.}

\myglsentry{Homomorphism}
\myglsdesc{mapping for which the following equation holds: the product of the codomains equals the codomain of its products. Homomorphism holds, for instance, for the addition of a residue class ring.}

\myglsentry{HXT}
\myglsdesc{Haskell-API for processing XML documents (also see $\nearrow$ HaXML).}

\myglsentry{Instance}
\myglsdesc{result of the $\nearrow$ instantiation.}

\myglsentry{Instantiation}
\myglsdesc{process turning a $\nearrow$ template with $\nearrow$ instantiation data into a $\nearrow$ instance.}

\myglsentry{Instantiation data evaluator}
\myglsdesc{part of the $\nearrow$ placeholder-plugin, which on request by the $\nearrow$ template engine issues $\nearrow$ instantiation data.}

\myglsentry{Instantiation Data}
\myglsdesc{denotes data sources, which are bound during an $\nearrow$ instantiation to $\nearrow$ slots.}

\myglsentry{Interleaving (Matching Rule)}
\myglsdesc{in terms of $\nearrow$ RelaxNG, denotes a non-deterministic matching rule.}

\myglsentry{Interpretation}
\myglsdesc{generally denotes the codomain of a function. $\nearrow$ Instances may be considered as an interpretation of a $\nearrow$ template.}

\myglsentry{Java Server Pages}
\myglsdesc{a $\nearrow$ template-language.}

\myglsentry{Canonisation}
\myglsdesc{recursively sorts all attributes within element nodes ascending by name.}

\myglsentry{Kleene's star operator}
\myglsdesc{denotes the star operator within regular expressions.}

\myglsentry{Combinator}
\myglsdesc{denotes an $\nearrow$ abstraction in the $\lambda$-calculus that has no free variables.}

\myglsentry{Commando-Tag}
\myglsdesc{denotes all tags in $\nearrow$ template-languages, which control the instantiation.}

\myglsentry{Competing}
\myglsdesc{Two non-terminals compete with each other if their sets of possible beginnings share at least one terminal.}

\myglsentry{Left-recursion}
\myglsdesc{causes the instantiation of a macro does not terminate.}

\myglsentry{Literal}
\myglsdesc{denotes text nodes and childless element nodes.}

\myglsentry{Local Tree Grammar}
\myglsdesc{is the weakest of all considered $\nearrow$ regular tree grammars, in which a terminal does not appear in any other rule.}

\myglsentry{Macro}
\myglsdesc{denotes in $\nearrow$ XTL a $\nearrow$ symbol.}

\myglsentry{Tag}
\myglsdesc{is for the distinction of text.}

\myglsentry{Matcher}
\myglsdesc{$\nearrow$ Graph-Matcher.}

\myglsentry{Memoisation}
\myglsdesc{denotes the $\nearrow$ call-by-need evaluation in programming languages.}

\myglsentry{Model-View-Controller}
\myglsdesc{architectural principle for the separation of concerns between model, view and control.}

\myglsentry{Monad}
\myglsdesc{denotes in Haskell an algebraic data type. Lists with \lit{\texttt{[]}} as neutral element represents a monad w.r.t. the associative operation \lit{$\pplus$}.}

\myglsentry{Pattern Style}
\myglsdesc{schema language whose syntax can be described by a regular expression -- rather than $\nearrow$ in a grammar style}.

\myglsentry{Non-deterministic Finite Automaton}
\myglsdesc{automaton, which recognises regular expressions and has a finite number of states.}

\myglsentry{Non-monotone and monotone operators}
\myglsdesc{Non-monotone operators alter the structure of passed (fragments of) documents. Monotone operators keep passed documents as is or extend those.}

\myglsentry{Non-strict Functions}
\myglsdesc{functions, which accept indefinite data structures as argument and do terminate after a finite leap of time.}

\myglsentry{OBDD}
\myglsdesc{ordered binary decision tree.}

\myglsentry{Partial Derivatives Algorithm}
\myglsdesc{Successive construction of an NFA from a regular expression. Partial derivatives for a particular terminal-symbol are evaluated $\nearrow$ in call-by-need mode.}

\myglsentry{Permutation}
\myglsdesc{$\nearrow$ commando-tag that swaps nodes.}

\myglsentry{Path Problem}
\myglsdesc{Theoretic questions regarding paths in a given graph.}

\myglsentry{Placeholder-Plugin}
\myglsdesc{a module evaluating specific requests formulated in a $\nearrow$ command language.}

\myglsentry{Precedence (of an operator)}
\myglsdesc{synonym for inverse operator priority. Operators with the lowest precedence have the highest priority over other operators.}

\myglsentry{Processing-Instruction}
\myglsdesc{XML-nodes containing a non-functional, but descriptive annotation to an XML node.}

\myglsentry{Source Language}
\myglsdesc{Formal Language $\nearrow$ instantiation data has to obey.}

\myglsentry{Redex}
\myglsdesc{denotes in $\lambda$-calculus terms, which may be reduced.}

\myglsentry{Referential Transparency}
\myglsdesc{A property that holds, when the evaluation of a function or an expression does not cause any side effects.}

\myglsentry{Regular Tree Automaton}
\myglsdesc{A tree automaton, that accepts regular $\nearrow$ tree languages.}

\myglsentry{Regular Tree Grammar}
\myglsdesc{A regular formal grammar whose terminal symbols extend to trees.}

\myglsentry{Regular Tree Language}
\myglsdesc{A formal language, which is generated by a $\nearrow$ regular tree.}

\myglsentry{RelaxNG}
\myglsdesc{XML-schema language, also see $\nearrow$ XSD.}

\myglsentry{Schema}
\myglsdesc{specifies an XML dialect for checking validity of an XML document.}

\myglsentry{Transformation}
\myglsdesc{Transforms a XML-document obeying one XML-schema to another XML-document possibly obeying another XML schema.}

\myglsentry{Schematron}
\myglsdesc{a XML schema language.}

\myglsentry{Safe Commands}
\myglsdesc{special $\nearrow$ commando-tags, which, when added to a certain schema, do not violate the closure property under union, intersection, minus. See also $\nearrow$ Syntactic Sugar.}

\myglsentry{Single-Type Grammar}
\myglsdesc{$\nearrow$ Regular tree grammar whose non-terminals inside of $\nearrow$ hedges do not compete with each other.}

\myglsentry{Slot}
\myglsdesc{part of a $\nearrow$ template, which is substituted/filled during $\nearrow $instantiation.}

\myglsentry{Slot-Markup Language}
\myglsdesc{$\nearrow$ template-language.}

\myglsentry{Structured Query Language}
\myglsdesc{the most popular standardised database query language.}

\myglsentry{String-grammar}
\myglsdesc{grammar whose derivations always generate words, also see $\nearrow$ word problem.}

\myglsentry{Stylesheet}
\myglsdesc{$\nearrow$ XSLT-$\nearrow$ template-document.}

\myglsentry{Substitution group}
\myglsdesc{denotes a $\nearrow$ symbol in $\nearrow$ XSD.}

\myglsentry{Super-combinator}
\myglsdesc{A specialised $\nearrow$ combinator, with all containing $\nearrow$ abstractions being super-combinator again. Combinators in imperative/procedural languages lead to modular programs. The introduction of new bindings, for example, turns $\lambda w. (\lambda x.x w)$ into $\lambda w.  (\lambda x y. x y) w$.}

\myglsentry{Symbol}
\myglsdesc{synonym for a symbol binding to a $\nearrow$ hedge.}

\myglsentry{Syntactic Sugar}
\myglsdesc{Constructs or idioms of a programming language improve usability, but they can be replaced by more complicated constructs from the same language.
Sugar does not extend functionality.
It may only increase expressibility. Idioms worsening the expressibility of a language are called Syntactic Salt.}

\myglsentry{Template}
\myglsdesc{a XML-document containing $\nearrow$ slots. It is used for $\nearrow$ instantiation and obeys the rules of a $\nearrow$ template-language.}

\myglsentry{Template Engine}
\myglsdesc{Software, which instantiates $\nearrow$ templates.}

\myglsentry{Template-Expansion}
\myglsdesc{$\nearrow$ instantiation.}

\myglsentry{Template-Language}
\myglsdesc{is described by $\nearrow$ command-tags, text nodes, element nodes.}

\myglsentry{Term-Evaluator}
\myglsdesc{Part of $\nearrow$ placeholder-plugins providing the $\nearrow$ template engine with formatted instantiation data.}

\myglsentry{Tracing}
\myglsdesc{tracking down errors by using a stack.}

\myglsentry{Type}
\myglsdesc{denotes a constraints on input and output parameters of a function in Haskell.}

\myglsentry{Typing Problem}
\myglsdesc{Theoretical question, if for a given $\lambda$.term $e$ a type $t$ may be inferred, s.t. $e::t$.}

\myglsentry{Type Isomorphism}
\myglsdesc{Equality of two $\nearrow$ types, allowing only renaming of type variables.}

\myglsentry{Type Constructors}
\myglsdesc{serve in Haskell for algebraic data type definitions.}

\myglsentry{Validation}
\myglsdesc{checks whether for a given $\nearrow$ template instantiation with previously unknown $\nearrow$ instantiation data returns a document obeying a given schema.}

\myglsentry{Vocabulary}
\myglsdesc{domain of valid XML namespaces.}

\myglsentry{Word Problem}
\myglsdesc{\nocite{InfDud:1993}  Theoretical question, if a given word is element of a formal language generated by a $\nearrow$ String-grammar.}

\myglsentry{XML-Entity}
\myglsdesc{denotes a special character in XML.}

\myglsentry{JXPath}
\myglsdesc{an XPath-reference implementation for Java.}

\myglsentry{XSD}
\myglsdesc{a XML-schema language, also see $\nearrow$ RelaxNG.}

\myglsentry{XSLT}
\myglsdesc{is a XML $\nearrow$ template-language.}

\myglsentry{Target Language}
\myglsdesc{denotes the language of all $\nearrow$ instance documents obtained after $\nearrow$ instantiation.}

\myglsentry{Membership-Function}
\myglsdesc{denotes a discrete mapping between domain and a real value between 0 and 1.}

\newpage


\section{APPENDIX A: Denotational Semantics}
\label{appendix:DenotationalSemantics}

\begin{large}\textbf{Instantiation}\end{large}\\

Source: \texttt{https://rhaber123.\-github\-.io/web-page/}

\begin{center}
  \parbox{8.9cm}{\input{SemanticInst}}
  \label{appendix:DenotationalSemanticInstantiation}
\end{center}

\newpage

\begin{large}\textbf{Validation}\end{large}\\

\begin{center}
  \parbox{8.7cm}{\input{SemanticVal}}
\end{center}

\newpage


\begin{center}
  \parbox{8.5cm}{\input{SemanticVal2}}
\end{center}


\newpage
\section{APPENDIX B: Partial-Derivatives algorithm}
\label{appendix:PartialDerivativesAlgorithm}

\begin{flushright}
\begin{tiny}
(Source: \cite{Ant:00})
\end{tiny}
\end{flushright}

Given: Regular expression $t=x^{*} \cdot \underbrace{(x\cdot x + y)^{*}}_{r}$ \\
To be found: NFA with $L(t)$?\\\\

Step 1: Determine linear form
\begin{flushleft}
\begin{tabular}{lcr}
	$lf(t)$ & = & $\underline{lf(x^{*})} \odot r \cup lf(r)  \nonumber$ \\
	   & = & $(\underline{lf(x)} \odot x^{*}) \odot r \cup lf(r)  \nonumber$ \\
	   & = & $\underline{(\{<x,\lambda>\} \odot x^{*})} \odot r \cup lf(r)  \nonumber$ \\
	   & = & $\underline{(\{<x,x^{*}>\}) \odot r} \cup lf(r)  \nonumber$ \\
	   & = & $\{<x,\underline{x^{*} \cdot r}>\} \cup lf(r)  \nonumber$ \\
	   & = & $\{<x,t>\} \cup lf(r)  \nonumber$ \\
	   & = & $\{<x,t>, <x,x \cdot r>, <y, r>\} \nonumber$
\end{tabular}
\end{flushleft}

\begin{flushleft}
\begin{tabular}{lcr}
	$lf(r)$ & = & $\underline{lf(x \cdot x + y)} \odot r \nonumber$ \\
		  & = & $(\underline{lf(x \cdot x)} \cup lf(y)) \odot r \nonumber$ \\
		  & = & $((lf(x) \odot x) \cup \underline{lf(y)}) \odot r \nonumber$ \\
		  & = & $((\underline{lf(x)} \odot x) \cup \{<y,\lambda>\}) \odot r \nonumber$ \\
		  & = & $(\underline{(<x,\lambda> \odot x)} \cup \{<y,\lambda>\}) \odot r \nonumber$ \\
		  & = & $\underline{(\{<x,x>\} \cup \{<y,\lambda>\})} \odot r \nonumber$ \\
		  & = & $\underline{\{<x,x>, <y,\lambda>\} \odot r} \nonumber$ \\
		  & = & $\{<x,x \cdot r>, <y,r>\} \nonumber$
\end{tabular}
\end{flushleft}

\begin{flushleft}
\begin{tabular}{lcr}
  $lf(x \cdot r)$ & = & $\underline{lf(x)} \odot r \nonumber$ \\
  & = & $\underline{\{<x,\lambda>\} \odot r} \nonumber$ \\
  & = & $\{<x,r>\} \nonumber$
\end{tabular}
\end{flushleft}

All linear forms are determined now for the second component.

Step 2: Apply the Partial-Derivatives algorithm:\\

\begin{tabular}{lcl}
  $<PD_{0}, \Delta_{0}, \tau_{0}>$ & := & $< \emptyset, \{t\}, \emptyset>$\\
  $PD_{1}$ & := & $PD_{0} \cup \Delta_{0}$ = $\{t\}$\\
  $\Delta_{1}$ & := & \parbox[t]{3.9cm}{$\cup_{p \in \Delta_{0}} \{ q \ |\ <x,q> \ \in \ lf(p) \ \wedge q \notin PD_{1}\}$ \ = \ $\{x \cdot r,r\}$}\\
  $\tau_{1}$ & := & \parbox[t]{3.9cm}{$\tau_{0} \cup \{<p,x,q> | p \in \Delta_{0} \wedge <x,q> \in lf(p)\}$}\\
  & = & \parbox[t]{3.9cm}{$\{<t,x,t>, <t,x,x \cdot r>, <t,y,r>\}$}\\\\
\end{tabular}

\begin{tabular}{lcl}
  $<PD_{1}, \Delta_{1}, \tau_{1}>$ & := & \parbox[t]{3.9cm}{$< \{t\},\ \{x \cdot r,r\},\ \{<t,x,t>,\ <t,x,x \cdot r>,\ <t,y,r>\} >$} \\
  $PD_{2}$ & = & $\{t, x \cdot r, r\}$ \\
  $\Delta_{2}$ & = &$\emptyset$\\
$\tau_{2}$ & = & \parbox[t]{3.9cm}{ $\{<t,x,t>, <t,x,x \cdot r>, <t,y,r>, <x \cdot r, x, r>, <r,x,x \cdot r>,$ \\ $\ <r,y,r>\}$ }\\\\  
\end{tabular}

\begin{tabular}{lcl}
  $<PD_{2}, \Delta_{2}, \tau_{2}>$ & := & \parbox[t]{3.9cm}{$< \{t,x \cdot r, r\}, \emptyset, \tau_{2} >$} \\
  $PD_{3}$ & = & $\{t, x \cdot r, r\}$ \\
  $\Delta_{3}$ & = & $\emptyset$\\
  $\tau_{3}$ & = & $\tau_{2} \rightarrow Halt!$\\\\
\end{tabular}

Final states:

\qquad $F\subseteq PD_{3}$ := $\{f|f \in PD_{3} \wedge f \in \tau_{1}\}$ = $\{r\}$\\

Remaining states:

\qquad $PD_{3} \setminus F$ = $\{t, x \cdot r\}$\\\\

Step 3: Building NFA:\\

\begin{minipage}{2cm}
  \xymatrixrowsep{30pt}
  \xymatrixcolsep{30pt}
  \xymatrix{
.   \ar[r] & *++[o][F-]\txt{$q_0$} \ar[dr]^y \ar[rr]^x^<(0.7){\fbox{$x \cdot r$}}^<(0.1){\fbox{$t$}} \ar@(l,d)[]_x && *++[o][F-]\txt{$q_1$}  \ar@/l1pc/[dl]_{x}\\
    && *++[o][F=]{\txt{$q_2$}} \ar@(l,d)[]_y_<(3){\fbox{$r$}}  \ar@/r1pc/[ru]_{x}
  }
\end{minipage}

\begin{flushleft}
 \begin{small}
   \parbox{10cm}{
 $\tau_{Reg0}$ .. terms that do \underline{not} contain $\varepsilon$\\
 $\tau_{Reg1}$ .. terms that do contain $\varepsilon$}
 \end{small}
\end{flushleft}


\end{document}

%% file: SemanticInst.tex

(S) \quad $\eval[Start]{\var{x}}{\var{s} \pi}$ :=
 $\eval{\var{x2}}{\var{s} \pi}$ \for \var{x2} = $\eval[r]{\var{x}}{}$ \\

(E) \quad \parbox[t]{14cm}{$\eval[r]{ElX \var{n} \var{a} \var{c}}{}$ := \\
\mbox{\quad \parbox{10cm}{
 \<let> \parbox[t]{14cm}{
		\var{attDefs} = $\func{filter}{2} ( \lambda \var{child}. \eval[MA]{\var{child}}{}) \var{c}$, \\
        \var{nodes} = $\func{filter}{2} ( \lambda \var{child}. \func{not}{1} \; \eval[MA]{\var{child}}{}) \var{c}$} \\
 \<in> \texttt{ElX \var{n} $(\func{qSort}{1} (\var{a} \pplus \var{attDefs})) \; \var{nodes}$}}
}}\\

(I1) \quad $\eval{XTxt \var{t}}{\_ \; \_}$ :=
  \texttt{XTxt \var{t}}
  
(I2) \quad $\eval{XAtt \var{n} \var{v}}{\_ \; \_}$ :=
  \texttt{XAtt \var{n} \var{v}}
  
(I3) \quad \parbox[t]{14cm}{$\eval{ElX \var{n} \var{a} \var{c}}{\var{s} \pi}$ := \\
\mbox{\quad \parbox{10cm}{
 \<let> \parbox[t]{14cm}{
		 \var{mdefs} = $\func{filter}{2} (\lambda \var{child}.\eval[MM]{\var{child}}{}) \var{c}$, \\
         \var{nodes} = $\func{filter}{2} (\lambda \var{child}.\func{not}{1} \; \eval[MM]{\var{child}}{}) \var{c}$} \\
  \<in> \texttt{ElX \var{n} \var{a}\\ $(\func{concatMap}{2} (\lambda \var{node}.\eval[\alpha]{\var{node}}({\var{s}, \var{mdefs}, \pi})) \;\\ \var{nodes})$}}
}}\\

(A1) \quad \parbox[t]{14cm}{$\eval[\alpha]{XIf \var{x} \var{gc}}{(\var{s},\mu,(\func{f}{1}_{0},\func{f}{2}_{0},\func{f}{3}_{0},\func{f}{4}_{0}))}$ :=\\ 
\mbox{\quad \parbox{10cm}{
	\<if> $(\func{f}{3}_{0} \var{x} \var{s})$ \<then>\\ $\func{concatMap}{2} (\lambda \var{c}.\eval[\alpha]{\var{c}}{}(\var{s},\mu,(\func{f}{1}_{0},\func{f}{2}_{0},\func{f}{3}_{0},\func{f}{4}_{0}))) \var{gc}$
  \\ \<else> \texttt{[ ]}}
}}\\
  
(A2) \quad \parbox[t]{14cm}{$\eval[\alpha]{XForEach \var{x} \var{gc}}{(\var{s},\mu,(\func{f}{1}_{0},\func{f}{2}_{0},\func{f}{3}_{0},\func{f}{4}_{0}))}$ := \\
\mbox{\quad  \parbox{10cm}{
 \<let> \var{sels} = $\func{f}{2}_{0} \var{x} \; \var{s}$ \\
 \<in> $\func{concatMap}{2} \; (\func{f3}{1}) \; \var{sels}$ \\
 \for $\func{f3}{1} = \lambda \var{c}.\\ \func{concatMap}{2} (\lambda \var{c2}. \eval[\alpha]{\var{c2}}{(\var{c}, \mu, (\func{f}{1}_{0},\func{f}{2}_{0},\func{f}{3}_{0},\func{f}{4}_{0}))}) \var{gc}$}
}}\\

(A3) \quad \parbox[t]{14cm}{$\eval[\alpha]{XCallMacro \var{m1}}{(\var{s},\mu,\pi)}$ := \\
\mbox{\quad  \parbox{10cm}{
 \<let> $\mu_{2}$ = $\func{getMacro}{1} \mu$ \\
 \<in> $\func{concatMap}{2} (\lambda \var{m}. \eval[\alpha]{\var{m}}{(\var{s},\mu,\pi)}) \; \mu_{2}$ \\
 \<where> \parbox[t]{7cm}{ 
    \texttt{getMacro [ ] = []} \\
    \texttt{getMacro (XCallMacro \var{m2} \var{c}):\var{xs}\\ | (\var{m1} == \var{m2}) = \var{c}} \\ \<otherwise> \texttt{getMacro xs}}}
}}\\

(A4) \quad $\eval[\alpha]{XTxt \var{t}}{(\var{s},\mu,(\func{f}{1}_{0},\_,\_,\_))}$ := \\
  \mbox{\qquad \texttt{[ XTxt $\func{f}{1}_{0} \; \var{t} \; \var{s}$ ]}}

(A5) \quad $\eval[\alpha]{XInclude \var{x}}{(\var{s},\mu,(\_,\_,\_,\func{f}{4}_{0}))}$ := \\
  \mbox{\qquad \texttt{[ $\func{f}{4}_{0} \; \var{x} \; \var{s}$ ]}}

(A6) \quad \parbox[t]{14cm}{$\eval[\alpha]{ElX \var{n} \var{a} \var{c}}({\var{s},\mu,\pi)}$ := \\
\mbox{\qquad  \parbox{10cm}{\texttt{[ ElX \var{n} \var{a}\\ $(\func{concatMap}{2} (\lambda \var{child}. \eval[\alpha]{\var{child}}{(\var{s},\mu,\pi)}) \var{c})$\\ ]}}
}}\\
  
(A7) \quad $\eval[\alpha]{TxtX \var{t}}{(\_,\_,\_)}$ := \texttt{[ TxtX \var{t} ]} \\

%% file: SemanticVal.tex


(S) \quad $\eval[Start]{\var{x}}{\var{s} \pi}$ :=
 $\eval{\var{x2}}{\var{s} \pi}$ \for \var{x2} = $\eval[r]{\var{x}}{}$\\

(E1) \quad $\eval{Epsilon, TxtR "\/"}{\mu}$ :=
 \texttt{True}
 
(E2) \quad $\eval{Epsilon, TxtR \_}{\mu}$ :=
 \texttt{False}
 
(E3) \quad $\eval{Epsilon, Epsilon}{\mu}$ :=
 \texttt{True}

(E4) \quad $\eval{Epsilon, ElR \_ \_ \_}{\mu}$ :=
 \texttt{False}

(E5) \quad $\eval{Epsilon, Star \_}{\mu}$ :=
 \texttt{True}

(E6) \quad $\eval{Epsilon, TextR \_}{\mu}$ :=
 \texttt{True}

(E7) \quad $\eval{Epsilon, Then \var{r1} \var{r2}}{\mu}$ := \\
\mbox{\qquad \qquad \texttt{$\eval{Epsilon, \var{r1}}{\mu} \wedge \eval{Epsilon, \var{r2}}{\mu}$}}\\\\

(Then1) \quad $\eval{Then \_ \_, Epsilon}{\mu}$ :=
 \texttt{False} \\[-0.3cm]

(Then2) \quad \parbox[t]{14cm}{$\eval{Then \var{r1} \var{r2}, TxtR \var{text}}{\mu}$ := \\
\mbox{\texttt{$\eval{\var{r1}, TxtR \var{text}}{\mu}$}}\\
$\wedge$
\mbox{$\eval{\var{r2}, Epsilon}{\mu}$}
}\\

(Then3) \quad \parbox[t]{14cm}{
$\eval{\parbox[t]{4.7cm}{Then\\ (ElR \var{name1} \var{atts1} \var{r1}) \var{r}$,\\ $ElR \var{name2} \var{atts2} \var{r2}}}{\mu}$ := \\ 

\mbox{\texttt{$\eval{   \parbox[t]{4.1cm}{ElR \var{name1} \var{atts1} \var{r1}, ElR \var{name2} \var{atts2} \var{r2}}  }{\mu}$}}\\
$\wedge$ \ \mbox{$\eval{\var{r}, Epsilon}{\mu}$}
}\\

(Then4) \quad $\eval{Then \_ \_, ElR \_ \_ \_}{\mu}$ :=
 \texttt{False} \\[-0.3cm]

(Then5) \quad \parbox[t]{14cm}{$\eval{Then \var{r1} \var{r2}, Star \var{s}}{\mu}$ := \\ 
\mbox{\quad \parbox[t]{10cm}{
 \mbox{\texttt{$\vee$[True|}} \\\mbox{\texttt{(\var{s1},\var{s2})$\leftarrow \func{frontSplits}{1}$(Then \var{r1} \var{r2})}}\\
$\wedge \ \eval{\var{s1}, \var{s}}{\mu}$\\
$\wedge \ \eval{\var{s2}, Star \var{s}}{\mu}$ \ ]}
}} \\

(Then6) \quad $\eval{\parbox[t]{3.5cm}{Then \var{r1} Epsilon,\\ Then \var{s1} \var{s2}}}{\mu}$ :=\\
\mbox{\qquad \qquad \qquad \quad $\eval{\var{r1}, Then \var{s1} \var{s2}}{\mu}$} \\[-0.3cm]

(Then 7) \quad \parbox[t]{14cm}{$\eval{Then \var{r1} \var{r2}, Then \var{s1} \var{s2}}{\mu}$ := \\ 
\mbox{\qquad \parbox[t]{6cm}{
 \texttt{$\vee$[True|\\(\var{t1},\var{t2})$\leftarrow \func{splits}{1}$(Then \var{r1} \var{r2})\\ \mbox{$\wedge \ \eval{\var{t1}, \var{s1}}{\mu} \ \wedge \ \eval{\var{t2}, \var{s2}}{\mu}$]}}}
}}\\

(Then 8) \quad $\eval{ \parbox[t]{4.7cm}{Then (TxtR \_) Epsilon,\\ TextR \_}}{\mu}$ :=\\
\mbox{\qquad \qquad \qquad \texttt{True}}\\[-0.3cm]

(Then 9) \quad $\eval{Then \_ \_, TextR \_}{\mu}$ :=\\
\mbox{\qquad \qquad \qquad \texttt{False}}\\


($\Phi$) \quad $\eval{\var{inst}, MacroR \var{mname}}{\mu}$:=\\
\parbox[t]{11cm}{
\mbox{\qquad \qquad \texttt{$\eval{\var{inst},\var{word}}{\mu}$}} \\
\mbox{\qquad \qquad \for \texttt{\var{word} = $\func{getMacro}{2}$ mname $\mu$}}}\\\\
 
($\Omega$) \quad $\eval{\var{inst}, Or \var{r1} \var{r2}}{\mu}$:=\\
\mbox{\qquad \qquad \texttt{$\eval{\var{inst}, \var{r1}}{\mu} \vee \eval{\var{inst}, \var{r2}}{\mu}$}}

%% file: SemanticVal2.tex

 (\#1) \quad \parbox[t]{14cm}{$\eval{TxtR \var{text}, Then \var{r1} \var{r2}}{\mu}$ :=\\
\mbox{\qquad \parbox{10cm}{
\texttt{$\vee[\ True\ |\\
(\var{s1},\var{s2})\leftarrow \func{splitText}{1} \var{text}\\
\wedge \ \eval{TxtR \var{s1}, \var{r1}}{\mu}$\\ \mbox{$\wedge \ \eval{TxtR \var{s2}, \var{r2}}{\mu}\ ]$}}}}}\\

(\#2) \quad $\eval{TxtR "\/", Epsilon}{\mu}$ :=
 \texttt{True}

(\#3) \quad $\eval{TxtR \_, Epsilon}{\mu}$ :=
 \texttt{False}

(\#4) \quad $\eval{TxtR "\/", Star \_}{\mu}$ :=
 \texttt{True}
         
(\#5) \quad \parbox[t]{14cm}{$\eval{TxtR \var{text}, Star \var{r}}{\mu}$ := \\ 
\mbox{\qquad \parbox{11cm}{
\<if> \texttt{($\vee$[\ True\ |\\(\var{s1},\var{s2})$\leftarrow \func{frontSplitText}{1} \var{text}\\
\wedge \ \eval{TxtR \var{s1}, \var{r}}{\mu}\\
\wedge \ \eval{TxtR \var{s2}, Star \var{r}}{} $] == True)\\
\<then> True \\
\<else> $\eval{TxtR \var{text}, Epsilon}{\mu}$}} \\
}}\\

(\#6) \quad $\eval{TxtR \var{text}, TextR \_}{\mu}$ :=
 \texttt{True}
  
(\#7) \quad $\eval{TxtR \var{text1}, TxtR \var{text2}}{\mu}$ :=\\
\mbox{\qquad \qquad \quad \texttt{text1 == text2}}
  
(\#8) \quad $\eval{TxtR \var{text}, ElR \_ \_ \_}{\mu}$ :=
 \mbox{\texttt{False}}\\

(ElR1) \quad \parbox[t]{5cm}{$\eval{\parbox[t]{4.1cm}{ElR \var{name1} \var{atts1} \var{r1},\\ ElR \var{name2} \var{atts2} \var{r2}}}{\mu}$ := \\ 
\mbox{\qquad \parbox{11cm}{
	\texttt{$(\var{name1} == \var{name2})\\
\wedge (\func{qSort}{1} \var{atts1} == \var{atts3})\\
\wedge \eval{\var{r1}, \var{r3}}{\mu}$}\\
          \for \parbox[t]{4.3cm}{\texttt{(ElR \_ \var{atts3} \var{r3}) =\\$\func{extractAttributes}{1}$\\ (ElR \var{name2} \var{atts2} \var{r2})}}}
}}\\

(ElR2) \quad \parbox[t]{5.5cm}{$\eval{\parbox[t]{4.5cm}{ElR \var{name1} \var{atts1} \var{r1}, Then (ElR \var{name2} \var{atts2} \var{r2}) \var{s}}}{\mu}$:=\\ 
\mbox{\qquad \parbox{11cm}{
 \texttt{$\eval{\parbox[t]{4.3cm}{ElR \var{name1} \var{atts1} \var{r1}, ElR \var{name2} \var{atts2} \var{r2}}}{\mu}\\
\wedge \eval{Epsilon, \var{s}}{\mu}$}}
}}\\
          
(ElR3) \quad \parbox[t]{5.5cm}{$\eval{\parbox[t]{4.5cm}{ElR \var{name1} \var{atts1} \var{r1}, Then (Or \var{s1} \var{s2}) \var{s}}}{\mu}$:= \\
\mbox{\qquad \parbox{11cm}{
 \texttt{$(\eval{\parbox[t]{4.5cm}{ElR \var{name1} \var{atts1} \var{r1},\\ Or \var{s1} \var{s2}}}{\mu}\\
\wedge \eval{Epsilon, \var{s}}{\mu})\\
\vee (\eval{Epsilon, Or \var{s1} \var{s2}}{\mu}\\
\wedge \eval{ElR \var{name1} \var{atts1} \var{r1}, \var{s}}{\mu})$}}\\
}}\\

(ElR 4) \quad \parbox[t]{5.5cm}{$\eval{\parbox[t]{4.5cm}{ElR \var{name1} \var{atts1} \var{r1}, Then (Star \var{s1}) \var{s}}}{\mu}$:= \\
\mbox{\qquad \parbox{11cm}{
\texttt{$(\eval{\parbox[t]{4.7cm}{ElR \var{name1} \var{atts1} \var{r1}, \var{s1}}}{\mu}\\
\wedge \eval{Epsilon, \var{s}}{\mu}) \ \vee \\ \eval{ElR \var{name1} \var{atts1} \var{r1}, \var{s}}{\mu}$}}
}}\\

(ElR 5) \quad \parbox[t]{5.5cm}{$\eval{\parbox[t]{4.5cm}{ElR \var{name1} \var{atts1} \var{r1},\\ Then (MacroR \var{m}) \var{s}}}{\mu}$:=\\
\mbox{\qquad \parbox{11cm}{
 \texttt{$(\eval{\parbox[t]{4.5cm}{ElR \var{name1} \var{atts1} \var{r1},\\ MacroR \var{m}}}{\mu}\\
\wedge \eval{Epsilon, \var{s}}{\mu}) \vee \\
(\eval{Epsilon, MacroR \var{m}}{\mu}\\
\wedge \eval{ElR \var{name1} \var{atts1} \var{r1}, \var{s}}{\mu})$}}
}}\\

(ElR6) \quad $\eval{ElR \_ \_ \_, Then \_ \_}{\mu}$ :=
 \mbox{\texttt{False}}

(ElR7) \quad $\eval{ElR \var{name} \var{atts} \var{r}, Star \var{s}}{\mu}$:=\\
 \mbox{\qquad \qquad \qquad \texttt{$\eval{ElR \var{name} \var{atts} \var{r}, \var{s}}{}$}}

(ElR8) \quad $\eval{ElR \_ \_ \_, TextR \_}{\mu}$ :=
 \texttt{False}

(ElR9) \quad $\eval{ElR \_ \_ \_, Epsilon}{\mu}$ :=
 \texttt{False}

(ElR10) \quad $\eval{ElR \_ \_ \_, TxtR \_}{\mu}$ :=
 \texttt{False}